\documentclass[preprint,12pt]{elsarticle}

\usepackage{amssymb}
\usepackage{amsmath}
\usepackage{amsthm}
\usepackage{graphicx} 
\usepackage{url}
\usepackage{amsmath}
\usepackage{listings}
\usepackage{multicol}
\usepackage{subcaption}
\usepackage{tikz}
\usetikzlibrary{arrows.meta, positioning}
\usepackage{xcolor} 

\journal{Journal of Computational Physics}

\begin{document}

\begin{frontmatter}



\title{A Stochastic Conservative Field Transfer Method for Black-box Multiscale and Multiphysics Coupling}
\author[a]{Abhiyan Paudel}
\author[b]{Cameron W. Smith}
\author[a,b]{Jacob S. Merson\corref{cor1}}

\cortext[cor1]{Corresponding author\\\textit{E-mail address:} mersoj2@rpi.edu}

\affiliation[a]{organization={Department of Mechanical, Aerospace, and Nuclear Engineering, Rensselaer Polytechnic Institute},
                addressline={110 8th St.},
                city={Troy},
                state={NY},
                postcode={12180},
                country={USA}}

\affiliation[b]{organization={Scientific Computation Research Center, Rensselaer Polytechnic Institute},
                addressline={110 8th St.},
                city={Troy},
                state={NY},
                postcode={12180},
                country={USA}}

\begin{abstract}
   This paper introduces a new method for performing field transfer operations in black-box coupling, when source discretization information is not available. This approach uses a stochastic approximation of the Galerkin projection which leads to a method that asymptotically provides conservation. Error in the accuracy and conservation has been compared to the mesh intersection method and radial basis functions on a simple domain, as well as on meshes of the LTX fusion reactor. For all cases tested, our new method provides higher accuracy and less conservation error than radial basis functions and can be used for black-box coupling, unlike the mesh-intersection method. Additionally, we demonstrate the implementation and performance of our method on an NVIDIA GeForce RTX 4060 GPU, showing that the cost is competitive with the mesh intersection method.
\end{abstract}




\end{frontmatter}

\section{Introduction} \label{sec:Intro}
Transferring field data between disparate discretizations is a critical component of multiscale and multiphysics workflows \cite{keyesMultiphysicsSimulationsChallenges2013} such as Fluid Structure Interactions (FSI) \cite{deboerComparisonConservativeConsistent2008,jaimanAssessmentConservativeLoad2005}, the Arbitrary Lagrangian-Eulerian (ALE) formulation \cite{shashkovRemappingMeshesIsoparametric}, and mesh adaptation \cite{ibanezCONFORMALMESHADAPTATION}. It has also been observed that there are stability and accuracy benefits to preserving physical constraints such as conservation of mass, energy, and momentum while performing field transfers \cite{deboerComparisonConservativeConsistent2008,Slattery2016,adamsProjectionMethodParticle2026}. Many commonly used field transfer techniques, such as interpolation, radial basis functions, and projections based on integrals over source or target elements, rely solely on pointwise field evaluations but generally fail to conserve integral quantities, such as mass or energy \cite{JiaoHeath2004,farrellConservativeInterpolationUnstructured2009}. In contrast, mesh intersection or supermesh-based methods provide high-fidelity conservative transfer but require explicit access to full source and target discretizations, limiting their applicability in black-box workflows \cite{Slattery2016,JiaoHeath2004,farrellConservativeInterpolationUnstructured2009,FarrellMaddison2011}. The need for black-box coupling arises naturally in workflows that seek to couple machine-learned surrogate models, which typically do not expose a traditional discretization, as well as in multiscale and multiphysics coupling frameworks.

The present work introduces a new strategy for performing projection-based conservative remapping relying only on point-wise field queries while maintaining similar levels of accuracy and conservation to mesh intersection methods. To accomplish this, we utilize a control variate Monte Carlo integration strategy that can avoid the assumptions of continuity that drive the use of mesh intersections. This work was inspired by our recent efforts to support unstructured mesh tallies for neutronics simulations \cite{hasanGPUAccelerationMonte2025, mersonSpatiallyContinuousFunctional}. We further note that the control variate is similar to integration with a two-level Multilevel Monte Carlo scheme \cite{gilesMultilevelMonteCarlo}; however, in this work we choose a cheap, correlated control function whose exact expectation can be computed with Gaussian quadrature.

This work enables conservative multiscale and multiphysics coupling in black-box coupling scenarios when source mesh topology is not available. Furthermore, it is easy to implement and can be used naturally on GPUs or other data-parallel systems. To place our method in the context of the literature, the remainder of this section provides a brief review of related field mapping methods.

Field remapping techniques can largely be categorized into those that carry out pointwise evaluations (e.g., nearest neighbor, or evaluating FEM interpolating polynomials), local reconstruction methods (e.g., radial basis functions or patch-based recovery), and variational methods such as the mesh intersection or supermesh methods. Pointwise evaluation methods are commonly deployed because the only additional component needed for coupling is point localization routines that can map global coordinates to local ones \cite{chourdakisPreCICEV2Sustainable2022,slatteryDataTransferKit2013, novakCoupledMonteCarlo2022}.

One of the main advantages of pointwise evaluation methods is that they are often computationally inexpensive.  For example, in a finite element library field evaluation is a standard operation that is routinely the focus of optimization. The main cost associated with direct pointwise evaluation is point localization, however if there is infrequent mesh adaptation, this cost can be amortized over coupling iterations. Additional advantages include that knowledge of the source discretization is only needed for localization and many applications provide APIs to perform evaluations based on global coordinates. A disadvantage of evaluation methods is that they do not provide a mechanism for preserving constraints, such as integral conservation \cite{Slattery2016,farrellConservativeInterpolationUnstructured2009}.

Local fitting methods such as radial basis functions and patch recovery methods have also found wide adoption for multiphysics coupling and error estimation \cite{chourdakisPreCICEV2Sustainable2022,zienkiewiczSuperconvergentPatchRecovery1992a,zienkiewiczSuperconvergentPatchRecovery1992b,bungartzPreCICEFullyParallel2016}. The main advantages of these methods is that they can be used in a black-box scenario. Additionally, these methods can be effectively parallelized on GPUs \cite{dingPerformanceEvaluationGPUAccelerated2018, schneiderDataParallelRadialBasisFunction2023, morricalAttributeAwareRBFsInteractive2023}, and can be used to provide a global conservation through the introduction of a  linear polynomial \cite{deboerComparisonConservativeConsistent2008,chourdakisPreCICEV2Sustainable2022}. In many problems of interest these local fitting methods require parameter tuning to obtain reasonable levels of accuracy and conservation.

Mesh intersection methods were introduced by Jiao and Heath in \cite{JiaoHeath2004} where they demonstrate a significant advantage when performing Galerkin projections over a common intersected mesh rather than over the source or target discretizations. The use of an intersected mesh removes approximation errors that arise from evaluating the integral of discontinuous functions with standard numerical integration schemes. 
These methods have been widely extended to efficiently support volume coupling \cite{farrellConservativeInterpolationUnstructured2009,FarrellMaddison2011} of unstructured meshes, polyhedral meshes, and curved meshes \cite{hermesHighorderSolutionTransfer2025}. Given the need for both the source and target discretization information, their applicability in black-box coupling is limited.  Likewise, the complexity of the intersection operations complicates GPU execution, and extending the methods to high-dimensional settings is prohibitive.

Key innovations described in this article include
\begin{itemize}
    \item Formulation of a conservative field transfer operator for multiscale and multiphysics coupling when only pointwise evaluation is available.
    \item Stochastic approximation of Galerkin projection with quantified conservation behavior.
    \item Comparison with mesh-intersection and interpolation-based field transfer methods.
    \item GPU-based implementation and performance characterization of the mesh-intersection, radial basis function, and Monte Carlo field transfer methods.
\end{itemize}

This paper is organized as follows: Section~\ref{sec:galerkin-projection} provides background on the Galerkin projection method. Section \ref{sec:mesh-intersection} describes our implementation of the mesh intersection algorithm. Section \ref{sec:monte-carlo-integration} gives an overview of our novel strategy for conservative coupling without source discretization information. Section \ref{sec:numerical-comparison} provides a set of computational experiments that describe the accuracy, conservation, and efficiency of our new method compared to the mesh intersection and radial basis function methods. Section \ref{sec:application} demonstrates the methods on fusion simulation data from the WEST tokamak. Lastly, section \ref{sec:conclusion} provides concluding remarks and future work.

\section{Conservative Galerkin Projection}\label{sec:galerkin-projection}
Let $\Omega \subset \mathbb{R}^d$ denote the physical domain of interest, 
where $d$ can be any dimension. We assume that the target discretization is a finite element mesh $\mathcal{M}_t$ with associated FE space $\mathcal{V}_t = \mathcal{V}(\mathcal{M}_t) 
   = \mathrm{span}\{\psi_i\}_{i=1}^{\mathcal{N}_t} \subset L^2(\Omega)$,
where $\mathcal{N}_t$ denotes the number of target degrees of freedom.
The target field $f^t \in \mathcal{V}_t$ is represented as
\begin{align}
    f^t(x) = \sum_{i=1}^{\mathcal{N}_t} f^{t}_{i}\, \psi_i(x), \qquad f^{t}_{i} \in \mathbb{R}.
\end{align}
On the source side, we do not assume a particular discretization. The field $f^s$ may be provided with discretization information (e.g., a source mesh $\mathcal{M}_s$) or accessed only through pointwise evaluations $f^s(x)$ for $x\in\Omega$. The conservative transfer seeks $f^t$ that is closest to $f^s$ in the $L^2$ norm:
\begin{equation}
  f^t = \arg\min_{f \in \mathcal{V}_t} \| f^s - f \|_{L^2(\Omega)} ,
  \label{eq:L2min}
\end{equation}
where $\| (\cdot)\|_{L^2(\Omega)}^2 = \int_\Omega (\cdot)^2\,\mathrm{d}\Omega$.
This $L^2$ minimization is equivalent to the Galerkin projection, obtained by requiring
the residual $f^t - f^s$ to be orthogonal to the target space,
\begin{equation}
  \langle f^t - f^s, \psi_k \rangle_{L^2(\Omega)} = 0,
  \qquad k = 1,\dots,\mathcal{N}_t,
  \label{eq:galerkin_cond}
\end{equation}
with the standard inner product $\langle u,v\rangle_{L^2(\Omega)} = \int_\Omega u\,v\,\mathrm{d}\Omega$.
Substituting the target expansion into Eq.~\eqref{eq:galerkin_cond} gives
\begin{align}
  \label{eq:galerkin_system}
  \mathbf{M}\,\mathbf{f}^t&=\mathbf{b}
\end{align}
where,
\begin{align}
  \label{eq:mass_matrix}
  M_{ki} &=\int_\Omega \psi_k\psi_i\,\mathrm{d}\Omega,  \text{ and}\\
  \label{eq:load_vector}
  b_k &=\int_\Omega f^s\psi_k\,\mathrm{d}\Omega.
\end{align}
Here, $\mathbf{M} \in \mathbb{R}^{\mathcal{N}_t \times \mathcal{N}_t}$ is the symmetric positive-definite sparse mass matrix in
$\mathcal{V}_t$, $\mathbf{f}^t = [f^{t}_{1}, \dots, f^{t}_{\mathcal{N}_t}]^T$ are the target coefficients, and
$\mathbf{b}$ contains the projection of the source field to the target basis. The key distinction is that $\mathbf M$ depends only on the target basis products, whereas $\mathbf b$ requires the product of the target basis and source fields.
If the constant function $1 \in \mathcal{V}_t$, then the conservation follows as 
\begin{align}
\label{eq:conservation}
  \int_\Omega f^t\,\mathrm{d}\Omega = \int_\Omega f^s\,\mathrm{d}\Omega.  
\end{align}
The formulation in Eq.~\eqref{eq:L2min}–\eqref{eq:conservation} is general for any dimension $d$ and provides
a conservative and $L^2$-optimal mapping from the source field to the target field. 

The mass matrix in Eq.~\eqref{eq:mass_matrix} involves the integral of the products of the target basis functions, which can be computed exactly on the target grid using Gaussian quadrature \cite{Dunavant1985,KEAST1986339}. In contrast, assembling the load vector in Eq.~\eqref{eq:load_vector} requires integrating the product of a target basis function and source field over $\Omega$, which is the primary challenge in conservative remapping. Next, we address the evaluation of the load vector using mesh intersections when source discretization information is available (Section~\ref{sec:mesh-intersection}) and sampling-based estimators in the absence of such information (Section~\ref{sec:monte-carlo-integration}). 

\section{Conservative Coupling with Discretization Information} \label{sec:mesh-intersection}

\subsection{Intersection-Based Galerkin Projection}
Several approaches have been proposed to compute the load vector using either source- or target-based discretization \cite{JiaoHeath2004}. However, these schemes can fail to simultaneously preserve accuracy and conservation in field transfer on non-matching meshes \cite{slatteryDataTransferKit2013}. This loss arises because integrating on one mesh requires evaluating quantities defined on the other mesh. In a source-based scheme, the target basis functions must be evaluated at source quadrature points, making the integrand piecewise-defined within each source element.
Conversely, integrating on the target mesh requires evaluating the source field at target quadrature points, making it piecewise-defined within each target element.
Because standard quadrature rules assume the integrand is sufficiently regular on each integration element, neither approach is suitable for accurately
constructing the load vector. This motivates integration over a common refinement or supermesh defined by geometric intersections of source and target elements.

In practice, the load vector is assembled elementwise. Using the target mesh,
\begin{equation}
 b_k \;=\; \sum_{t\in\mathcal M_t} b_k^t,
 \qquad
 b_k^t \;=\; \int_{\Omega_t} f^s(x)\,\psi_k(x)\,\mathrm d\Omega.
 \label{eq:bk_elementwise_mi}
\end{equation}
Although an equivalent elementwise assembly can be performed on $\mathcal{M}_s$, we perform all integrations and accumulations over target elements $t\in\mathcal{M}_t$ in this work.

Because $f^s$ is not regular on a target element $t$ when the meshes are non-matching, the integral on $\Omega_t$ must be decomposed into integrals over the geometric intersections of $t$ with the source elements:
\begin{equation}
\int_{\Omega_t} f^{s}(x)\,\psi_k(x)\,\mathrm d\Omega
  \;=\;
  \sum_{s\in \mathcal{S}(t)}
  \int_{\Omega_t \cap \Omega_s} f^{s}(x)\,\psi_k(x)\,\mathrm d\Omega.
\label{eq:partition_ts}
\end{equation}
where $\mathcal{S}(t) = \{\, s_i : |\Omega_t \cap \Omega_{s_i}| > 0 \,\}$ denotes the set of source elements that geometrically intersect $t$. Therefore, there are two clear challenges in computing the load vector:
\begin{enumerate}
    \item identifying all nonempty intersections $t \cap s$ (i.e., $\mathcal S(t)$); and

    \item evaluating the integrals over these intersection regions.
\end{enumerate}
The following section provides an overview of our GPU-based implementation of the mesh intersection method.
\subsection{Common-Refinement/Supermesh Overview}
\begin{figure}[htbp]
    \centering
    \begin{subfigure}[b]{0.30\textwidth}
        \centering
        \includegraphics[clip=true, trim=85 40 100 30,width=\linewidth]{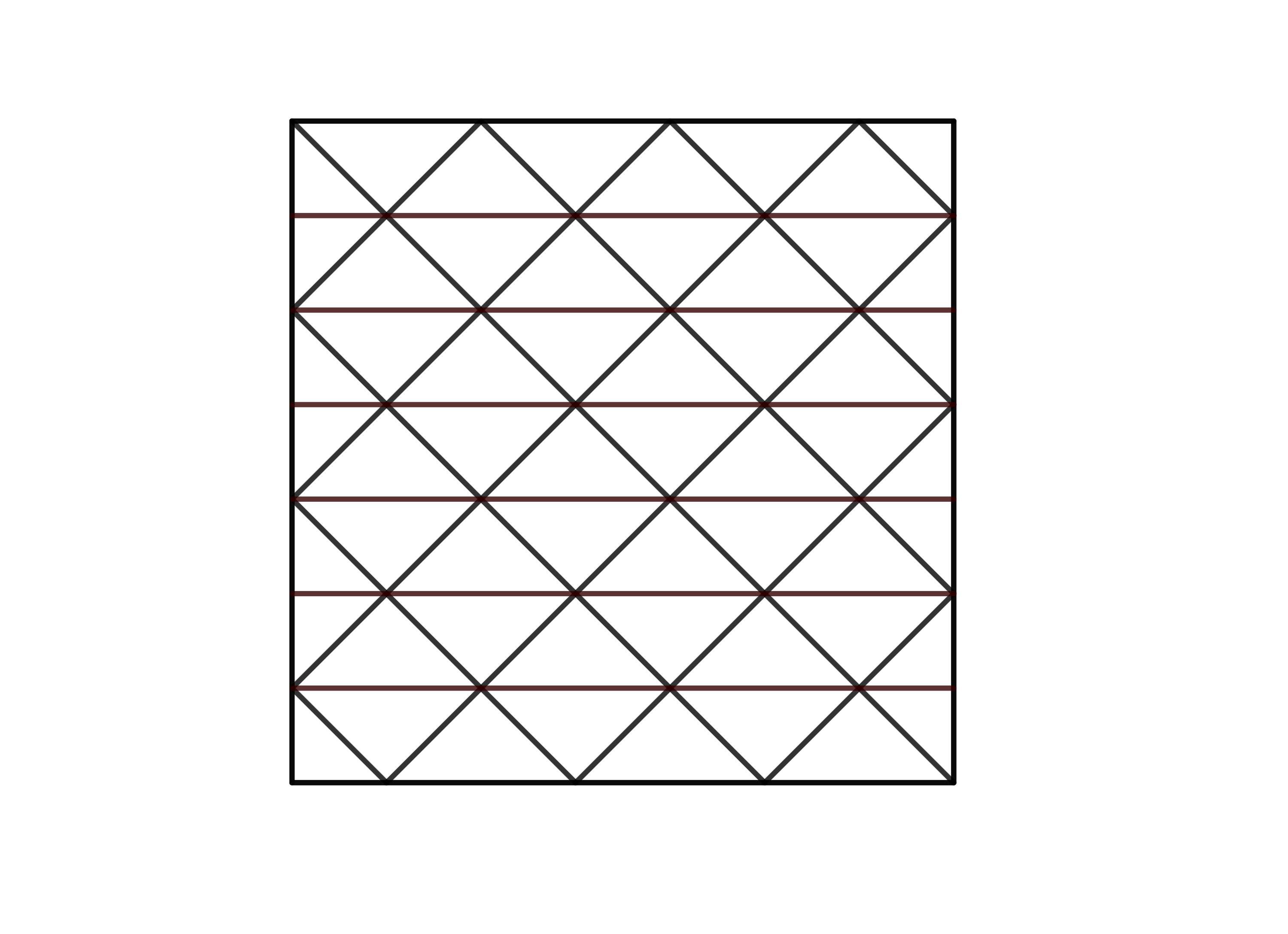}
         \caption{Source mesh $\mathcal{M}_s$}
        \label{fig:source_mesh}
    \end{subfigure}
    \hfill
    \begin{subfigure}[b]{0.30\textwidth}
        \centering
        \includegraphics[clip=true, trim=85 40 100 30, width=\linewidth]{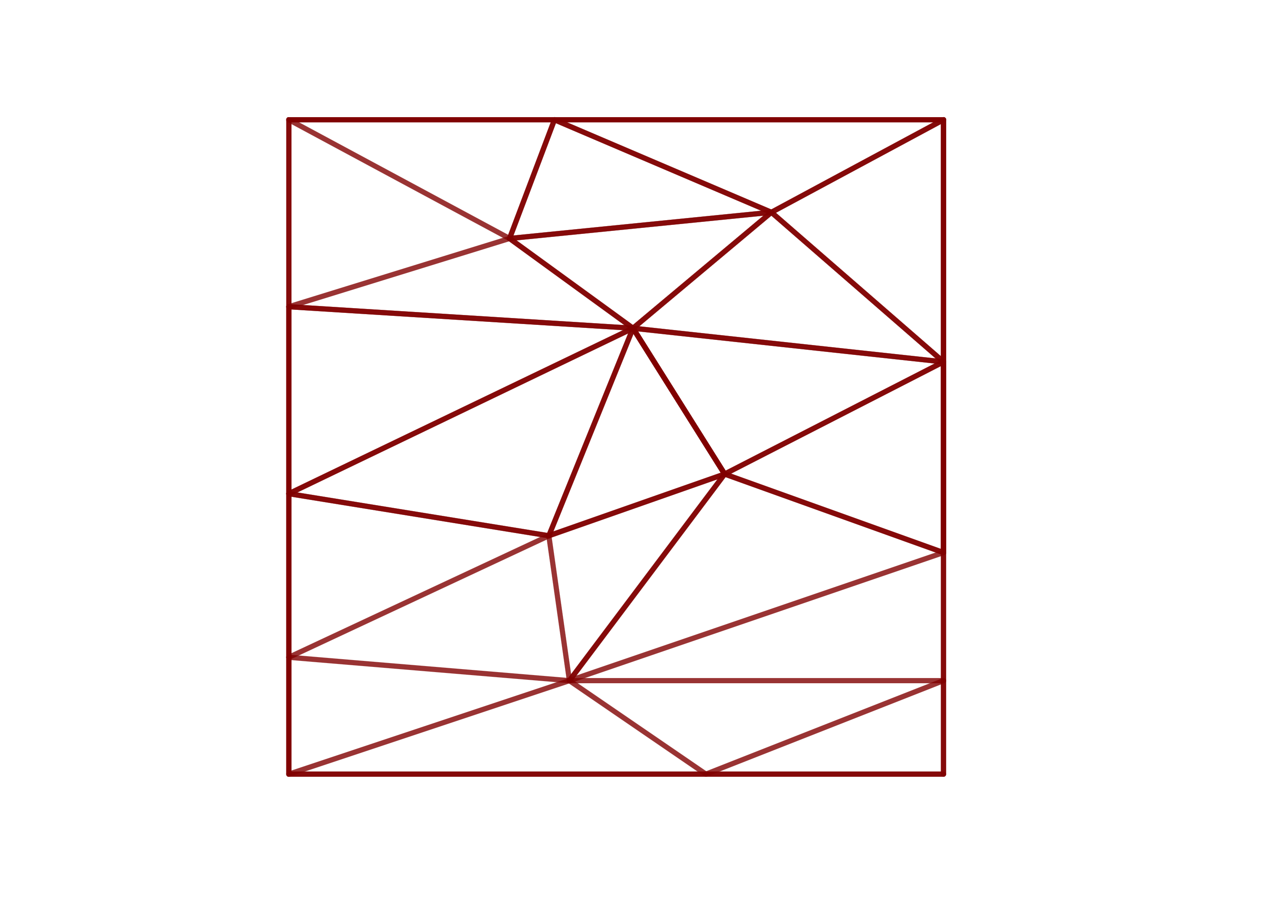}
        \caption{Target mesh $\mathcal{M}_t$}
        \label{fig:target_mesh}
    \end{subfigure}
    \hfill
    \begin{subfigure}[b]{0.30\textwidth}
        \centering
        \includegraphics[clip=true, trim=85 40 100 30,width=\linewidth]{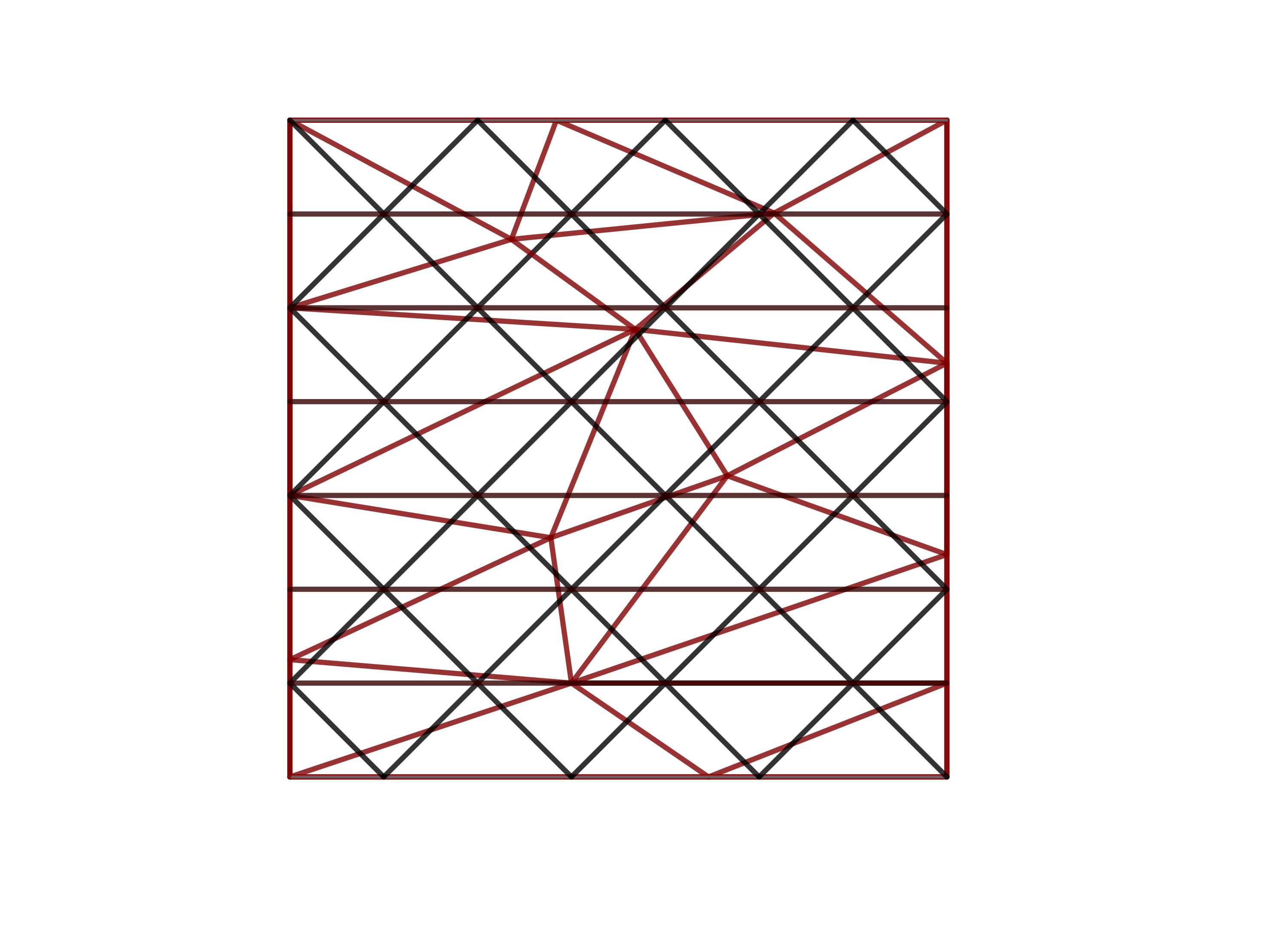}
         \caption{Supermesh $\mathcal{M}_t \cap \mathcal{M}_s$}
        \label{fig:supermesh}
    \end{subfigure}
 \caption{The supermesh (c) is constructed from geometric intersections of source (a) and target (b) elements.}
  \label{fig:supermesh_overview}
\end{figure}

One of the earliest general frameworks for conservative field transfer on non-matching meshes is the \textit{common-refinement} method introduced by Jiao and Heath \cite{JiaoHeath2004,JiaoHeathCommonRefinement}. Their approach constructs a global intermediate mesh by projecting the points of one surface onto the other and defining all edge intersections. This results in a globally consistent overlay mesh; however, the method is applicable only to 2D surface meshes. Extending such a global refinement strategy to 3D volume meshes is prohibitively expensive as the required global overlay and intersection computations results in substantial memory usage and computational overhead. Farrell et al.~\cite{FarrellMaddison2011} addressed this limitation by introducing the \textit{supermesh} method. The supermesh is defined as a collection of all non-empty intersections:
$$
\mathcal{M}_{ts} = \{ t \cap s : t \in \mathcal{M}_t,\; s \in \mathcal{M}_s,\; |\Omega_t \cap \Omega_s| > 0 \}
$$
as illustrated in Figure~\ref{fig:supermesh_overview}.
Unlike the global common refinement approach, the supermesh is constructed locally. For each target element, intersecting source elements are identified, and each intersection region $t \cap s$ is meshed using Eberly's clipping algorithm \cite{Eberly2006}. Although this local construction avoids building a global overlay mesh, it still requires forming and storing an explicit mesh for every intersection region, which can become memory-intensive for large problems. Moreover, the implementation in \cite{FarrellMaddison2011} is CPU-based and relies on classical geometric-clipping procedures.

\subsection{R3D-Based Intersection and GPU Integration}
In the current work, the construction of the local supermesh is handled using the R3D algorithm \cite{POWELL2015340, Powell2015R3D} as implemented  in the \texttt{Omega\_h} library \cite{ibanezCONFORMALMESHADAPTATION}. This approach is naturally suited to the GPU execution. Instead of constructing and storing an explicit intersection mesh for every non-empty region $t \cap s$, the present method computes all geometric intersections and performs numerical integration \emph{on-the-fly}. R3D represents each clipped region $t \cap s$ not as a triangulated mesh but as a convex polytope that stores the number of vertices, their coordinates, and the neighboring vertex lists defining the polygonal face. The next section describes the procedure for identifying intersecting source elements, which serve as inputs to the R3D clipping and integration stages.
\begin{figure}[htbp]
  \centering

  \begin{subfigure}[b]{0.19\textwidth}
    \centering
    \includegraphics[clip=true, trim=150 80 300 85, width=\linewidth]{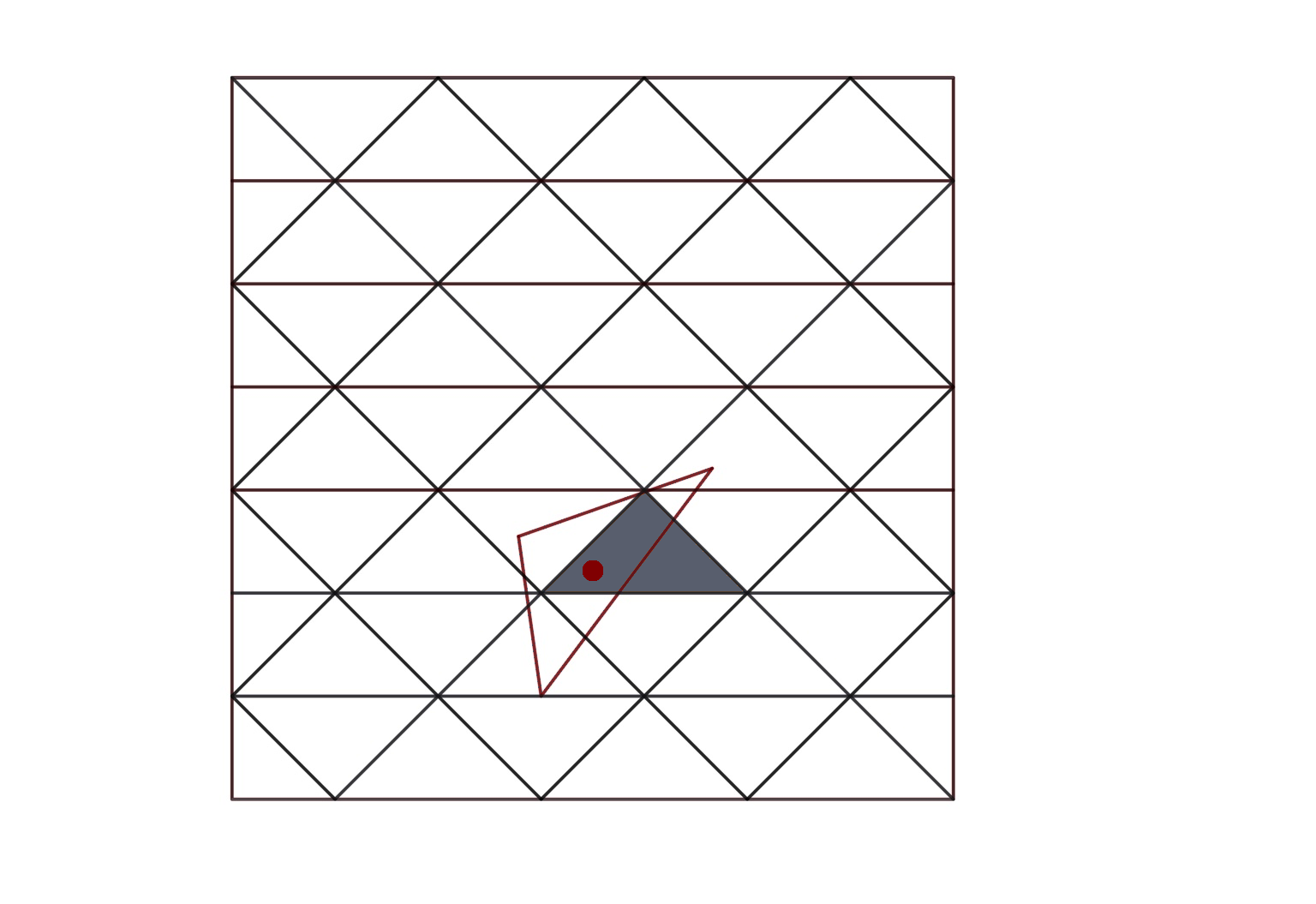}
    \caption*{(a)}
  \end{subfigure}\hfill
  \begin{subfigure}[b]{0.19\textwidth}
    \centering
    \includegraphics[clip=true,trim=520 220 800 140,width=0.84\linewidth, height=0.835\linewidth]{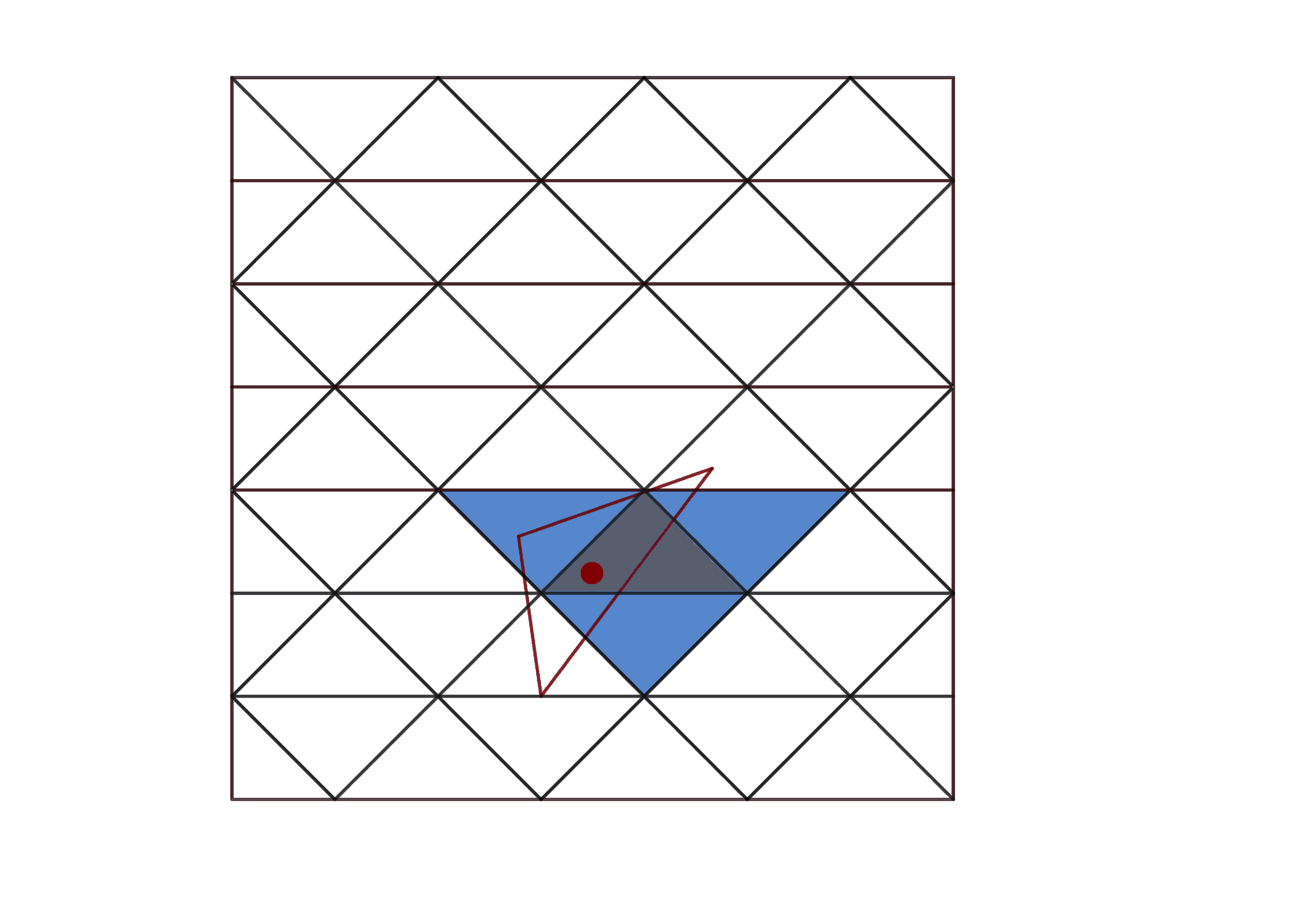}
    \caption*{(b)}
  \end{subfigure}\hfill
  \begin{subfigure}[b]{0.19\textwidth}
    \centering
    \includegraphics[clip=true, trim=200 80 320 60,width=0.84\linewidth, height=0.82
    \linewidth]{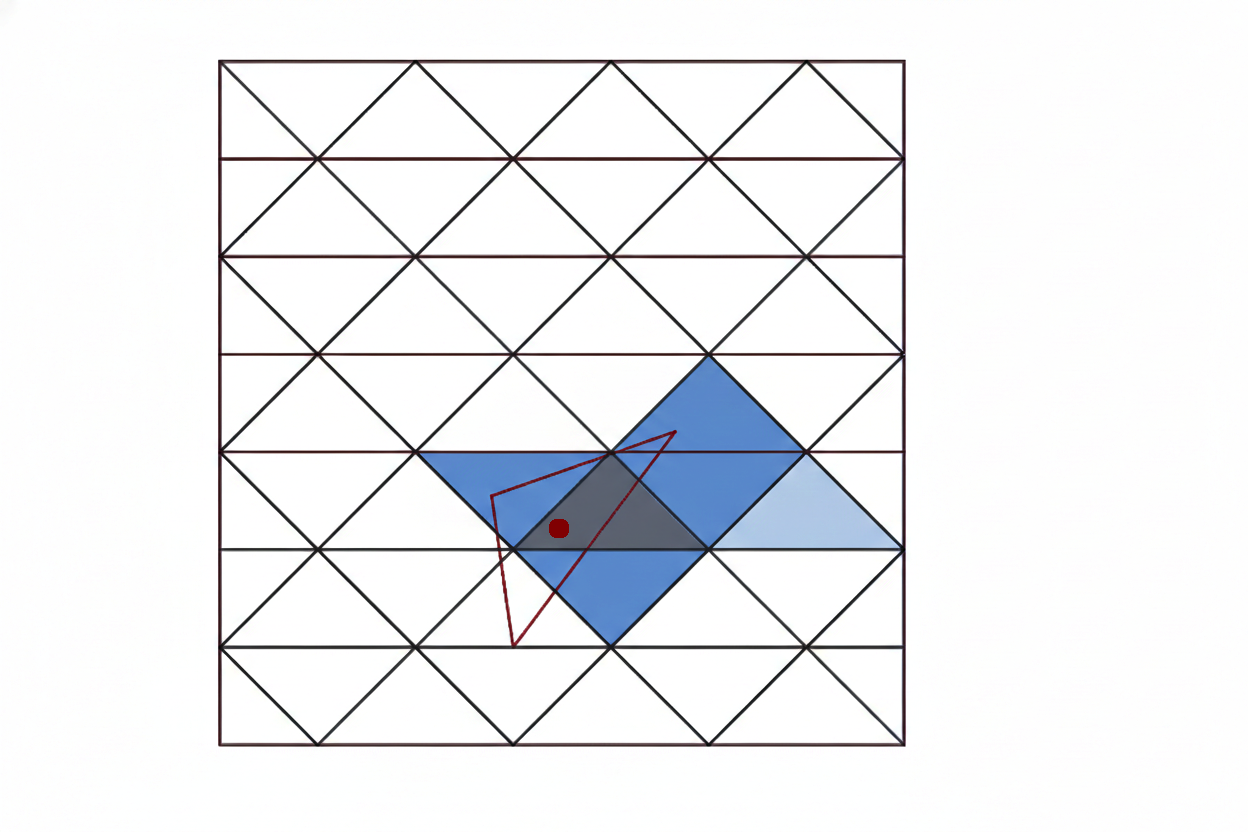}
    \caption*{(c)}
  \end{subfigure}\hfill
  \begin{subfigure}[b]{0.19\textwidth}
    \centering
    \includegraphics[clip=true, trim=200 80 320 60,width=0.84\linewidth, height=0.82
    \linewidth]{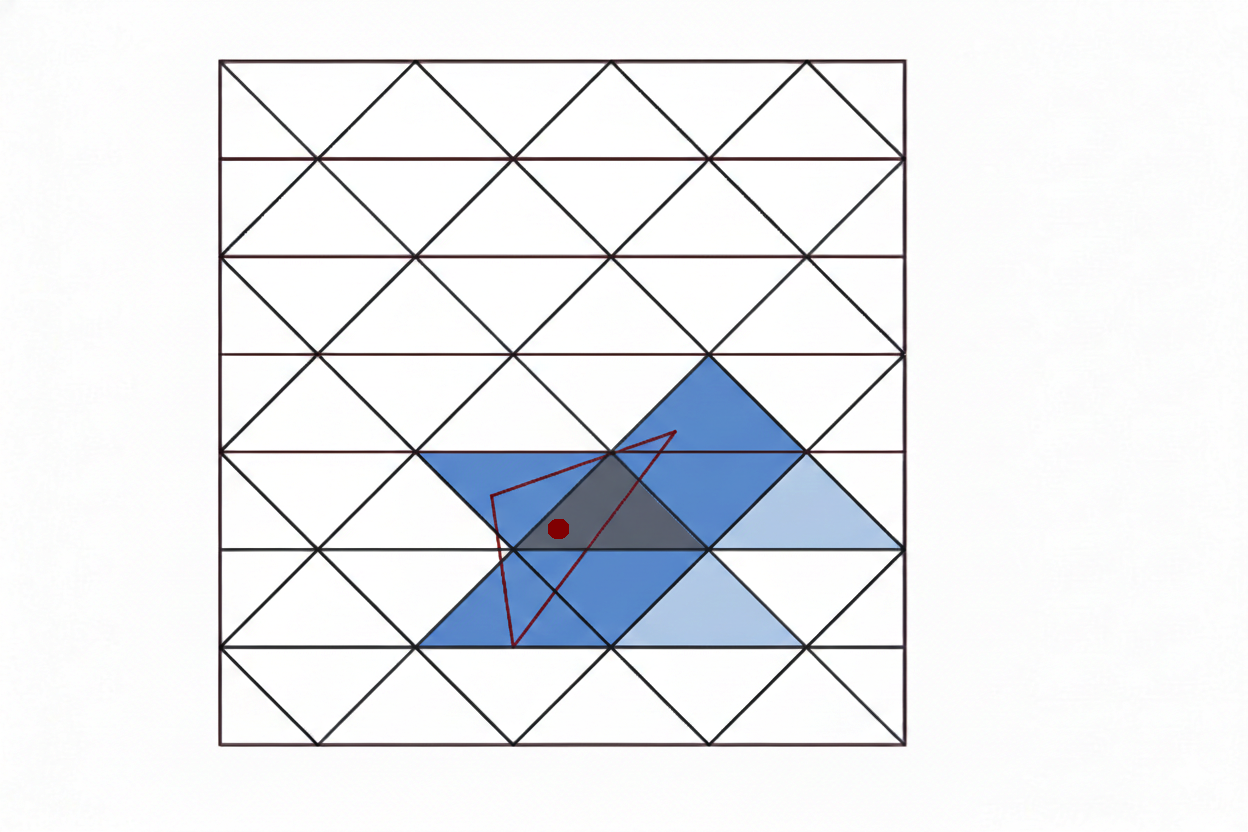}
    \caption*{(d)}
  \end{subfigure}\hfill
  \begin{subfigure}[b]{0.19\textwidth}
    \centering
    \includegraphics[clip=true, trim=200 80 320 60,width=0.84\linewidth, height=0.82
    \linewidth]{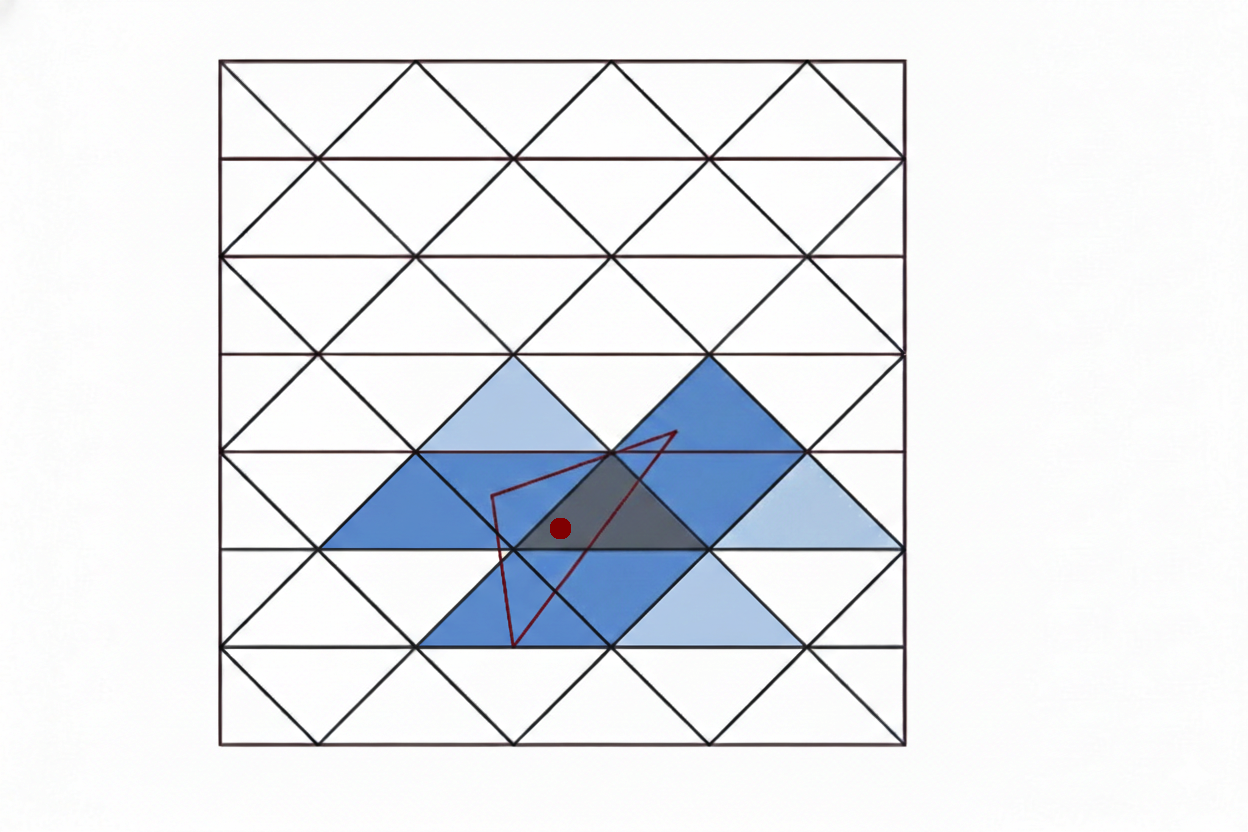}
    \caption*{(e)}
  \end{subfigure}

  \vspace{0.5em}

  \begin{subfigure}[b]{0.19\textwidth}
    \centering
    \includegraphics[clip=true, trim=200 80 320 60,width=0.84\linewidth, height=0.82
    \linewidth]{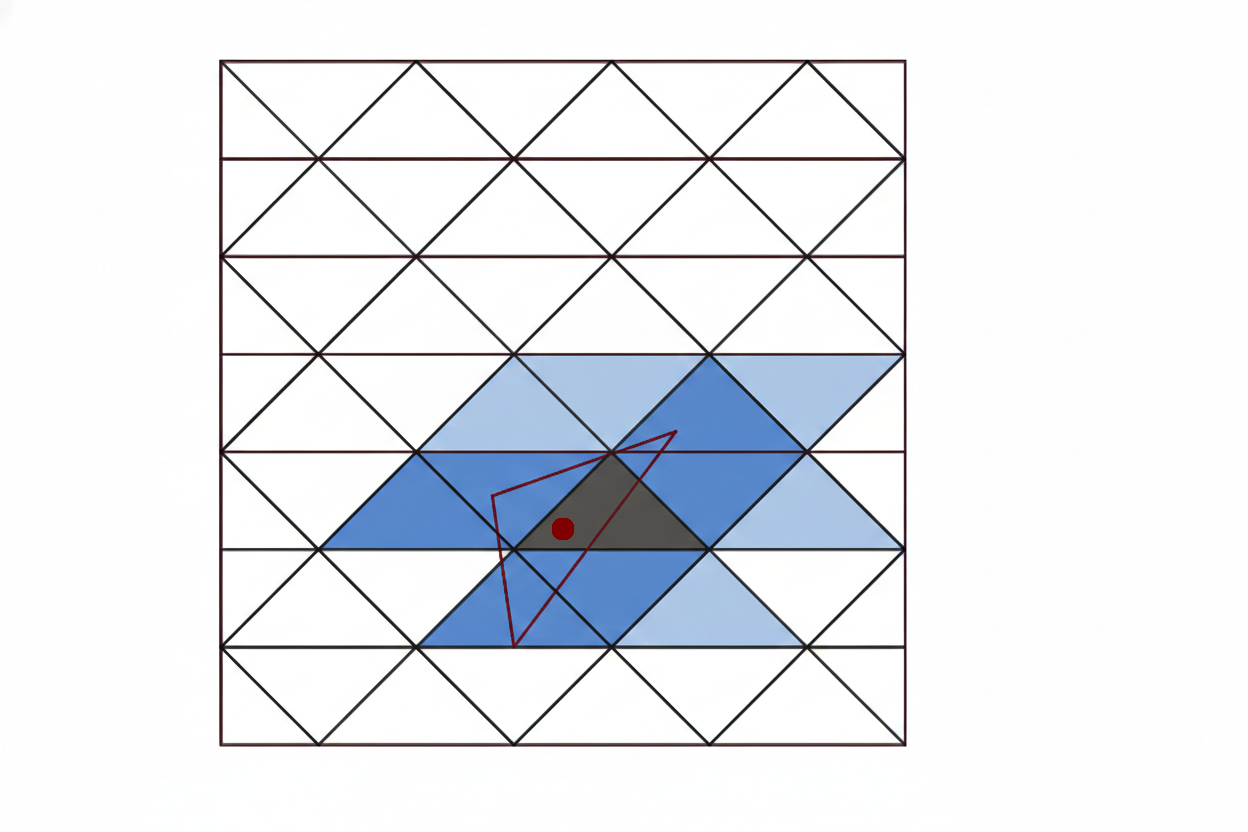}
    \caption*{(f)}
  \end{subfigure}\hfill
  \begin{subfigure}[b]{0.19\textwidth}
    \centering
    \includegraphics[clip=true, trim=200 80 320 60,width=0.86\linewidth, height=0.82
    \linewidth]{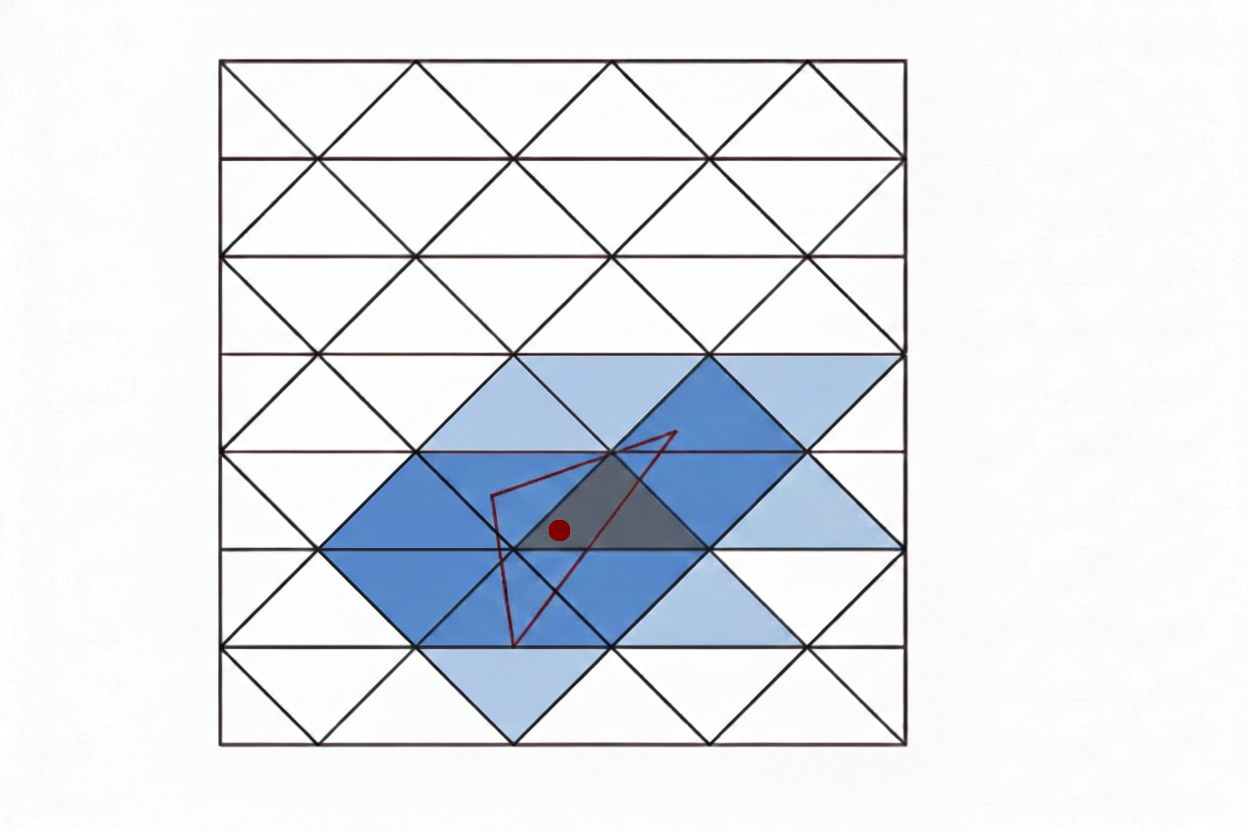}
    \caption*{(g)}
  \end{subfigure}\hfill
  \begin{subfigure}[b]{0.19\textwidth}
    \centering
    \includegraphics[clip=true, trim=200 80 320 60,width=0.84\linewidth, height=0.82
    \linewidth]{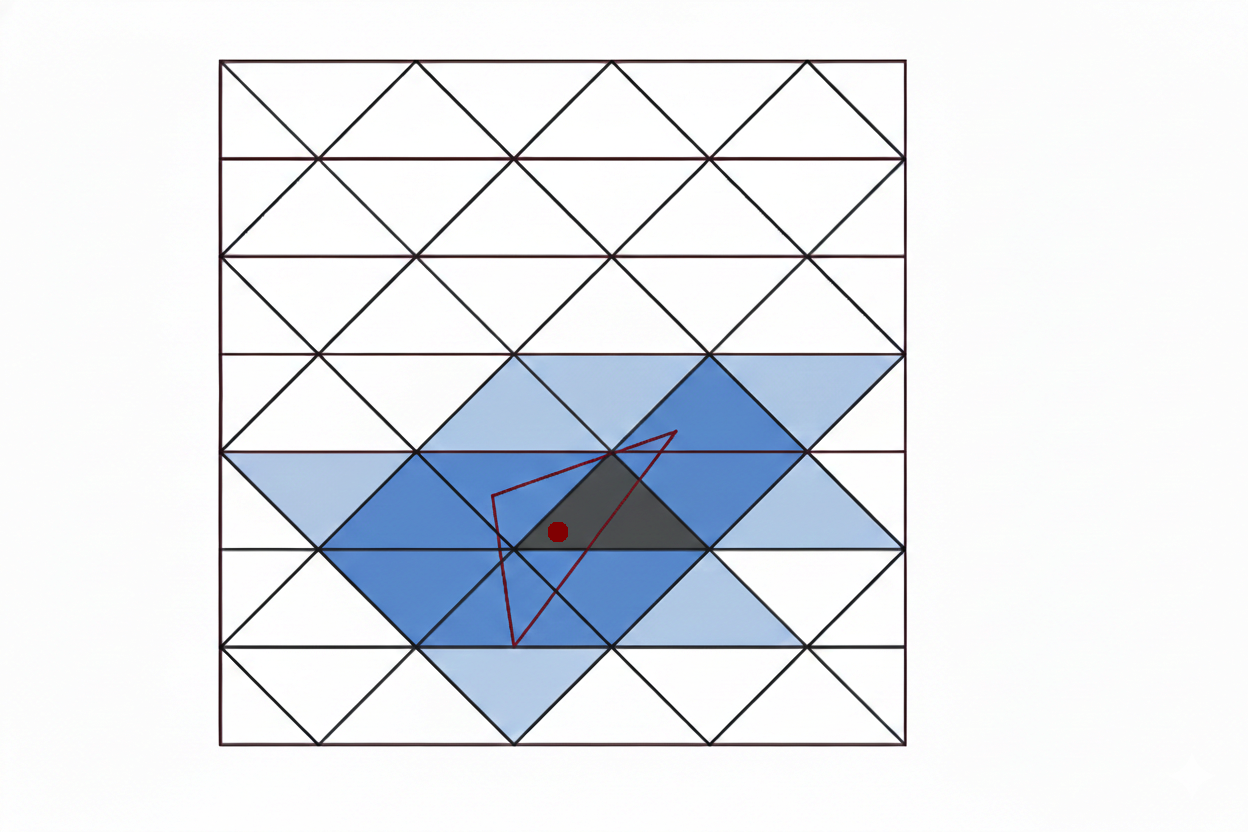}
    \caption*{(h)}
  \end{subfigure}\hfill
  \begin{subfigure}[b]{0.19\textwidth}
    \centering
    \includegraphics[clip=true, trim=200 80 320 60,width=0.84\linewidth, height=0.82
    \linewidth]{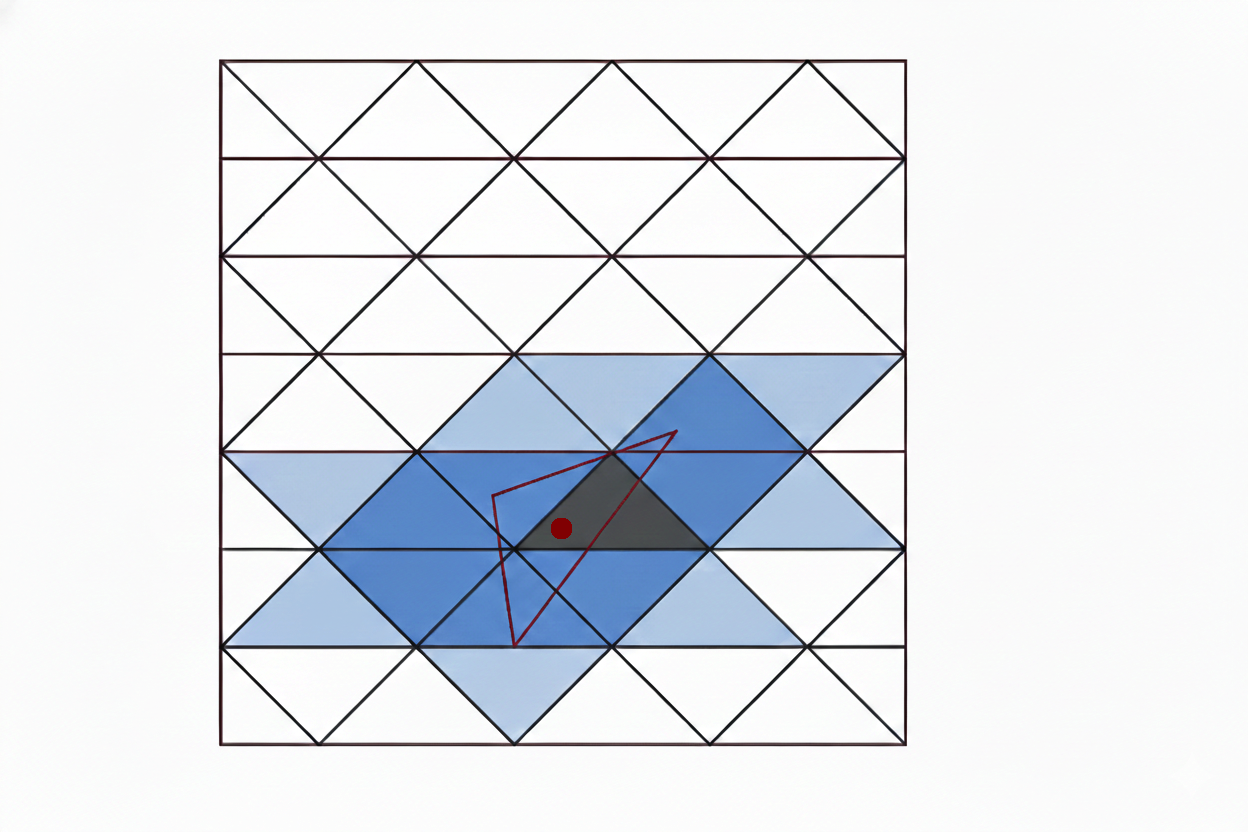}
    \caption*{(i)}
  \end{subfigure}\hfill
  \begin{subfigure}[b]{0.19\textwidth}
    \centering
    \includegraphics[clip=true, trim=200 80 320 60,width=0.84\linewidth, height=0.82
    \linewidth]{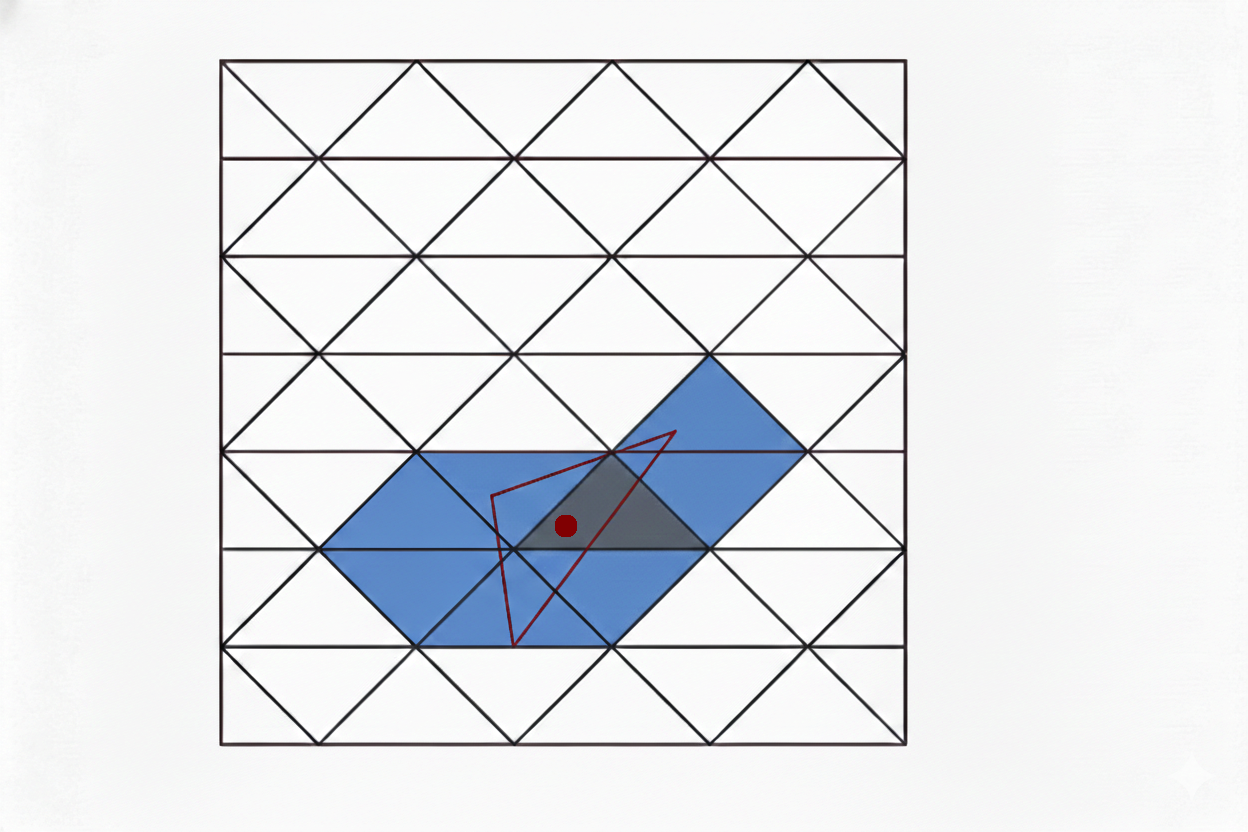}
    \caption*{(j)}
  \end{subfigure}

  \caption{
    Adjacency-based identification of intersecting source elements for a single target element (red) and its centroid (red dot). Figures (a)–(j) illustrate the progressive expansion of the search front which is expanded from the gray element which is identified through localization of the target element centroid. The dark blue elements represent the expanded set of search elements in each step which constitute the integration regions for the load-vector assembly; light blue elements indicate source elements that are traversed during the search but do not intersect the target element of interest.
  }
  \label{fig:intersection-search-sequence}
\end{figure}

\subsubsection{Adjacency-Based Intersection Search}
To construct a local supermesh, the intersecting pair of elements of $\mathcal{M}_s$ and $\mathcal{M}_t$ must be identified. The algorithm applied in this study is shown in Figure  ~\ref{fig:intersection-search-sequence} and summarized below.
\begin{enumerate}
    \item  For a given target element $t \in \mathcal{M}_t$, the centroid of $t$ (depicted with red dot) is mapped onto the source mesh $\mathcal{M}_s$ through point localization, yielding the initial seed source element $s_0$ as highlighted in grey color   (Figure~\ref{fig:intersection-search-sequence}a).
    
    \item A breadth-first traversal of the source-mesh adjacency graph is initiated from $s_0$. Each visited source element $s$ is tested for geometric intersection with $t$ using R3D \cite{POWELL2015340}. Neighbors of intersecting source elements are added to the search front, while non-intersecting elements are discarded. The search expands until no new neighbors remain to be tested. (Figure~\ref{fig:intersection-search-sequence}b--Figure~\ref{fig:intersection-search-sequence}i)
    
    \item The output is the set $\mathcal{S}(t)$ of source elements intersecting $t$. The corresponding regions $t \cap s_i$ form the integration subdomains for load vector assembly (Figure~\ref{fig:intersection-search-sequence}j).
\end{enumerate}
The same adjacency-based traversal is performed for every target element $t \in \mathcal{M}_t$ to identify its intersecting source elements. 
\begin{figure}[htbp]
  \centering
  \begin{subfigure}[b]{0.3\textwidth}
    \centering
    \includegraphics[clip=true,trim = 130 120 120 60, width=\linewidth]{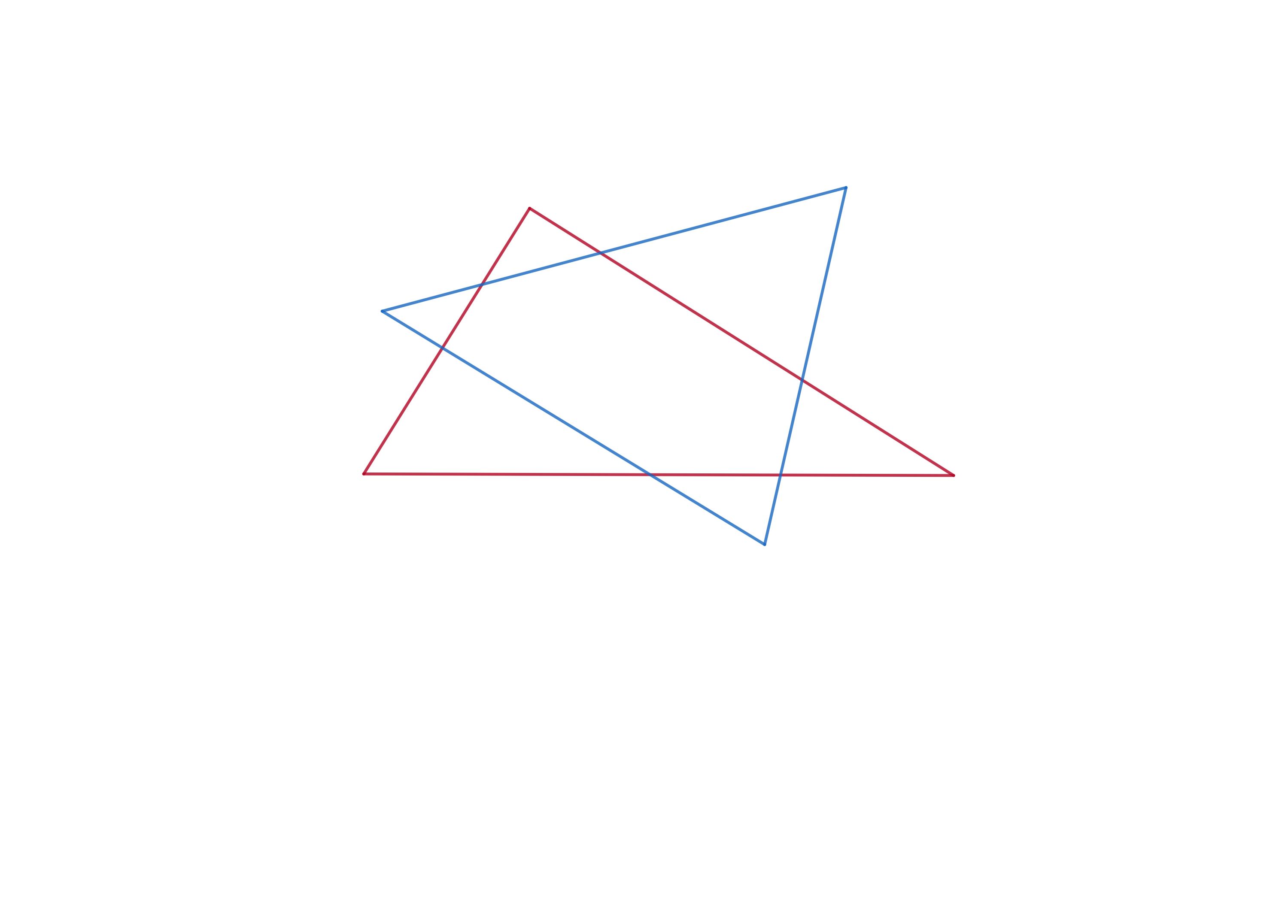}
    \caption{Intersection of target (red) and source (blue) elements}
  \end{subfigure}
  \hfill
  \begin{subfigure}[b]{0.3\textwidth}
    \centering
    \includegraphics[clip=true,trim = 130 120 120 60,width=\linewidth]{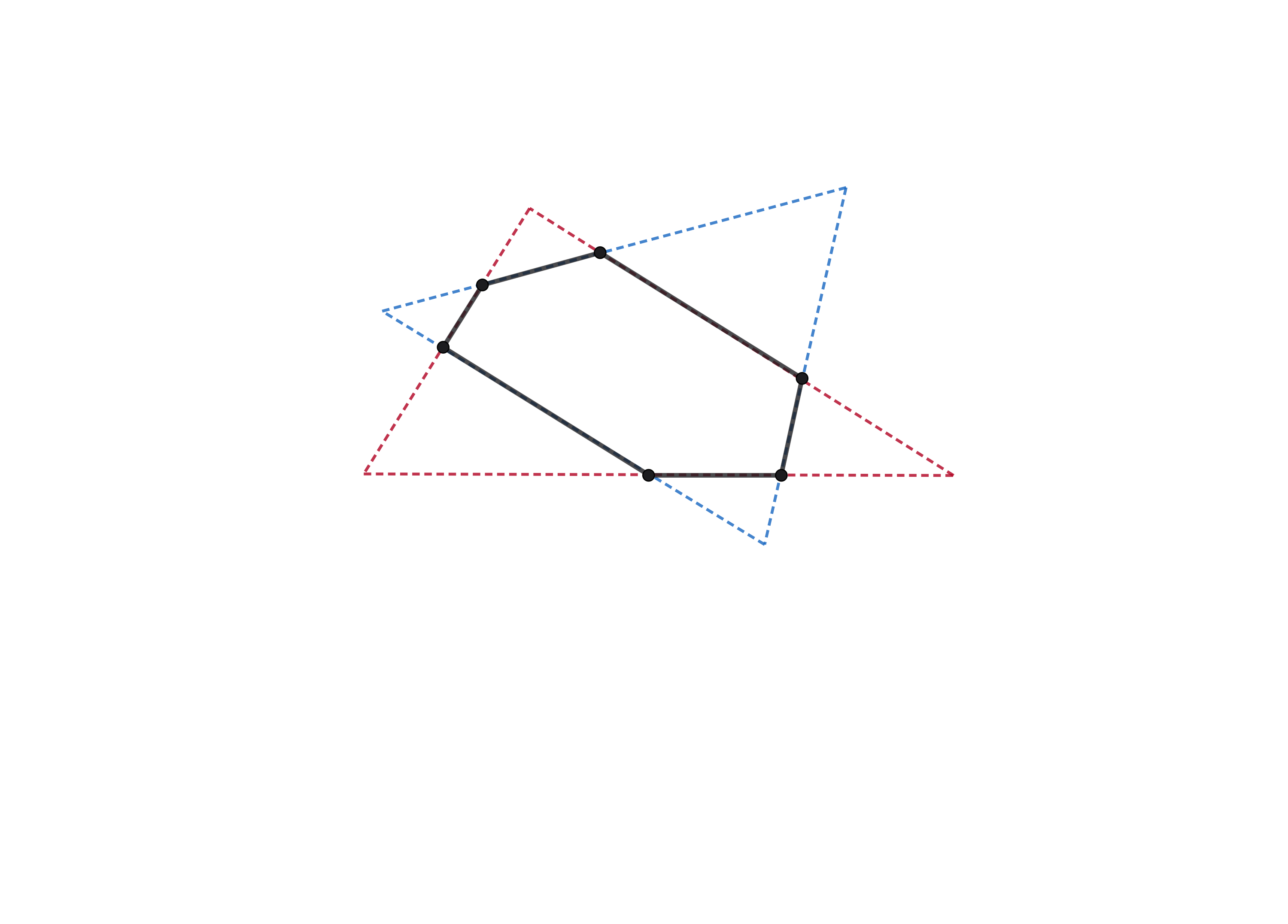}
    \caption{Polygonal intersection region $P_{ts}$}
  \end{subfigure}
  \hfill
  \begin{subfigure}[b]{0.3\textwidth}
    \centering
    \includegraphics[clip=true,trim = 160 160 170 60,width=\linewidth]{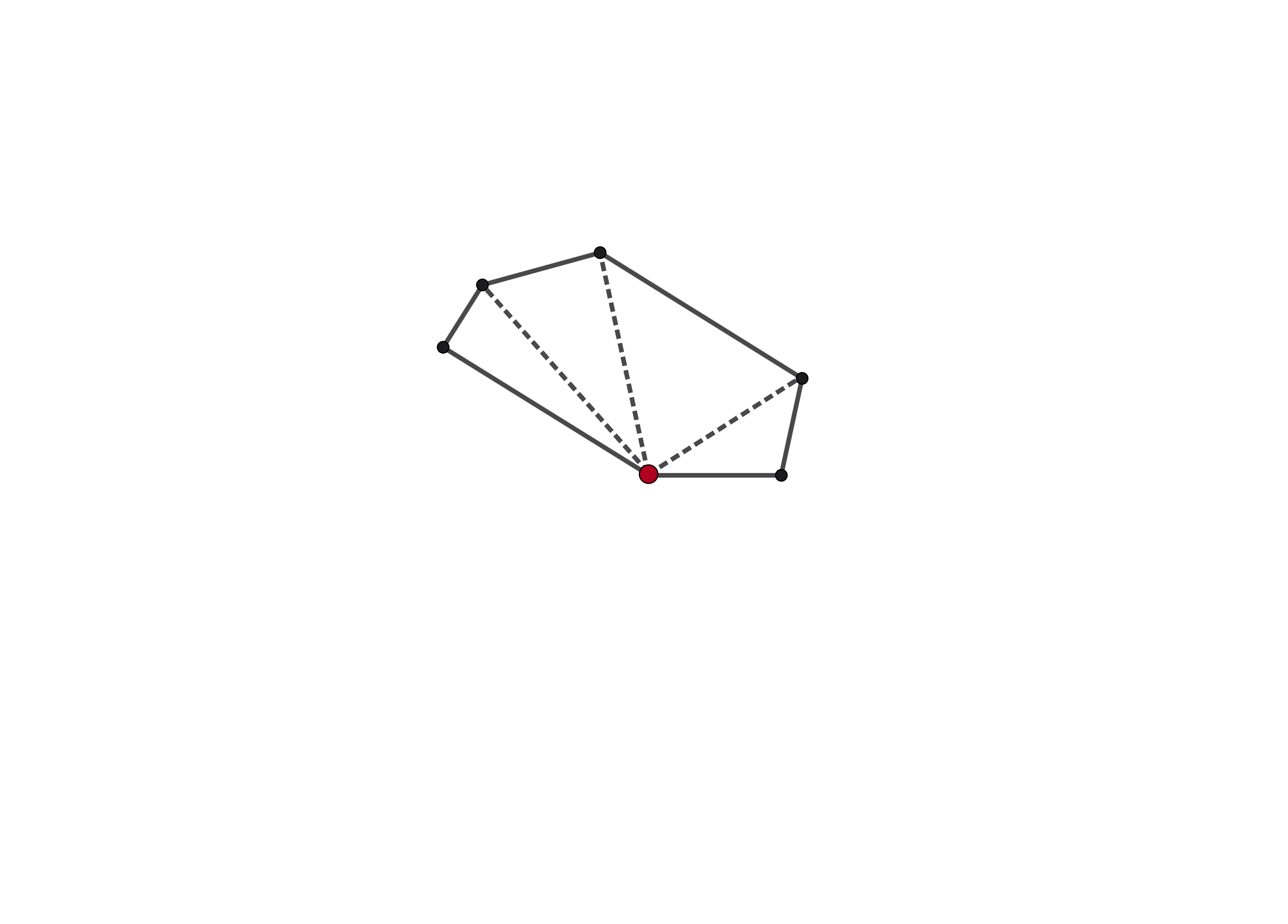}
    \caption{Simplicial decomposition using anchor vertex (red)}
  \end{subfigure}
  \caption{Intersection integration procedure: geometric clipping, polytope
  construction, and simplicial decomposition for numerical quadrature.}
  \label{fig:intersection-integration}
\end{figure}
\subsubsection{Intersection Integration}
Once the intersection set is identified for a target element $t$, the contribution to the load vector is assembled by integrating over each intersection region $t \cap s$. Unlike supermesh-based approaches, no explicit intersection mesh is stored. Instead, each intersection region is generated, decomposed, and used for integration on the fly. The overall procedure is
illustrated in Figure ~\ref{fig:intersection-integration} and proceeds as follows:

\begin{enumerate}
  \item For each target element $t \in \mathcal{M}_t$ processed independently and in parallel on the GPU, initialize its local load vector $b_k^t$.

  \item For every intersecting source element $s \in \mathcal{S}(t)$, compute the geometric intersection $P_{ts} = t \cap s$ using the R3D clipping algorithm. As shown in Figure ~\ref{fig:intersection-integration}a, this operation takes the target element (red) and source element (blue) and produces a convex polygon representing their overlap.

  \item The resulting intersection region $P_{ts}$ is represented as a convex polytope via its planar graph, storing only vertex coordinates and adjacency information. Figure~\ref{fig:intersection-integration}b illustrates the polygonal intersection region obtained after clipping.

  \item The vertices of each polygon of $P_{ts}$ are sorted into a consistent counterclockwise (CCW) ordering to enable a stable simplicial decomposition.

  \item A simplicial decomposition is then performed on the fly by selecting an anchor vertex (shown in red in Figure ~\ref{fig:intersection-integration}c) and forming triangles (in 2D) by connecting the anchor to successive vertex pairs. 

  \item Numerical quadrature is applied on each simplex to accumulate the local contribution to the load vector associated with $t$.

  \item After all $s \in \mathcal{S}(t)$ have been processed, the contributions from all simplices of all intersection polytopes $P_{ts}$ are summed into  $b_k^t$. The global load vector $b_k$ is then assembled from the element-local vectors $b_k^t$.
\end{enumerate}
This approach is fully local, memory efficient, and naturally suited to GPU parallelism. 

\section{Conservative Coupling without Discretization Information} \label{sec:monte-carlo-integration}
\subsection{Control-Variate Monte Carlo Approximation of the Galerkin Projection}
\label{subsec:cv_mc_galerkin_projection}

In this section, we consider the black-box coupling regime, in which the source discretization information is unavailable and the source field \(f^s(x)\) is accessible only through pointwise evaluations. In this setting, the mass matrix in the Galerkin system in Eq.~\eqref{eq:galerkin_system} is assembled deterministically on the target mesh, whereas the load-vector entries in Eq.~\eqref{eq:load_vector} must be approximated.

Define the exact element integral operator
\begin{equation}
    I_t[v]
    =
    \int_{\Omega_t} v(x)\,d\Omega ,
    \label{eq:exact_element_integral_operator}
\end{equation}
where \(\Omega_t\) is a target element. Let \(p_t(x)\) be a probability density function on \(\Omega_t\), satisfying
\begin{equation}
    p_t(x)\ge 0,
    \qquad
    \int_{\Omega_t} p_t(x)\,d\Omega = 1 .
\end{equation}
Let \(X_{t,1},X_{t,2},\ldots,X_{t,N}\) be independent and identically distributed samples drawn from \(p_t(x)\). For a function \(v\), the corresponding Monte Carlo approximation of \(I_t[v]\) is defined as
\begin{equation}
    \widehat I_{t}[v]
    =
    \frac{1}{N}
    \sum_{j=1}^{N}
    \frac{v(X_{t,j})}{p_t(X_{t,j})}.
    \label{eq:mc_element_integral_operator_general}
\end{equation}
This operator is unbiased because
\begin{align}
    \mathbb E[\widehat I_{t}[v]]
    &=
    \frac{1}{N}
    \sum_{j=1}^{N}
    \mathbb E
    \left[
        \frac{v(X_{t,j})}{p_t(X_{t,j})}
    \right]
    \notag \\
    &=
    \frac{1}{N}
    \sum_{j=1}^{N}
    \int_{\Omega_t}
    \frac{v(x)}{p_t(x)}
    p_t(x)\,d\Omega
    \notag \\
    &=
    I_t[v].
    \label{eq:mc_operator_unbiased}
\end{align}
For uniform sampling within the target element, \(p_t(x)=1/|\Omega_t|\), and Eq.~\eqref{eq:mc_element_integral_operator_general} becomes
\begin{equation}
    \widehat I_{t}[v]
    =
    \frac{|\Omega_t|}{N}
    \sum_{j=1}^{N}
    v(X_{t,j}) .
    \label{eq:mc_element_integral_operator_uniform}
\end{equation}
For the remainder of this work, we use uniform sampling on each target element.
The assembled Monte Carlo operator over the target mesh is
\begin{equation}
    \widehat I[v]
    =
    \sum_{t\in\mathcal M_t}
    \widehat I_{t}[v].
    \label{eq:global_mc_operator}
\end{equation}
The exact Galerkin load-vector entry can be written using the exact element integral operator as
\begin{equation}
    b_i
    =
    \sum_{t\in\mathcal M_t}
    I_t[f^s\psi_i].
    \label{eq:exact_load_operator_mc}
\end{equation}

Let \(g\) be a surrogate field obtained by interpolating the source field onto the target mesh,
\begin{equation}
    g = I_h f^s ,
    \label{eq:cv_control_function}
\end{equation}
where \(I_h\) denotes the interpolation operator on the target finite element space \(\mathcal V_t\). Since \(g\) is represented on the target mesh \(\mathcal M_t\), integrals involving \(g\) and the target basis functions can be evaluated deterministically using standard quadrature rule.

On each target element \(\Omega_t\), the centered control-variate integrand is defined as
\begin{equation}
    z(x)
    =
    f^s(x)
    -
    \alpha
    \left(
    g(x)-\mathbb E_{\Omega_t}[g]
    \right),
    \label{eq:cv_centered_integrand}
\end{equation}
where \(\alpha\) is the control-variate parameter and
\begin{equation}
    \mathbb E_{\Omega_t}[g]
    =
    \frac{1}{|\Omega_t|}
    \int_{\Omega_t} g(x)\,d\Omega .
    \label{eq:cv_element_mean_g}
\end{equation}
This preserves the element mean of the source field as
\begin{align}
    \mathbb E_{\Omega_t}[z]
    &=
    \mathbb E_{\Omega_t}[f^s].
    \label{eq:cv_mean_preserving}
\end{align}
Thus, the control variate does not change the expected element average, but it can reduce variance when \(g\) is correlated with \(f^s\). Applying the Monte Carlo approximation operator in Eq.~\eqref{eq:mc_element_integral_operator_uniform} to the centered
control-variate integrand \(z\), we obtain
\begin{align}
    \widehat I_{t}[z]
    &=
    \frac{|\Omega_t|}{N}
    \sum_{j=1}^{N}
    \left[
    f^s(X_{t,j})
    -
    \alpha
    \left(
    g(X_{t,j})-\mathbb E_{\Omega_t}[g]
    \right)
    \right]
    \notag \\
    &=
     \alpha |\Omega_t| \mathbb E_{\Omega_t}[g] + 
    \frac{|\Omega_t|}{N}
    \sum_{j=1}^{N}
    \left[
    f^s(X_{t,j})
    -
    \alpha g(X_{t,j})
    \right]
    .
    \label{eq:control_variate_approximation}
\end{align}
Substituting Eq.~\eqref{eq:cv_element_mean_g} into
Eq.~\eqref{eq:control_variate_approximation} gives
\begin{align}
    \widehat I_{t}[z]
    &=  \int_{\Omega_t}  \alpha g(x)\,d\Omega + \frac{|\Omega_t|}{N}
    \sum_{j=1}^{N}
    \left[
    f^s(X_{t,j})
    -
    \alpha g(X_{t,j})
    \right] \notag \\
    &= I_t[\alpha g] + \widehat I_t[ f^s - \alpha g].
    \label{eq:control_variate_residual_plus_control}
\end{align}
For Galerkin load-vector assembly, the same decomposition is applied to
the integrand weighted by the target basis function \(\psi_i\). Therefore, on
each target element,
\begin{equation}
    \widehat b_i^t
    =
    I_t[\alpha g\psi_i]
    +
    \widehat I_{t}
    \left[
    \left(f^s-\alpha g\right)\psi_i
    \right].
    \label{eq:cv_element_load_residual_operator}
\end{equation}
Thus, the deterministic part of the load vector is assembled from the control
field contribution \(\alpha g \psi_i\), while the Monte Carlo operator is applied only to the residual
integrand \((f^s-\alpha g)\psi_i\). Summing over all target elements gives the global control-variate Monte Carlo
load-vector approximation
\begin{equation}
    \widehat b_i=I\left[\alpha g\psi_i\right]
    +\widehat I
    \left[
    \left(f^s-\alpha g\right)\psi_i
    \right].
    \label{eq:cv_load_operator}
\end{equation}
Using the definition of \(I\) and \(\widehat I\), Eq.~\eqref{eq:cv_load_operator} becomes
\begin{equation}
    \widehat b_i
    =
    \int_{\Omega}
    \alpha g(x)\psi_i(x)\,d\Omega
    +
    \sum_{t\in\mathcal M_t}
    \frac{|\Omega_t|}{N}
    \sum_{j=1}^{N}
    \left(
    f^s(X_{t,j})-\alpha g(X_{t,j})
    \right)
    \psi_i(X_{t,j}) .
    \label{eq:cv_load_expanded}
\end{equation}
The control-variate estimator is unbiased because
\begin{align}
    \mathbb E[\widehat b_i]
    &=
    \int_{\Omega}
    \alpha g(x)\psi_i(x)\,d\Omega
    +
    \mathbb E
    \left[
    \widehat I
    \left[
    \left(f^s-\alpha g\right)\psi_i
    \right]
    \right]
    \notag \\
    &=
    \int_{\Omega}
    \alpha g(x)\psi_i(x)\,d\Omega
    +
    \int_{\Omega}
    \left(f^s(x)-\alpha g(x)\right)\psi_i(x)\,d\Omega
    \notag \\
    &=
    \int_{\Omega}
    f^s(x)\psi_i(x)\,d\Omega
    =
    b_i .
    \label{eq:cv_load_unbiased}
\end{align}
Therefore,
\begin{equation}
    \mathbb E[\widehat{\mathbf b}]
    =
    \mathbf b .
    \label{eq:expectation_of_b}
\end{equation}
The transferred coefficient vector \(\widehat{\mathbf f}^{\,t}\) is obtained from
\begin{equation}
    \mathbf M
    \widehat{\mathbf f}^{\,t}
    =
    \widehat{\mathbf b}.
    \label{eq:cv_galerkin_system}
\end{equation}
Since \(\mathbf M\) is deterministic,   using Eq.~\eqref{eq:expectation_of_b} and Eq.~\eqref{eq:galerkin_system}
\begin{equation}
    \mathbb E[\widehat{\mathbf f}^{\,t}]
    =
    \mathbf M^{-1}
    \mathbb E[\widehat{\mathbf b}]
    =
    \mathbf M^{-1}\mathbf b
    =
    \mathbf f^t .
    \label{eq:cv_solution_unbiased}
\end{equation}
Hence, the control-variate Monte Carlo Galerkin projection is unbiased. The variance reduction arises because the Monte Carlo operator is applied to the residual integrand \(\left(f^s-\alpha g\right)\psi_i\) instead of the original integrand \(f^s\psi_i\). For a suitable choice of \(\alpha\), the residual \(f^s-\alpha g\) can have smaller variance than \(f^s\), particularly when \(g\) is strongly correlated with \(f^s\). This reduces the sampling error in the load-vector approximation. The detailed error analysis is presented in \ref{app:accuracy_error_bound} and \ref{app:conservation_error_bound}. 
The process involved in approximating the Galerkin projection using the control-variate Monte Carlo method is summarized below.
\begin{enumerate}
    \item Construct the control function \(g=I_hf^s\) on the target mesh by evaluating the source field at the target degrees of freedom and interpolating those values in the target finite element space.

    \item Evaluate the deterministic control contribution to the load vector for each element,
    \[
        b_{i,g}^t
        =
        \int_{\Omega_t}
        \alpha g(x)\psi_i(x)\,d\Omega ,
    \]
    using standard target finite element assembly.

    \item For each target element \(t\in\mathcal M_t\), choose a sample count \(N\). For \(j=1,\dots,N\), draw two independent scalars
    \(\xi_j\sim\mathrm{Uniform}(0,1)\) and \(\eta_j\sim\mathrm{Uniform}(0,1)\), convert \((\xi_j,\eta_j)\) to area-uniform barycentric weights on the reference triangle using the mapping in~\cite{osada2002shape}, and map the samples to physical coordinates \(X_{t,j}\in\Omega_t\).

    \item Perform point localization for each sample point \(X_{t,j}\) to evaluate the source field \(f^s(X_{t,j})\). Also evaluate the control function \(g(X_{t,j})\) and the target basis functions \(\psi_i(X_{t,j})\) associated with element \(t\).

    \item For each basis function \(\psi_i\) supported on \(t\), compute the element-local residual Monte Carlo contribution
    \[
        \widehat b_{i,\mathrm{res}}^{\,t}
        =
        \frac{|\Omega_t|}{N}
        \sum_{j=1}^{N}
        \left(
        f^s(X_{t,j})-\alpha g(X_{t,j})
        \right)
        \psi_i(X_{t,j}) .
    \]

    \item Assemble the element-local deterministic term \(b_{i,g}^t\) into global deterministic contribution \(b_{i,g}\) and the residual contribution \(\widehat b_{i,\mathrm{res}}^{\,t}\) into the global residual load vector \(b_{i,\mathrm{res}}\) and combine them to obtain the total load vector
    \[
        \widehat b_i
        =
        b_{i,g}
        +
        \widehat b_{i,\mathrm{res}} .
    \]

    \item After all element contributions are assembled, solve the target Galerkin system
    \[
        \mathbf M
        \widehat{\mathbf f}^{\,t}
        =
        \widehat{\mathbf b}.
    \]
\end{enumerate}
 
\section{Numerical Comparison} \label{sec:numerical-comparison}
In the present work, our objective is to compare the accuracy, conservation and  performance properties of three field-transfer techniques:
\begin{enumerate}
    \item a non-conservative Radial Basis Function (RBF)–based method,
    \item a deterministic, conservative Mesh–Intersection (MI)–based method, and
    \item an asymptotically conservative Monte Carlo (MC)–based method.
\end{enumerate}
All three methods transfer a scalar field from one mesh to a non-matching mesh on a shared geometric domain.

The RBF method used here is a local weighted polynomial fitting method. The corresponding minimization problem is stated as
\begin{equation}
    \underset{c}{\text{min}} ||\phi \cdot (Ac - b)||_2^2 + \lambda||c||_2^2,
\end{equation}
where \(A\) is the Vandermonde matrix constructed from the set source points within a radius $r$ of the target point, \(\phi\) is the diagonal weight matrix, \(\lambda\) is the regularization parameter, \(b\) is the vector of source field values. All presented tests use the C4 basis function. The complete details of the C4 basis function and source set construction are described in \cite{merson2025pcms}. For this work, we expand the radius so that each target element has at least the minimum number of source supports required for linear fitting. 

\subsection{Convergence Analysis}
An important property of supermesh-based Galerkin projection, as
highlighted by Farrell and Maddison~\cite{FarrellMaddison2011}, is that the
projection error can be evaluated exactly up to quadrature and roundoff on the
supermesh. Since each intersection region forms a geometric subdomain on which
both the source and target finite element spaces are exactly representable, the
error field
\begin{align}
    e = f^s - f^t
\end{align}
is itself a well-defined function on every cell of the supermesh. This enables
the exact evaluation of the error by integrating \(e\) over \(\mathcal{M}_{ts}\)
in any desired norm. To quantify the quality of the field transfer, we evaluate
two relative integral norms that measure accuracy and global conservation.

\paragraph{Continuous supermesh-based accuracy error}
The relative \(L^2(\Omega)\) accuracy error, evaluated using the
supermesh-based integration, is defined as
\begin{equation}
E_{L^2}
  =
  \frac{
    \| f^s - f^t \|_{L^2(\Omega)}
  }{
    \| f^s \|_{L^2(\Omega)}
  }
  =
  \frac{
    \left(
      \int_{\Omega}
        \bigl( f^s(x) - f^t(x) \bigr)^2
      \,\mathrm{d}\Omega
    \right)^{1/2}
  }{
    \left(
      \int_{\Omega}
        \bigl( f^s(x) \bigr)^2
      \,\mathrm{d}\Omega
    \right)^{1/2}
  } .
\label{eq:relative_L2_error}
\end{equation}

\paragraph{Continuous supermesh-based conservation error}
To examine how well the global integral of the field is preserved, we use the
relative conservation error evaluated on the supermesh:
\begin{equation}
E_{\mathrm{mass}}^{\mathrm{SM}}
  =
  \frac{
    \left|
      \int_{\Omega} f^s(x)\,\mathrm{d}\Omega
      -
      \int_{\Omega} f^t(x)\,\mathrm{d}\Omega
    \right|
  }{
    \left|
      \int_{\Omega} f^s(x)\,\mathrm{d}\Omega
    \right|
  } .
\label{eq:relative_conservation_error}
\end{equation}

\begin{figure}[htbp]
    \centering
    \begin{subfigure}[b]{0.45\textwidth}
        \centering
        \includegraphics[width=\linewidth]{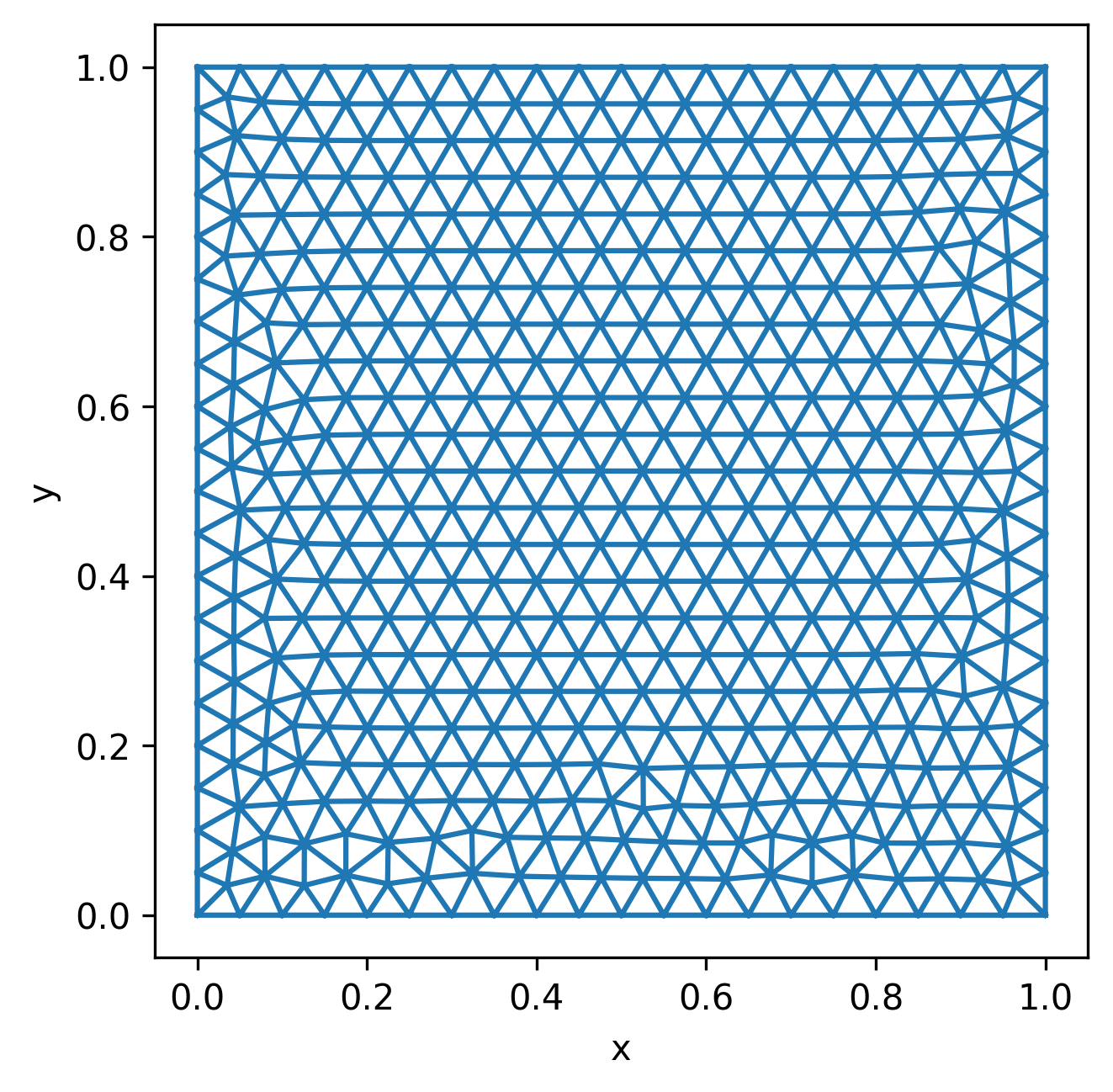}
        \caption{Source mesh}
        \label{fig:source_mesh_sq}
    \end{subfigure}
    \hfill
    \begin{subfigure}[b]{0.45\textwidth}
        \centering
        \includegraphics[width=\linewidth]{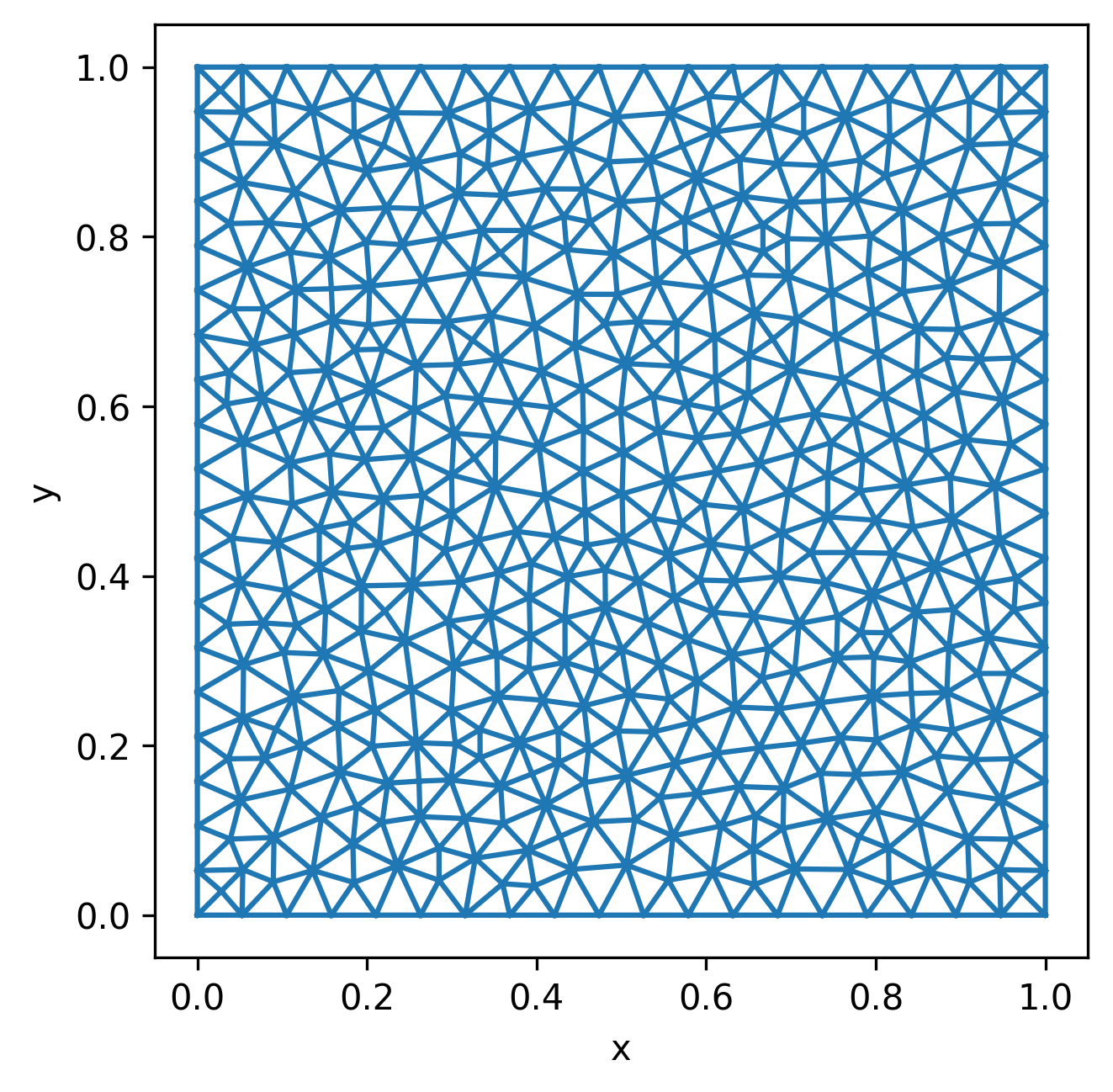}
        \caption{Target mesh}
        \label{fig:target_mesh_sq}
    \end{subfigure}
    \caption{Representative source and target meshes used in the convergence analysis.}
    \label{fig:meshes_sq}
\end{figure}

For this study, we generated a sequence of target/source mesh pairs with
progressively smaller characteristic mesh size \(h\) using gmsh
\cite{geuzaine2009gmsh}. The target and source meshes are topologically
different even though their characteristic sizes match at each refinement level,
as used in supermesh-based convergence studies~\cite{FarrellMaddison2011}.
The field transfer is computed from the source space to the target space for
these meshes, and both error metrics are evaluated and recorded. All fields use
linear \((p=1)\) Lagrange shape functions.

For the Monte Carlo integration, \(N=20\) samples are drawn independently for each
target element, generated uniformly on the reference triangle and mapped to the
physical element by the affine element map. The uniform sampling on the reference element induces
area-uniform sampling on the physical element. The same sample count is used for all refinement levels and all functions. In the control-variate Monte Carlo method, the parameter \(\alpha\)
controls the amount of deterministic control contribution included in the
load-vector approximation. The case \(\alpha=0\) corresponds to direct Monte
Carlo sampling of the original source field contribution, whereas
\(\alpha=1\) corresponds to deterministically assembling the interpolated
control field and sampling only the residual \(f^s-g\).

For the convergence study, the following functions are interpolated onto the source mesh
\begin{align}
    f_1(x,y) &= 5y^3 + x^2 + 2y + 3, \notag \\
    f_2(x,y) &= \sin(x)\cos(y) + 2, \notag \\
    f_3(x,y) &= e^{x^2} + 2y .
    \label{eq:test_functions}
\end{align}


The interpretation of the convergence results follows from the error estimates
derived in Eq.~\eqref{eq:error_decomposition_norm} and
Eq.~\eqref{eq:conservation_bound} which rely on adequate regularity of the source field. The accuracy estimate separates the expected
squared \(L^2\) error into a deterministic finite-element approximation term
and a stochastic sampling term:
\begin{align}
    \mathbb E\left[
    \|f^s-\widehat f^t\|_{L^2(\Omega)}^2
    \right]
    &\lesssim
    h^{2p+2}
    \|f^s\|^2_{H^{p+1}(\Omega)}
    \notag \\
    &\quad
    +
    \frac{h^d}{N}
    \|K_h\|_{L^\infty(\Omega)}
    \left[
    (1-\alpha)^2
    \|f^s\|^2_{L^2(\Omega)}
    +
    \alpha^2 h^{2p+2}
    \|f^s\|^2_{H^{p+1}(\Omega)}
    \right].
    \label{eq:accuracy_bound_interpretation}
\end{align}
Here \(a\lesssim b\) denotes \(a\le Cb\) for a constant \(C>0\) independent of \(h\), \(N\), and \(\alpha\). The first term corresponds to the deterministic target-space approximation
error, while the second term corresponds to the Monte Carlo sampling error in
the control-variate load-vector approximation. For \(p=1\), the deterministic
accuracy contribution gives an \(L^2\)-error rate of order
\(\mathcal O(h^2)\). Although the bound also contains a sampling term, for the
control-variate choice \(\alpha=1\), the leading stochastic contribution
proportional to \((1-\alpha)^2\) vanishes. The remaining sampling contribution in
Eq.~\eqref{eq:accuracy_bound_interpretation} is of the same order as the deterministic contribution ($\|K_h\|_{L^\infty(\Omega)}\propto h^{-d}$). However, the size of the contributition is limited due to the \(1/N\) scaling with the number of sample points. Therefore, with the fixed
sample count \(N=20\), the leading accuracy behavior is expected to be
approximately second order for the linear target finite elements demonstrated here.

Similarly, Eq.~\eqref{eq:conservation_bound} gives the expected squared
conservation error as
\begin{equation}
    \mathbb E\left[
    E_{\mathrm{cons}}^2
    \right]
    \lesssim
    \frac{h^d}{N}
    \left[
    (1-\alpha)^2
    \|f^s\|_{L^2(\Omega)}^2
    +
    \alpha^2 h^{2p+2}
    \|f^s\|_{H^{p+1}(\Omega)}^2
    \right].
    \label{eq:conservation_bound_interpretation}
\end{equation}
For \(\alpha=1\), the leading non-control term vanishes, and the remaining
conservation error is governed by the higher-order residual approximation
term. In two dimensions with linear elements \((p=1)\), this gives the RMS
scaling
\[
    E_{\mathrm{cons}}
    =
    \mathcal O\left(\frac{h^3}{\sqrt N}\right).
\]
Thus, with \(N=20\), the theory predicts approximately second-order accuracy
convergence and approximately third-order conservation convergence when the
control field accurately represents the source field.

\begin{figure}[htbp]
  \centering
  \begin{subfigure}[b]{0.45\textwidth}
    \centering
    \includegraphics[width=\linewidth]{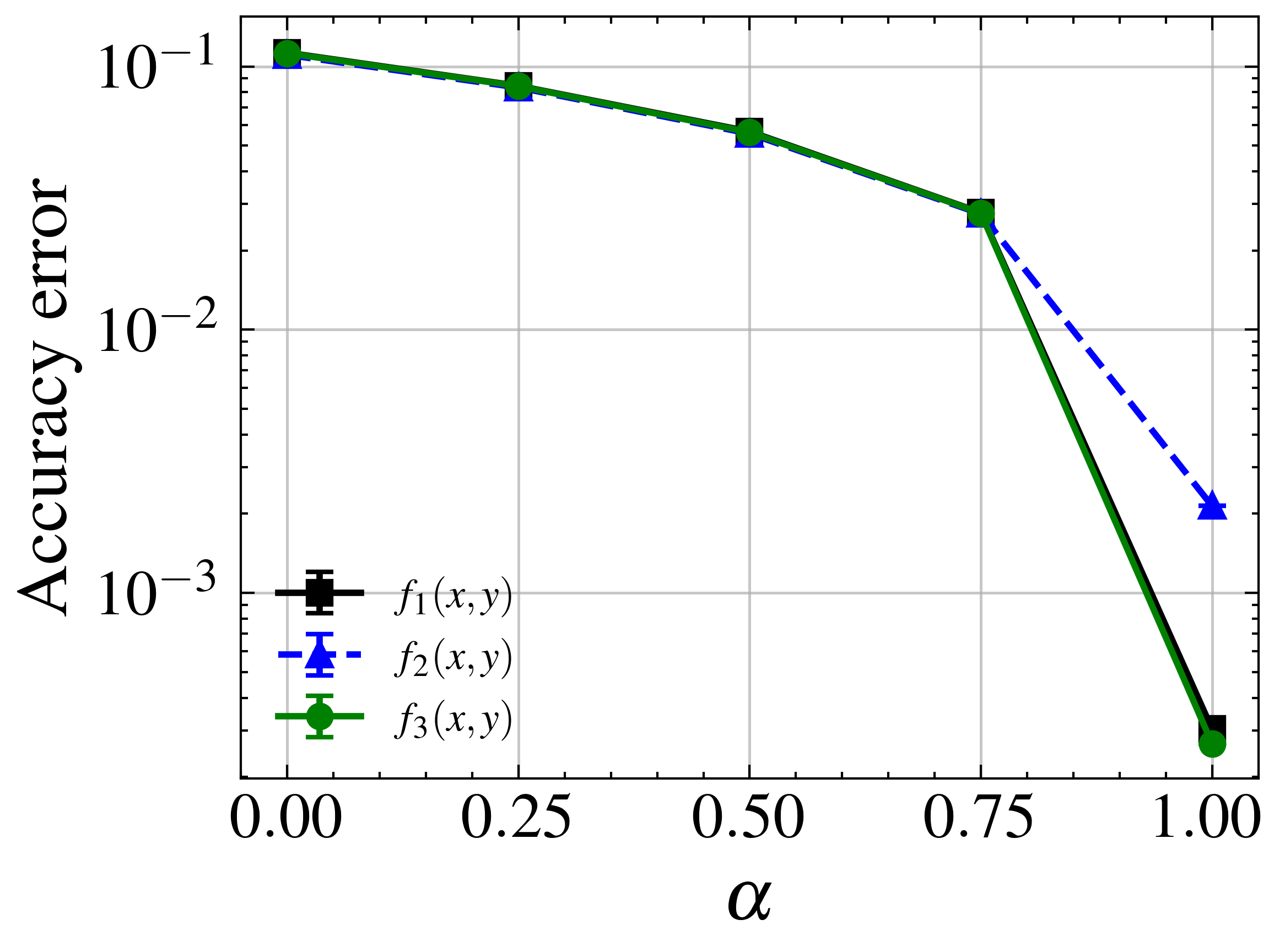}
    \caption{Accuracy error vs. \(\alpha\)}
    \label{fig:alpha-accuracy-error}
  \end{subfigure}
  \hfill
  \begin{subfigure}[b]{0.45\textwidth}
    \centering
    \includegraphics[width=\linewidth]{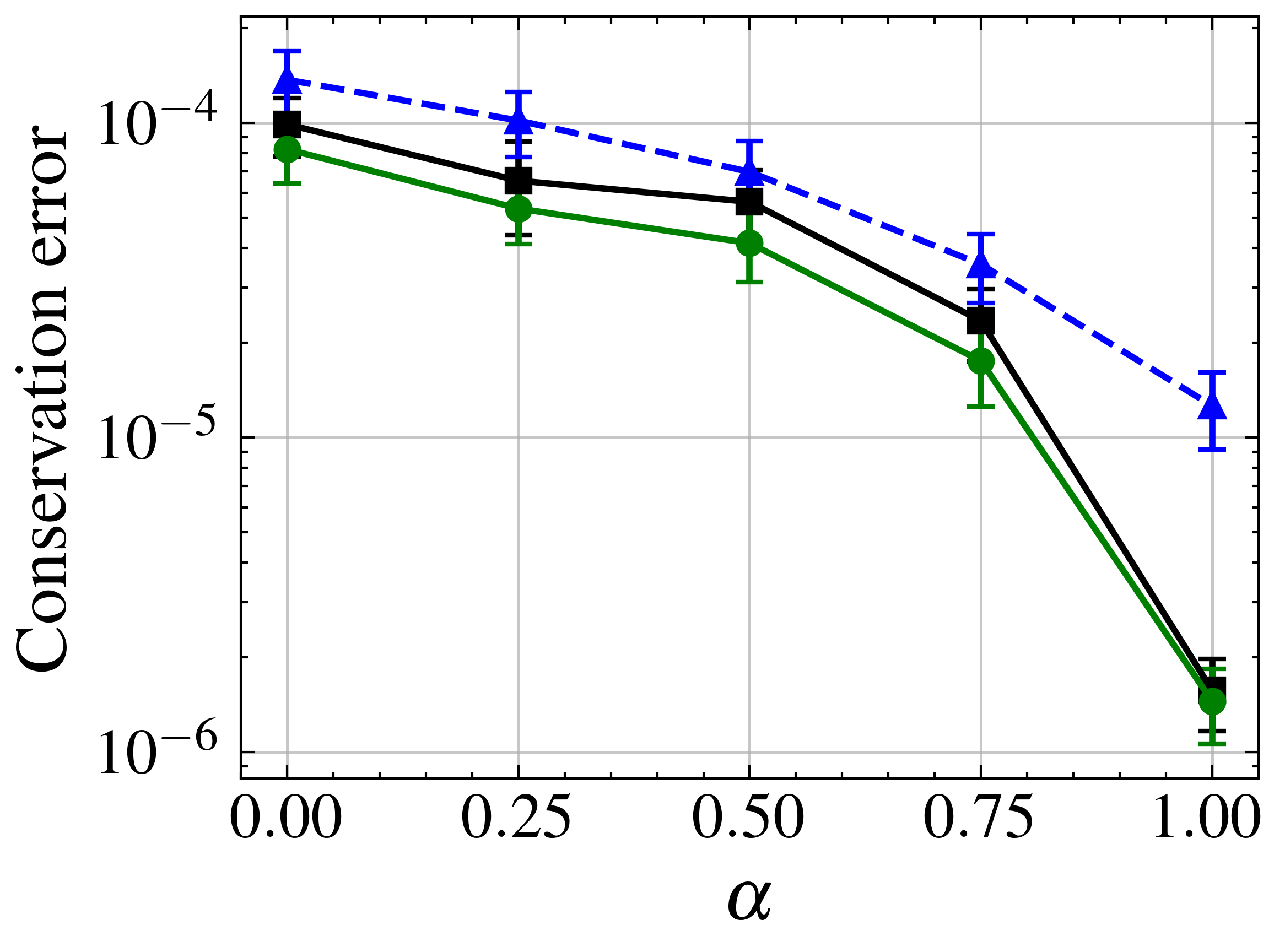}
    \caption{Conservation error vs. \(\alpha\)}
    \label{fig:alpha-conservation-error}
  \end{subfigure}
  \caption{Effect of the control-variate coefficient \(\alpha\) on accuracy and conservation errors for the three smooth test functions.}
  \label{fig:alpha-error-study}
\end{figure}

Figure~\ref{fig:alpha-error-study} shows the effect of the control-variate
coefficient \(\alpha\) on the accuracy (Figure~\ref{fig:alpha-accuracy-error}) and conservation (Figure~\ref{fig:alpha-conservation-error}) errors for the three
functions in Eq.~\eqref{eq:test_functions}. Using the same source--target
mesh pair for all three functions, the study is performed on the
third refinement mesh level from the convergence study. For all cases, both the accuracy and
conservation errors decrease as \(\alpha\) increases from \(0\) to \(1\). This
behavior is consistent with Eq.~\eqref{eq:accuracy_bound_interpretation} and
Eq.~\eqref{eq:conservation_bound_interpretation}. When \(\alpha=0\), the
sampling error contains the leading contribution associated with the source
field itself. As \(\alpha\) approaches one, this leading contribution is
removed and the Monte Carlo estimator is applied only to the residual
\(f^s-g\). The reduction is especially pronounced between \(\alpha=0.75\) and
\(\alpha=1\), where the sampled residual becomes much smaller than the original
source-field contribution. The \(f_1(x,y)\) and \(f_3(x,y)\) show nearly
identical behavior in the accuracy plot, while \(f_2(x,y)\)
retains a slightly larger accuracy error at \(\alpha=1\). In the conservation
plot, all three functions show the same decreasing trend, with the
\(f_2(x,y)\) remaining somewhat larger than \(f_1(x,y)\) and \(f_3(x,y)\) cases. These differences are consistent with the fact that the
quality of the control variate depends on how well the interpolated target-space
field \(g\) represents the source field \(f^s\) on the selected mesh. Since \(g\) is obtained by interpolating the source field into the target finite
element space, it is strongly correlated with \(f^s\), and the sampled residual
has much smaller variance than the original source field and consequently,
\(\alpha=1\) gives the smallest observed errors for all three functions in this
study and is used for the remaining mesh-refinement tests.

\begin{figure}[htbp]
  \centering
  \begin{subfigure}[b]{0.45\textwidth}
    \centering
    \includegraphics[width=\linewidth]{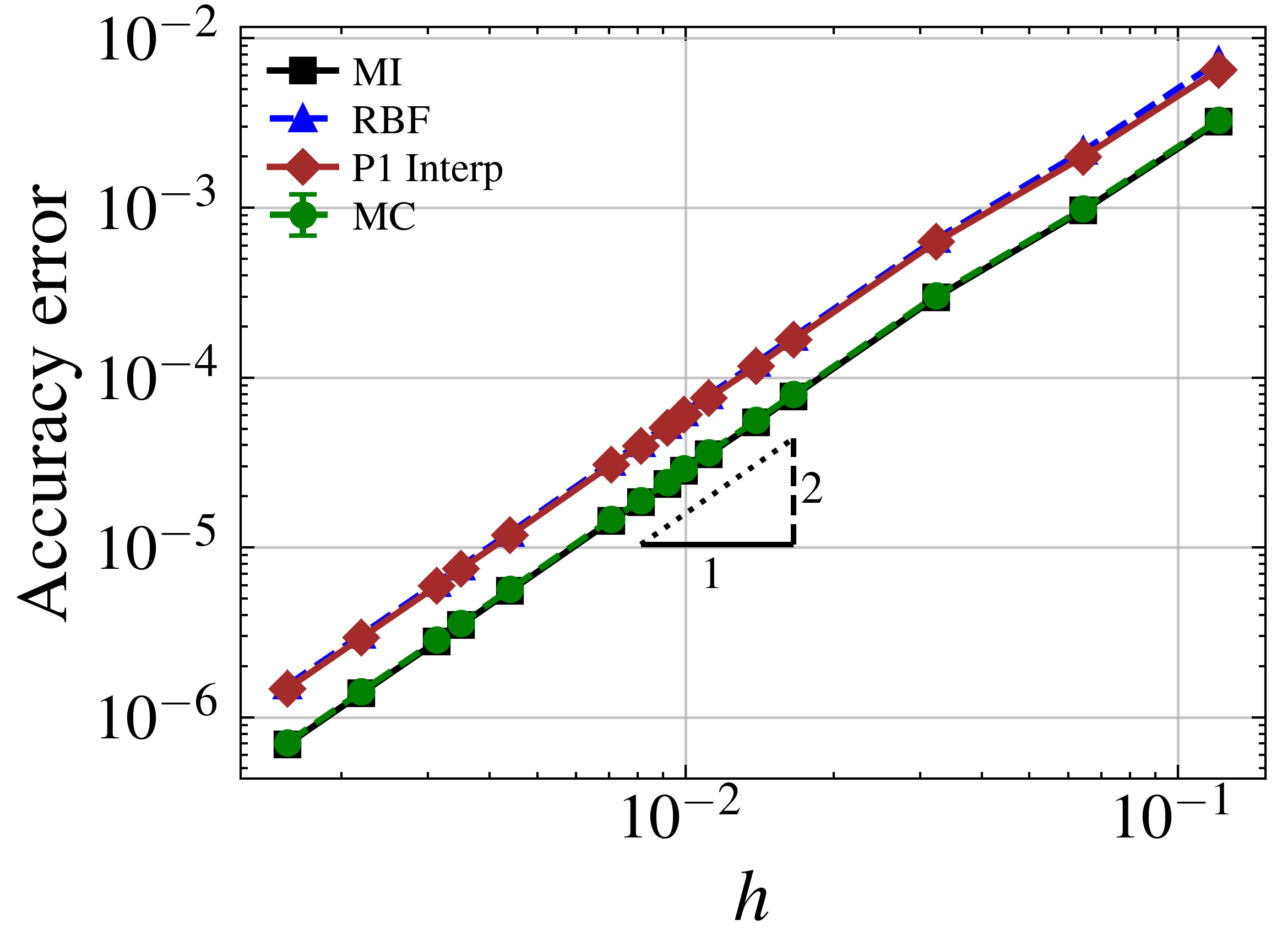}
    \caption{Accuracy error vs. mesh size}
    \label{fig:accuracy-error-cubic}
  \end{subfigure}
  \hfill
  \begin{subfigure}[b]{0.45\textwidth}
    \centering
    \includegraphics[width=\linewidth]{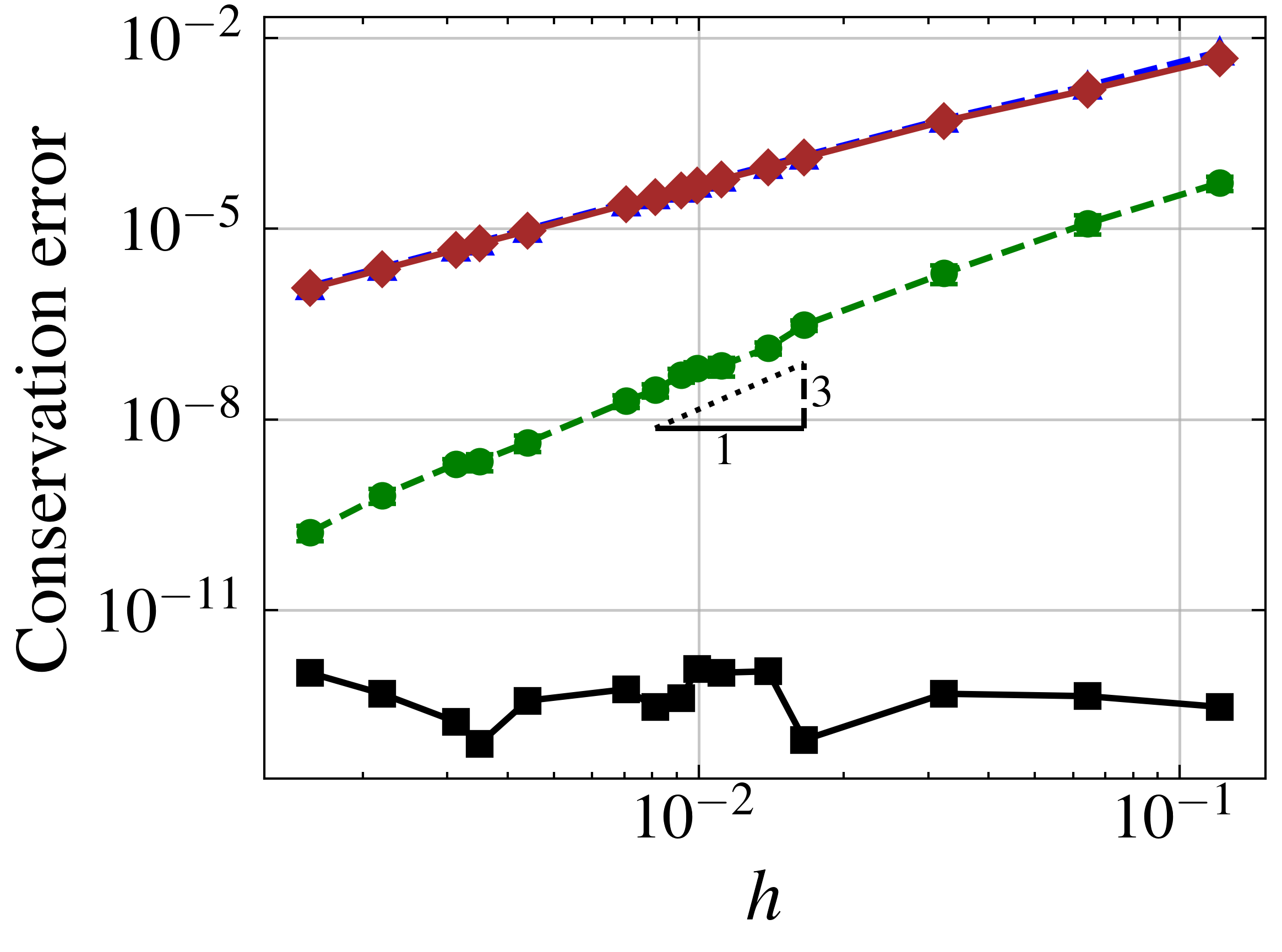}
    \caption{Conservation error vs. mesh size}
    \label{fig:conservation-error-cubic}
  \end{subfigure}
  \caption{Convergence of accuracy and conservation errors for $f_1(x,y)$.}
  \label{fig:convergence-errors-cubic}
\end{figure}

\begin{figure}[htbp]
  \centering
  \begin{subfigure}[b]{0.45\textwidth}
    \centering
    \includegraphics[width=\linewidth]{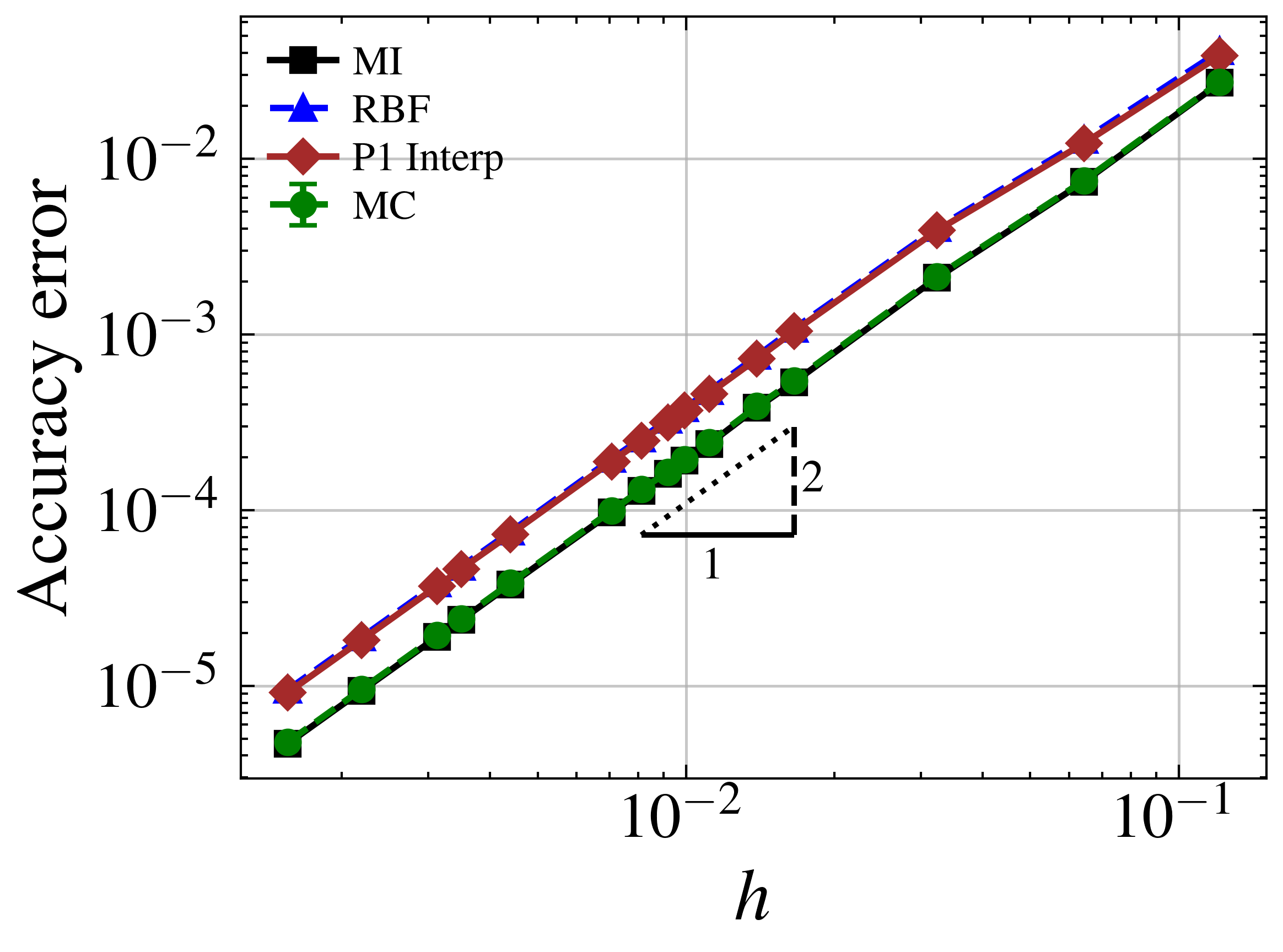}
    \caption{Accuracy error vs. mesh size}
    \label{fig:accuracy-error-trig}
  \end{subfigure}
  \hfill
  \begin{subfigure}[b]{0.45\textwidth}
    \centering
    \includegraphics[width=\linewidth]{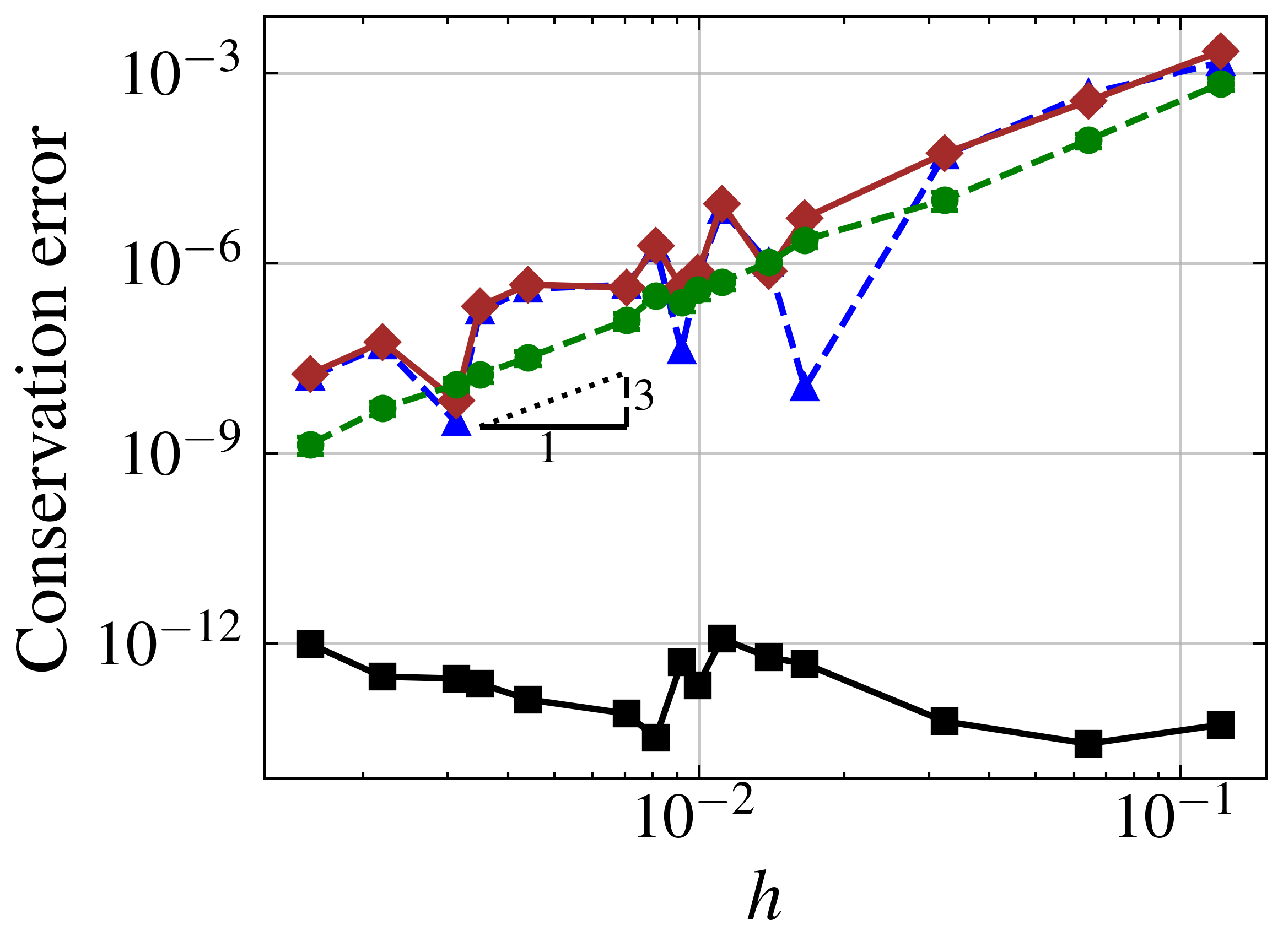}
    \caption{Conservation error vs. mesh size}
    \label{fig:conservation-error-trig}
  \end{subfigure}
  \caption{Convergence of accuracy and conservation errors for $f_2(x,y)$.}
  \label{fig:convergence-errors-trig}
\end{figure}

\begin{figure}[htbp]
  \centering
  \begin{subfigure}[b]{0.45\textwidth}
    \centering
    \includegraphics[width=\linewidth]{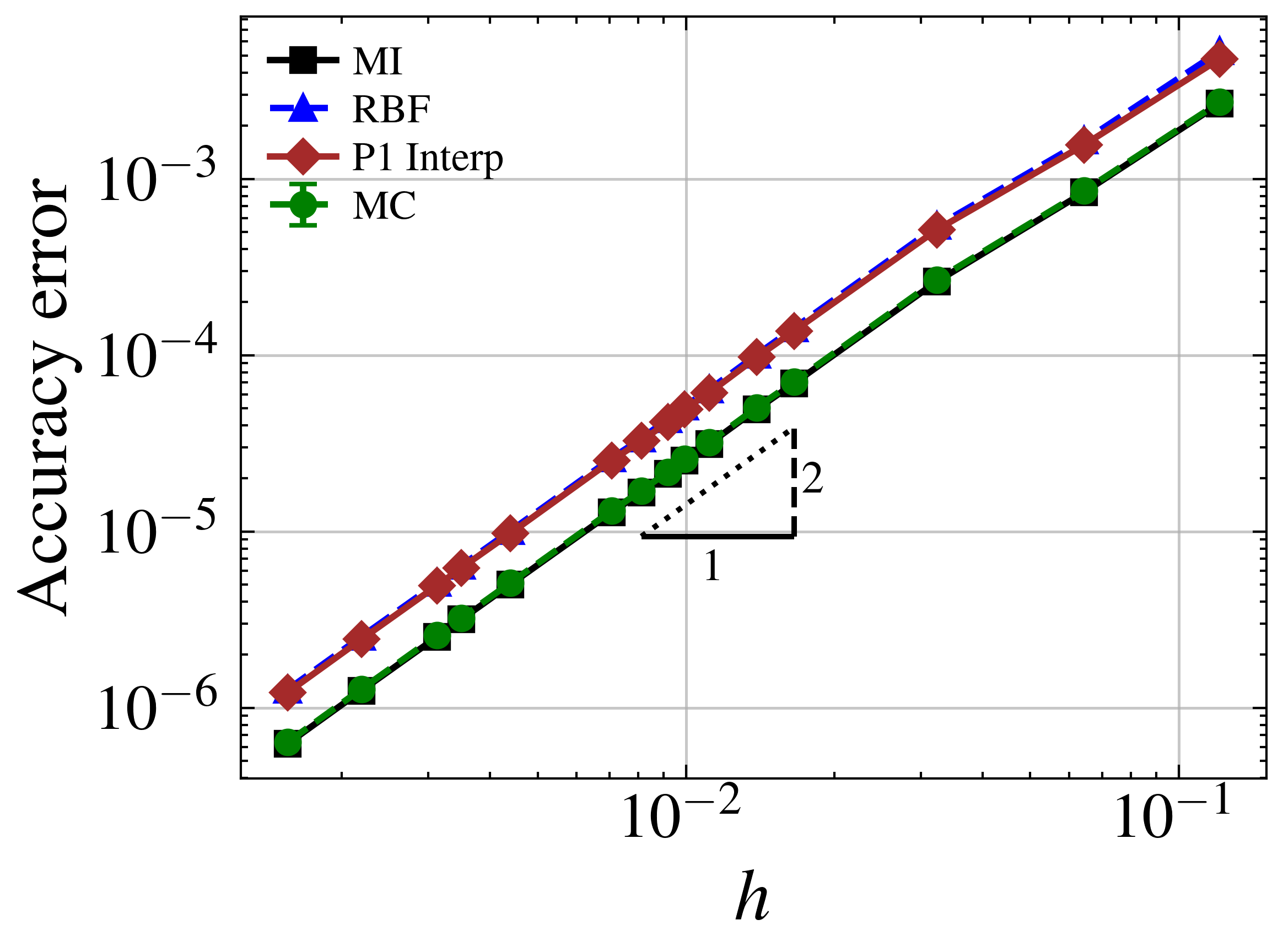}
    \caption{Accuracy error vs. mesh size}
    \label{fig:accuracy-error-exp}
  \end{subfigure}
  \hfill
  \begin{subfigure}[b]{0.45\textwidth}
    \centering
    \includegraphics[width=\linewidth]{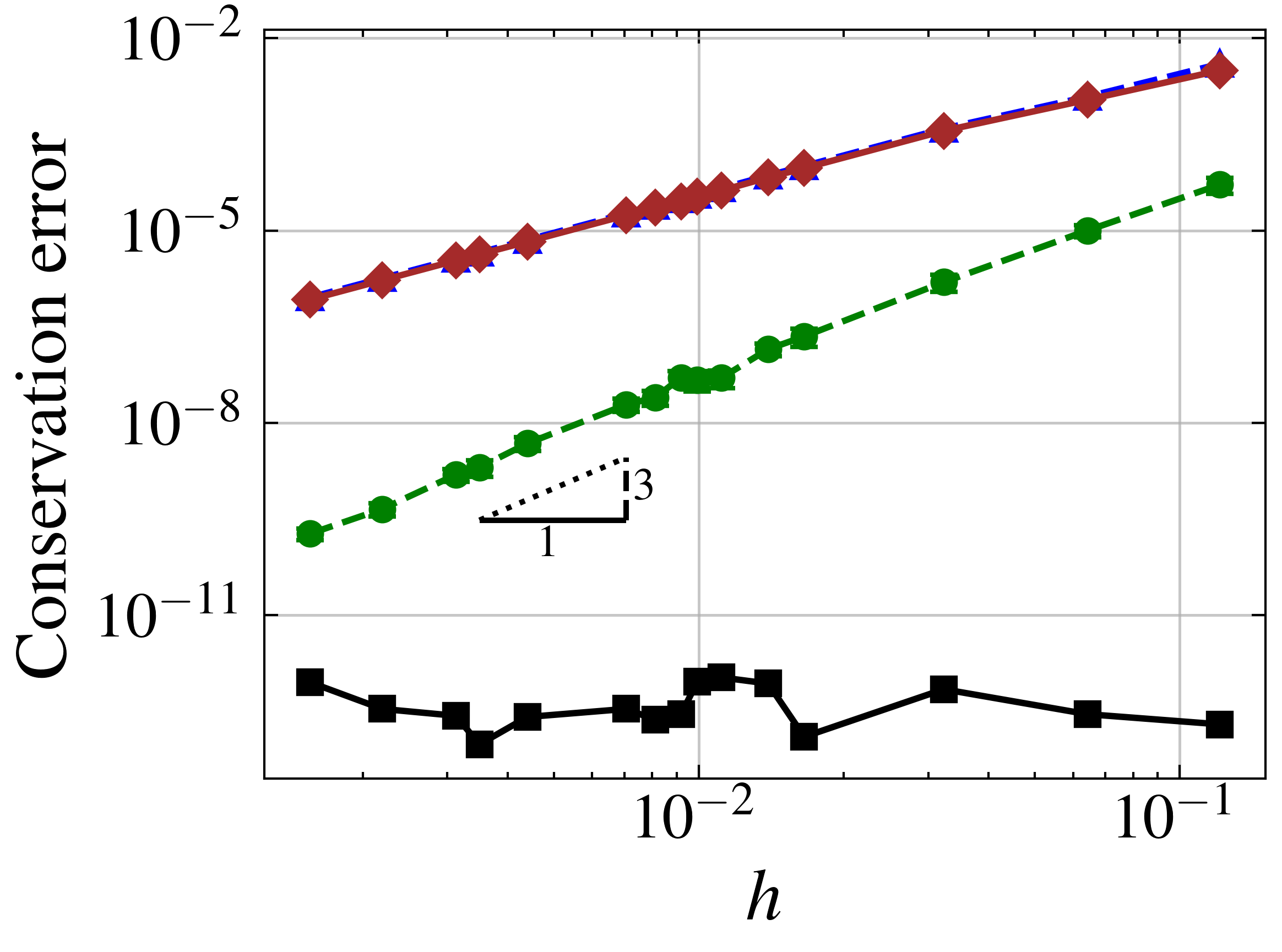}
    \caption{Conservation error vs. mesh size}
    \label{fig:conservation-error-exp}
  \end{subfigure}
  \caption{Convergence of accuracy and conservation errors for $f_3(x,y)$.}
  \label{fig:convergence-errors-exp}
\end{figure}

Figures~\ref{fig:convergence-errors-cubic},
\ref{fig:convergence-errors-trig}, and
\ref{fig:convergence-errors-exp} show the accuracy and conservation errors for
\(f_1(x,y)\), \(f_2(x,y)\) and \(f_3(x,y)\) respectively. The accuracy results show that mesh intersection (MI), radial basis functions (RBF), first-order finite element interpolation (P1), and control variate Monte Carlo (MC) all exhibit an
approximately second-order convergence trend with respect to the mesh size
\(h\). This agrees with Eq.~\eqref{eq:accuracy_bound_interpretation} that for linear
finite elements \((p=1)\), the deterministic approximation contribution scales
as \(\mathcal O(h^{p+1})=\mathcal O(h^2)\) in the \(L^2\) norm. The MC curves
follow the same asymptotic order, showing that the control-variate Monte Carlo
approximation does not change the finite-element approximation order of the
target space. Instead, the control variate reduces the stochastic contribution
to the load-vector error by applying Monte Carlo sampling only to the residual
\(f^s-g\).

Across all three functions, the MC accuracy error is below the RBF and P1
interpolation errors and follows the same asymptotic trend as the
mesh-intersection result. The MC method approaches this behavior because the deterministic control term captures the leading
target-space contribution and only the residual is sampled.

The conservation-error results show a stronger separation between the methods.
The MI method preserves the global integral to machine precision level, as expected
from supermesh-based conservative Galerkin projection. In contrast, RBF and P1
interpolation do not enforce global conservation by construction, and their
conservation errors remain several orders of magnitude larger than those of MI
and MC. For \(f_2(x,y)\), the RBF and P1 interpolation curves show
small fluctuations at some refinement levels. These fluctuations
are likely due to the sensitivity of interpolation-based transfers, especially for
an oscillatory field. In contrast, the MC curve remains smoother and
below the interpolation-based errors, indicating that the control-variate
Galerkin formulation provides a more stable approximation across the tested
mesh sequence. Similarly, the MC conservation errors decrease systematically with mesh refinement and
show an approximately third-order trend, as indicated by the reference slope in
the conservation plots. This observation is consistent with the conservation error estimate Eq.~\eqref{eq:conservation_bound_interpretation}.  Therefore,
the numerical results demonstrate that the reduced regularity of the source finite element space does not significantly degrade the estimated convergence rates.

\subsection{Iterative Analysis}
This numerical experiment examines the long-term behavior of each transfer operator under repeated remapping. We initiate from a reference field on the first mesh and
apply a round-trip map (first mesh $\rightarrow$ second mesh $\rightarrow$ first mesh) for a prescribed number of iterations without introducing any additional modifications to the field. After each round-trip, both the accuracy and conservation errors are evaluated on the first mesh relative to the initial (reference) field values. This analysis provides insight into whether repeated
application of a transfer operator introduces accumulating errors in the
solution and/or integral quantities.  Unlike the convergence study, which
evaluates continuous norms using supermesh integration, the metrics below are computed on the first mesh: accuracy is measured in the discrete
$\ell^2$ norm of the degrees of freedom, and conservation is evaluated by quadrature of the reconstructed field on the first mesh. 
\begin{figure}[htbp]
  \centering
  \begin{subfigure}[t]{0.3\linewidth}
    \centering
    \includegraphics[scale = 0.2,width=\linewidth]{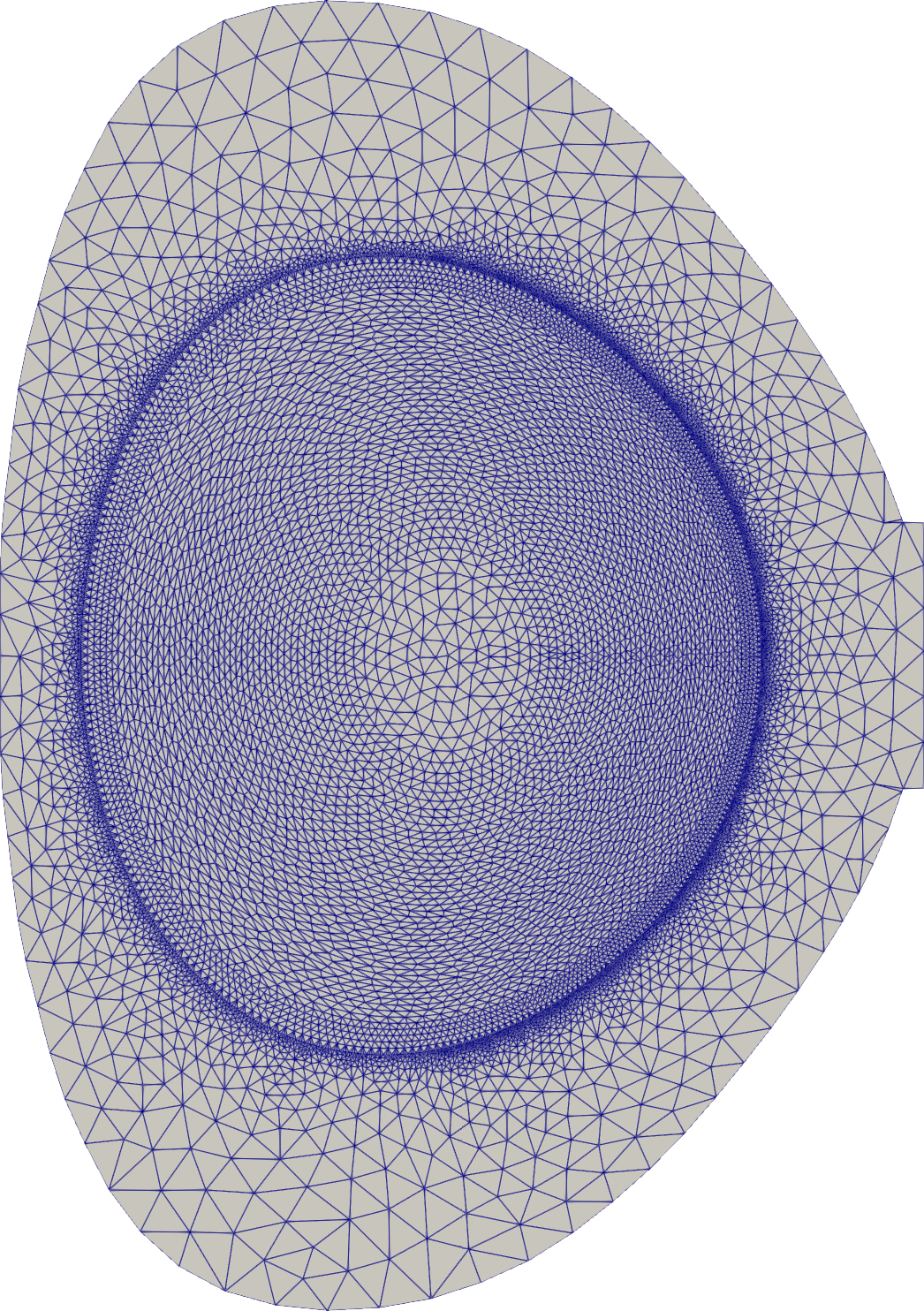}
    \caption{Field following mesh with 20319 elements of the LTX reactor used by the XGC code for evolving the plasma profiles.}
    \label{fig:ltx-xgc-mesh}
  \end{subfigure}
  \hfill
  \begin{subfigure}[t]{0.3\linewidth}
    \centering
    \includegraphics[scale = 0.2, width=\linewidth]{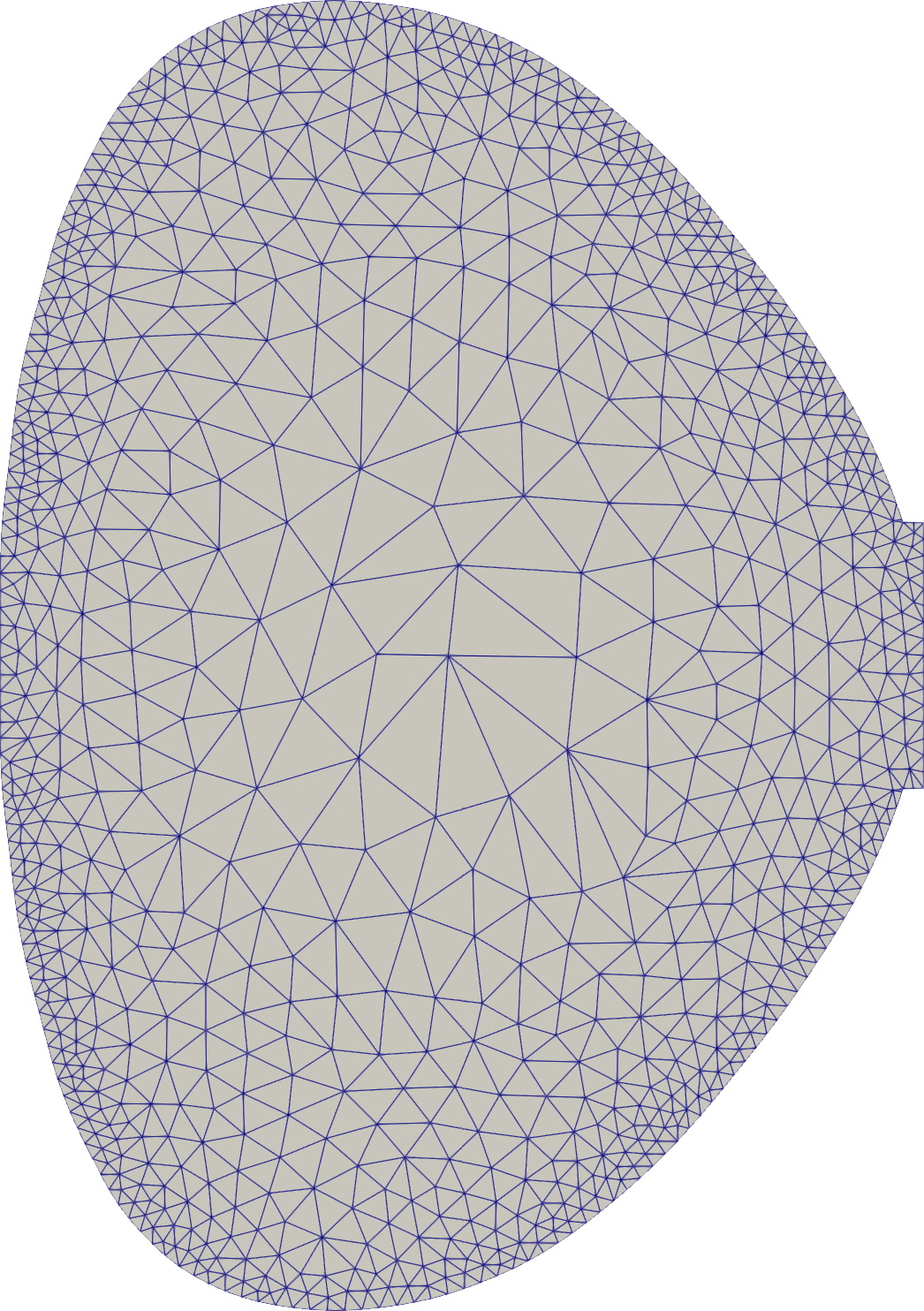}
    \caption{Mesh with 1784 elements of the LTX reactor used by the DEGAS2 code to evolve the neutral particle profiles.}
    \label{fig:ltx-degas2-mesh}
  \end{subfigure}
  \caption{Meshes of the LTX reactor}
  \label{fig:ltx-meshes}
\end{figure}

\paragraph{Discrete (DoF-based) accuracy error}
The relative $\ell^2$ accuracy error is computed as
\begin{equation}
E_{\ell^2}
=
\frac{\|f^{\text{approx}} - f^{\text{ref}}\|_2}{\|f^{\text{ref}}\|_2}
=
\frac{
\left(\sum_{i=1}^{\mathcal{N}}\left(f_i^{\text{approx}}-f_i^{\text{ref}}\right)^2\right)^{1/2}
}{
\left(\sum_{i=1}^{\mathcal{N}}\left(f_i^{\text{ref}}\right)^2\right)^{1/2}
},
\end{equation}
where $\mathcal{N}$ is the total number of degrees of freedom (control points) on the
first mesh, $f^{\text{ref}}$ denotes the initial (reference) DoF values, and $f^{\text{approx}}$ denotes the DoF values after a given number of round-trip iterations.

\paragraph{Discrete (mesh-based) conservation error}
The relative conservation error is computed as
\begin{equation}
E_{\text{mass}}^{\mathrm{M1}}
=
\frac{
\left|
\int_{\Omega} f^{\text{approx}}(x)\,\mathrm{d}\Omega -
\int_{\Omega} f^{\text{ref}}(x)\,\mathrm{d}\Omega
\right|
}{
\left|
\int_{\Omega} f^{\text{ref}}(x)\,\mathrm{d}\Omega
\right|
},
\end{equation}
where the integrals are evaluated using a quadrature rule on the first mesh.

For this study we use the LTX reactor configuration, where the reference field \(f_2(x,y)\) is defined
on the XGC mesh, shown in Figure~\ref{fig:ltx-xgc-mesh}, and repeated mapping is performed through the
DEGAS2 mesh, shown in Figure~\ref{fig:ltx-degas2-mesh}. The same transfer operators considered in the convergence study are used here: mesh intersection (MI), radial-basis-function interpolation (RBF), and the proposed control-variate Monte Carlo method (MC). For the MC-based transfer, \(N=20\) samples are used per target element.
\begin{figure}[htbp]
  \centering
  \begin{subfigure}[b]{0.45\textwidth}
    \centering
    \includegraphics[width=\linewidth]{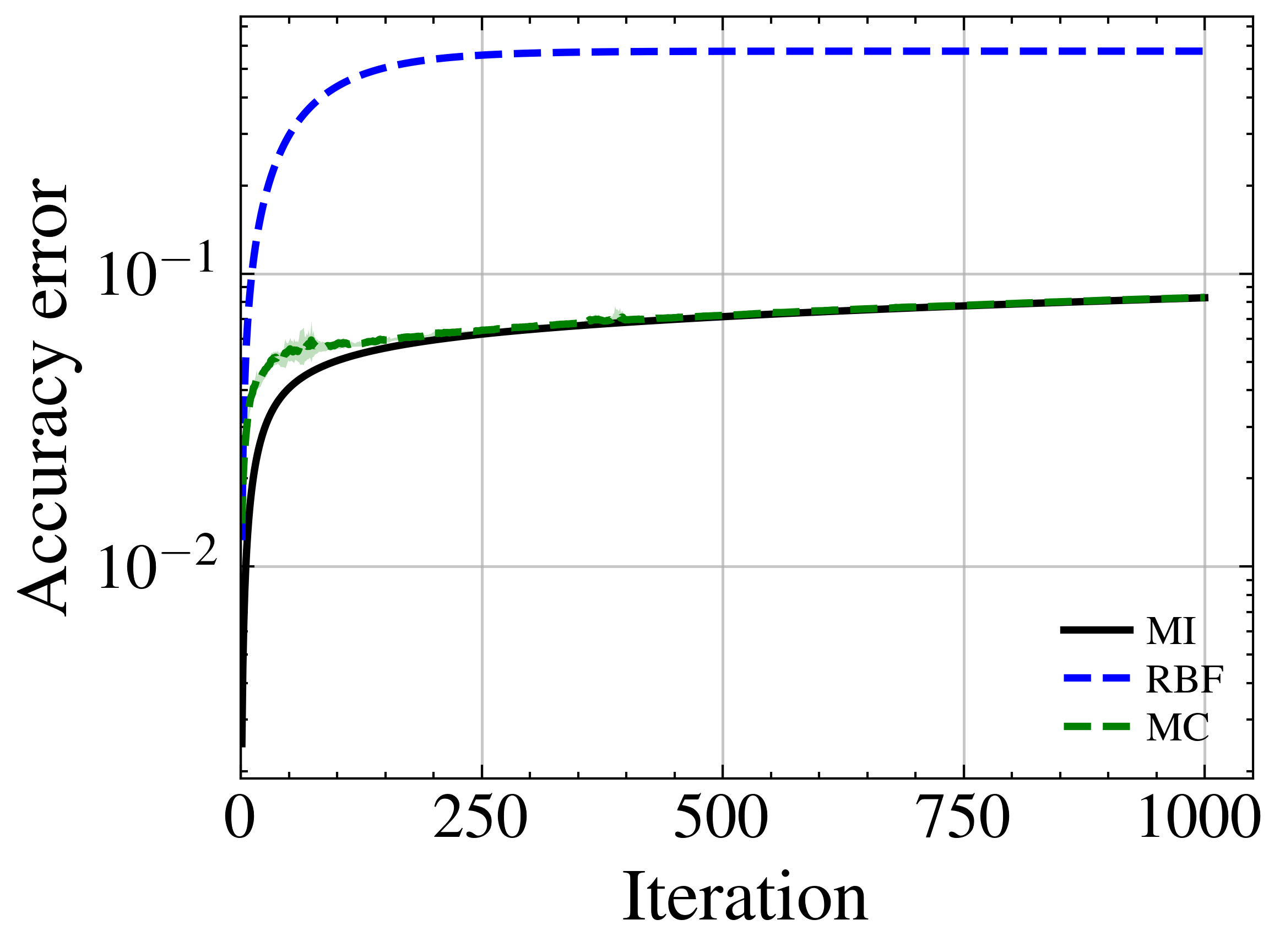}
    \caption{Accuracy error vs. iteration}
    \label{fig:accuracy-vs-iteration}
  \end{subfigure}
  \hfill
  \begin{subfigure}[b]{0.45\textwidth}
    \centering
    \includegraphics[width=\linewidth]{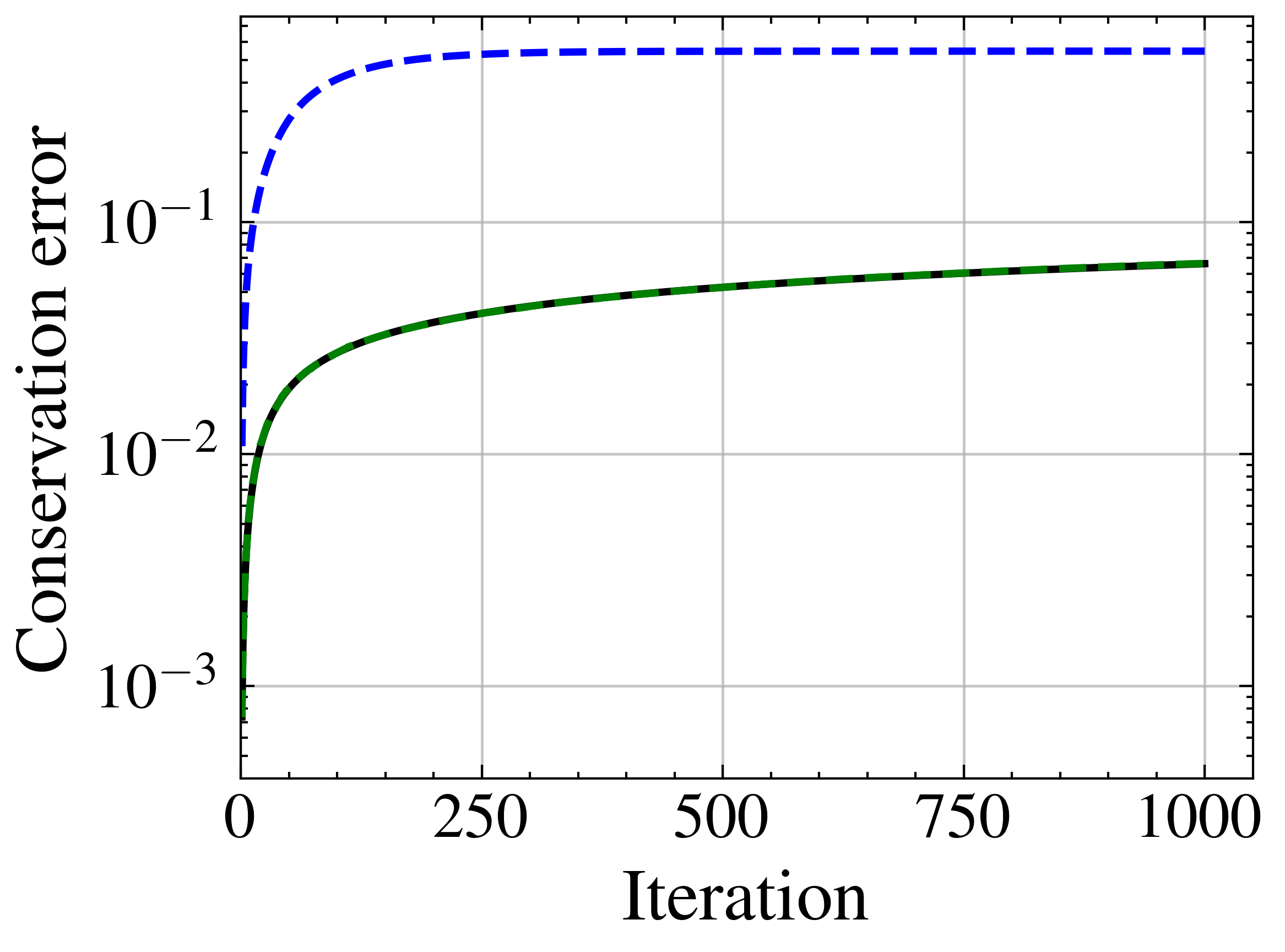}
    \caption{Conservation error vs. iteration}
    \label{fig:conservation-vs-iteration}
  \end{subfigure}
   \caption{Evolution of accuracy and conservation errors under repeated round-trip transfer.}
  \label{fig:errors-vs-iteration}
\end{figure}
 Figures~\ref{fig:accuracy-vs-iteration} and~\ref{fig:conservation-vs-iteration} depict the
evolution of accuracy and conservation errors with iteration. In both metrics, the errors
increase rapidly during the first few round-trip transfers and then transition to a slower
growth. This behavior indicates that most of the transfer-induced error occurs
during the initial remapping steps, after which the repeated application of the operators
approaches a more slowly varying error state.

From Figure~\ref{fig:accuracy-vs-iteration}, the MI and MC methods show nearly identical
long-term accuracy behavior and remain at substantially lower error levels than RBF throughout
the full iteration range. The MC curve closely follows the MI curve, indicating that the
control-variate Monte Carlo transfer preserves the dominant behavior of the conservative
Galerkin projection even under repeated transfer. The small fluctuations and shaded variability
in the MC curve are due to stochastic sampling in the residual contribution, but these variations
are minor compared with the separation between MC and RBF. In contrast, the RBF-based transfer
exhibits a much larger accuracy error. The error rises sharply during the early iterations and
then levels off at a significantly higher value than both MI and MC.

The conservation behavior in Figure~\ref{fig:conservation-vs-iteration} shows a similar
separation between the conservative projection methods and RBF. The MI and MC curves are
almost indistinguishable over the full iteration range, indicating that the control-variate
Monte Carlo method preserves the global integral behavior of the MI transfer very closely in
this repeated remapping experiment. Although the conservation error grows with iteration, the
growth for MI and MC remains much smaller than for RBF. In contrast, the RBF conservation
error increases rapidly and remains well above the MI and MC results, as expected because RBF
interpolation is not conservative by construction and therefore accumulates integral error
under repeated transfers. The iterative experiment reinforces the
conclusion from the convergence study which is that the control-variate Monte Carlo method closely tracks
the conservative MI behavior whereas RBF exhibits substantially larger accuracy and conservation errors under repeated transfer applications.
\subsection{Performance Analysis}
In this section, we examine the scaling properties of the different field transfer methods with respect to the problem size. The sequence of meshes are created with the number of elements ranging from 1000 to 2.3 million using gmsh \cite{geuzaine2009gmsh}. An example of the meshes used for this study are shown in Figure~\ref{fig:meshes_sq}. 

For the mesh-intersection (MI) method, the dominant costs are:
(i) the adjacency-based search used to identify candidate intersecting source
elements for each target element, (ii) R3D clipping to construct the
intersection polytopes \(P_{ts}\), (iii) on-the-fly simplicial decomposition
and numerical integration over these regions, and (iv) a global solve of
Eq.~\eqref{eq:galerkin_system} during each online coupling iteration.
In contrast, the control-variate Monte Carlo method avoids geometric intersection operations; its cost is dominated by
(i) generating sample points on the reference element and mapping them to
physical target elements, (ii) localizing these sample points in the source
mesh, (iii) evaluating the source and control fields at the sample points to
assemble the residual contribution, (iv) assembling the deterministic control
contribution on the target mesh, and (v) a global solve of
Eq.~\eqref{eq:galerkin_system} during each online coupling iteration.
The cost of the RBF method is dominated by (i) a search to identify source
support points around each target point during initialization and (ii) a local
polynomial fitting and evaluation routine during each online coupling
iteration.

The MI and MC methods incur a per-target-element cost, but the nature of the work differs: intersection search, geometric clipping and integration for MI versus repeated point-localization and function evaluation for MC. The RBF method incurs a per node cost.

\begin{figure}[htbp]
  \centering
  \begin{subfigure}[b]{0.48\textwidth}
    \centering
    \includegraphics[width=\linewidth]{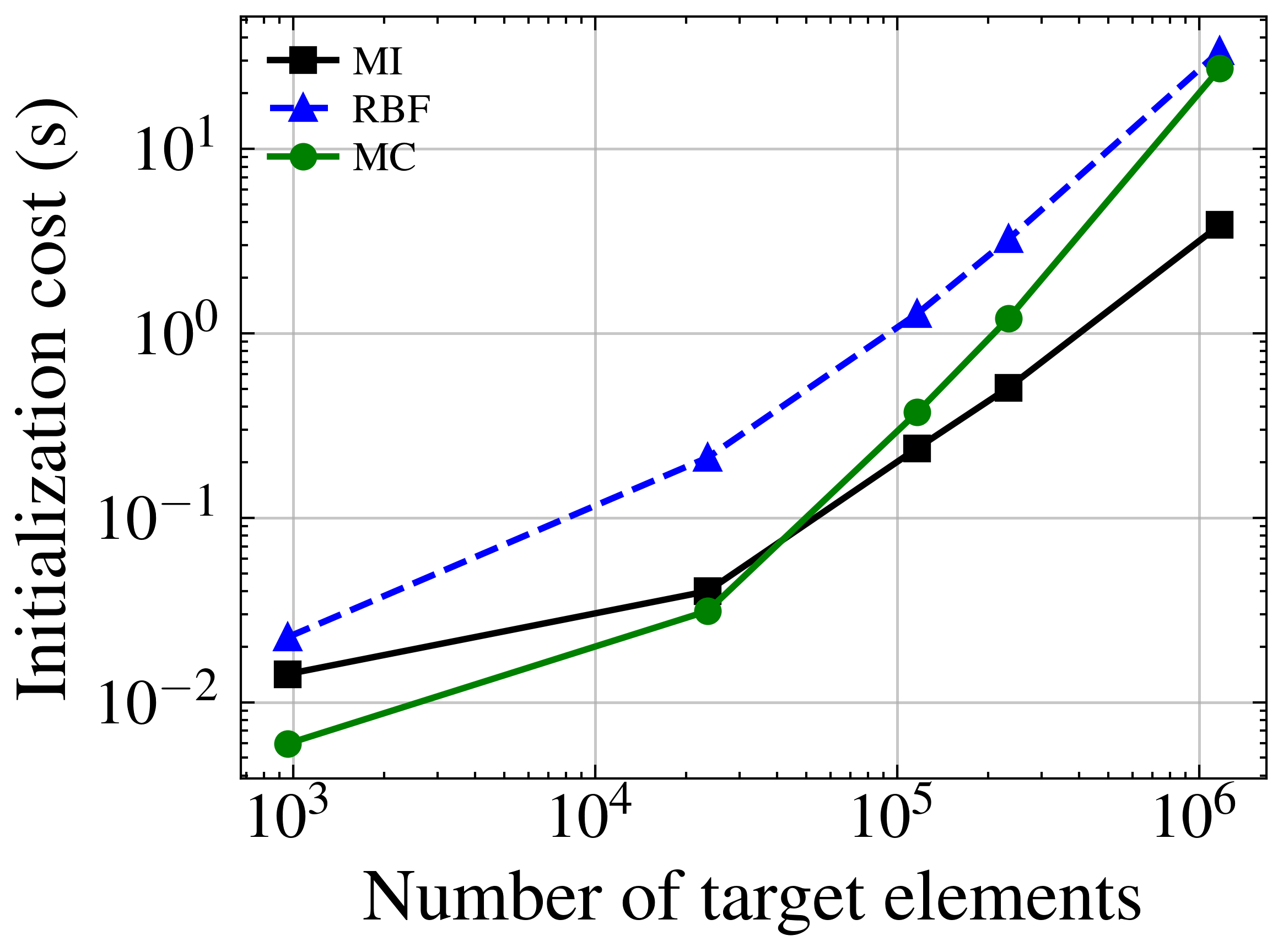}
    \caption{Initialization time}
    \label{fig:initialization-cost}
  \end{subfigure}
  \hfill
  \begin{subfigure}[b]{0.48\textwidth}
    \centering
    \includegraphics[width=\linewidth]{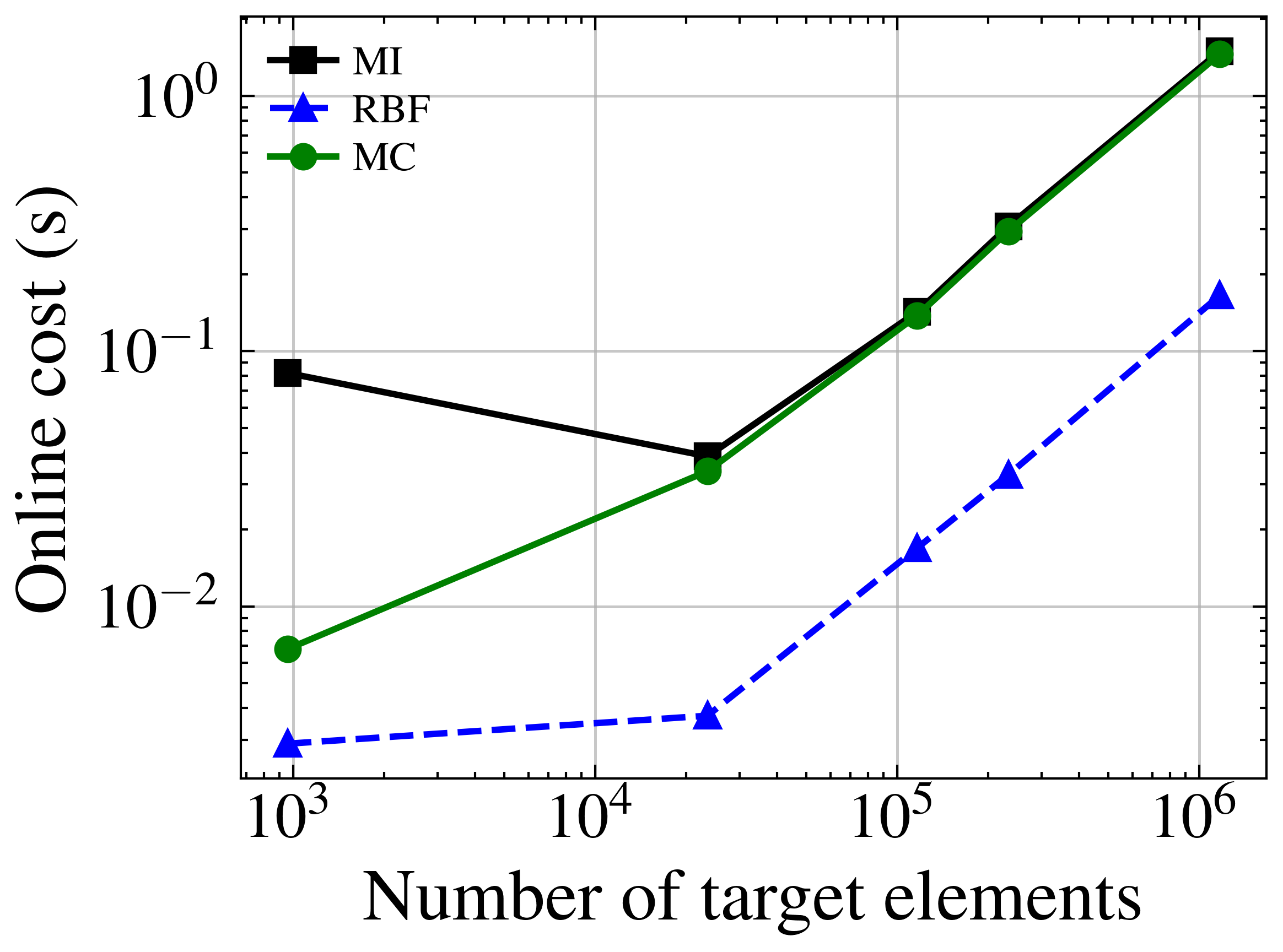}
    \caption{Online time}
    \label{fig:runtime-cost}
  \end{subfigure}
  \caption{Timing results of performing field transfer using MI, MC, and RBF methods on an NVIDIA GeForce RTX 4060 GPU.}
  \label{fig:cost-comparisons}
\end{figure}

All the profiling experiments were run on an NVIDIA GeForce RTX 4060 GPU and the timings reported below correspond to GPU time only. The timings reported in Figures~\ref{fig:initialization-cost} and \ref{fig:runtime-cost} reflect initialization (setup), and online costs, respectively. Initialization is the cost that must be paid once for each unique mesh (intersection computation, localization, and source point set identification). Online costs are those that must be paid for each coupling iteration such as evaluation of integrals and global solve for MI and MC and local solve and evaluation for RBF.

For localization, we make use of a uniform grid search implemented in PCMS \cite{merson2025pcms} with the number of grid cells in each direction given by \(nx=ny=\text{int}(L\sqrt{\# \text{elements}})\), where \(L\) is the characteristic size of the domain. The localization grid size was chosen to balance performance against memory use. The localization method is critical to achieving performance during initialization, however in a black-box coupling scenario, where no source discretization is available, there is limited opportunity to control the localization procedures. Furthermore, some black-box source fields such as machine-learned surrogates may not require explicit global-to-local mapping. 

\begin{figure}
    \centering
   \includegraphics[width=0.6\linewidth]{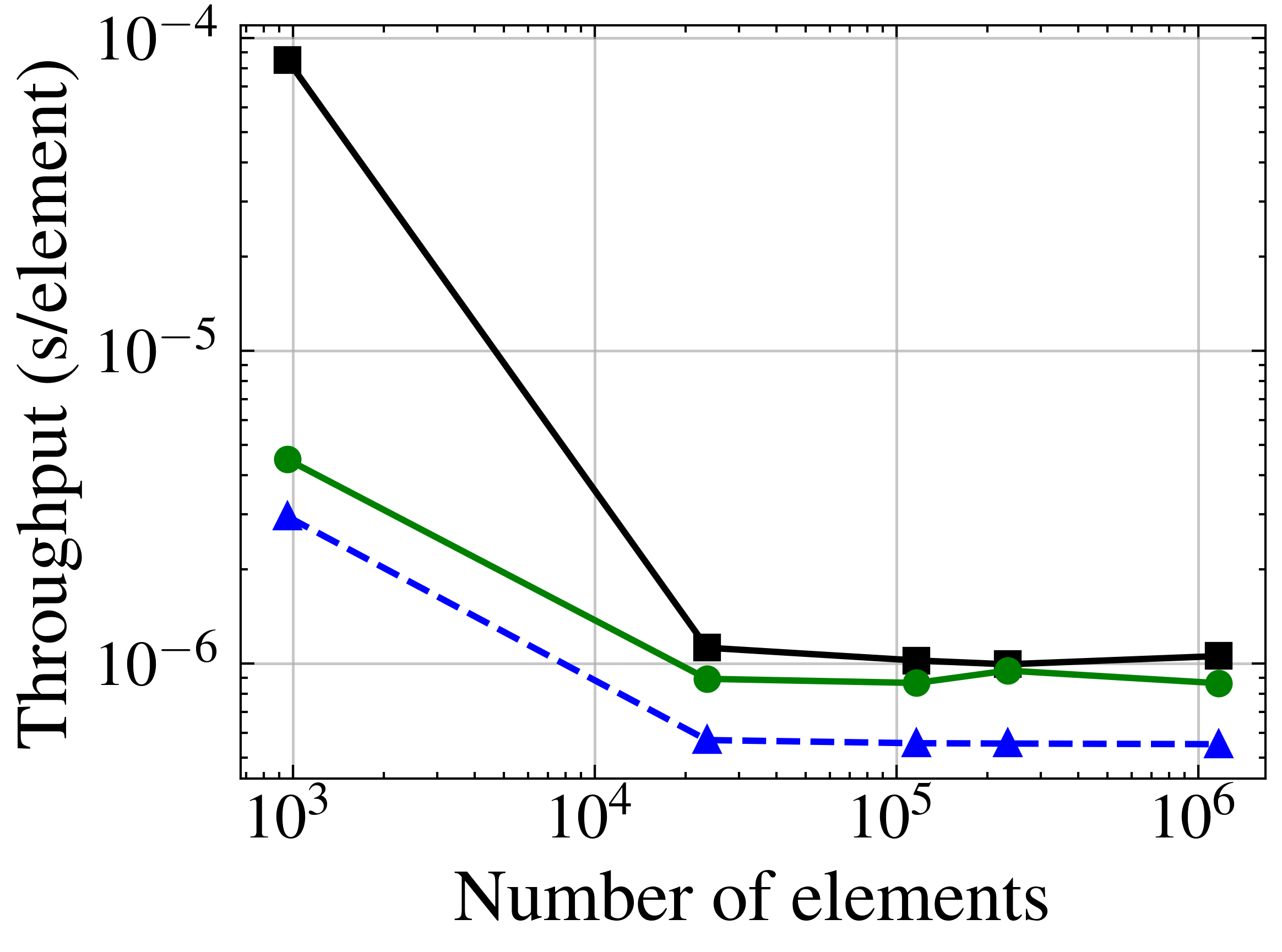}
    \caption{Online cost normalized by number of elements.}
    \label{fig:throughput}
\end{figure}

The initialization cost for each method (Figure~\ref{fig:initialization-cost}) scales with the number of elements. The initialization cost for the RBF method is high due to the iterative method for radius adaptation. Although this increases the initialization cost, minimizing the number of source points reduces the online cost. The initialization cost for the MI and MC cases is similar for the smaller meshes, but the MC initialization cost becomes higher as the mesh size increases. This is because the control-variate Monte Carlo method requires localizing the sample points associated with each target element in the source mesh, as well as localization of the target nodes to evaluate the control field \(g\) needed for the residual construction. The MC results are reported for a single realization as there is minimal variation due to the nearly uniform distribution of target points and sample points in the source mesh. Since the initialization only needs to be done once for a pair of fixed meshes, this cost is amortized across coupling steps.

The online cost (Figure~\ref{fig:runtime-cost}) is the cost of performing a single field transfer operation once the initialization is complete. Due to the local nature of the solution methodology, the RBF method incurs the lowest cost for all mesh sizes. For most mesh sizes, the control-variate Monte Carlo method has an online cost comparable to the mesh-intersection method, with MI slightly higher at the largest mesh size. Both MI and MC maintain similar scaling with the number of elements, since both methods require assembling a Galerkin load contribution and solving the target system during each coupling iteration. The online cost of the MC method does not vary significantly across realizations.

The data from Figure~\ref{fig:runtime-cost} are replotted normalized by the number of elements in Figure~\ref{fig:throughput}. This shows an initial drop in the cost per element as the amount of parallel work increases, which helps amortize the latencies associated with launching GPU work. For larger meshes, the cost per element reaches a plateau, indicating approximately linear scaling with the number of elements. The onset of the plateau starts at around \(10^5\) elements for all three methods. In this regime, RBF has the lowest online cost per element, while MI and MC have comparable per-element costs, with MC remaining slightly lower than MI for the largest mesh sizes.

\section{Application}\label{sec:application}

\begin{figure}[htbp]
  \centering

  \begin{subfigure}[b]{0.44\textwidth}
    \centering
    \includegraphics[width=\linewidth, trim={250pt 50pt 200pt 10pt},
            clip]{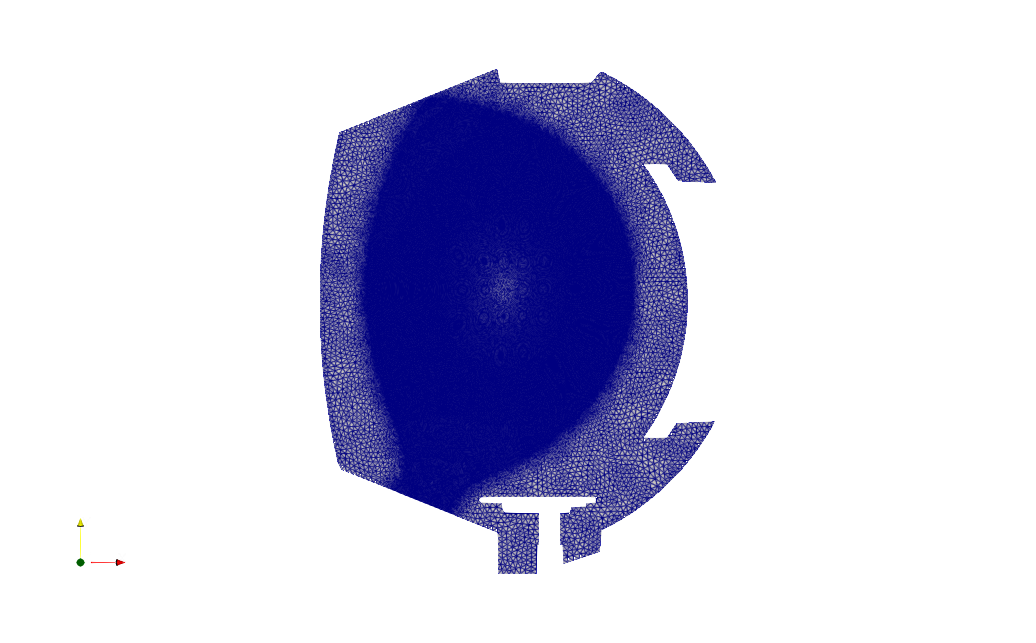}
    \caption{Source mesh}
    \label{fig:source-mesh-llcp-case}
  \end{subfigure}
  \hfill
  \begin{subfigure}[b]{0.44\textwidth}
    \centering
    \includegraphics[width=\linewidth, trim={200pt 50pt 200pt 10pt},
            clip]{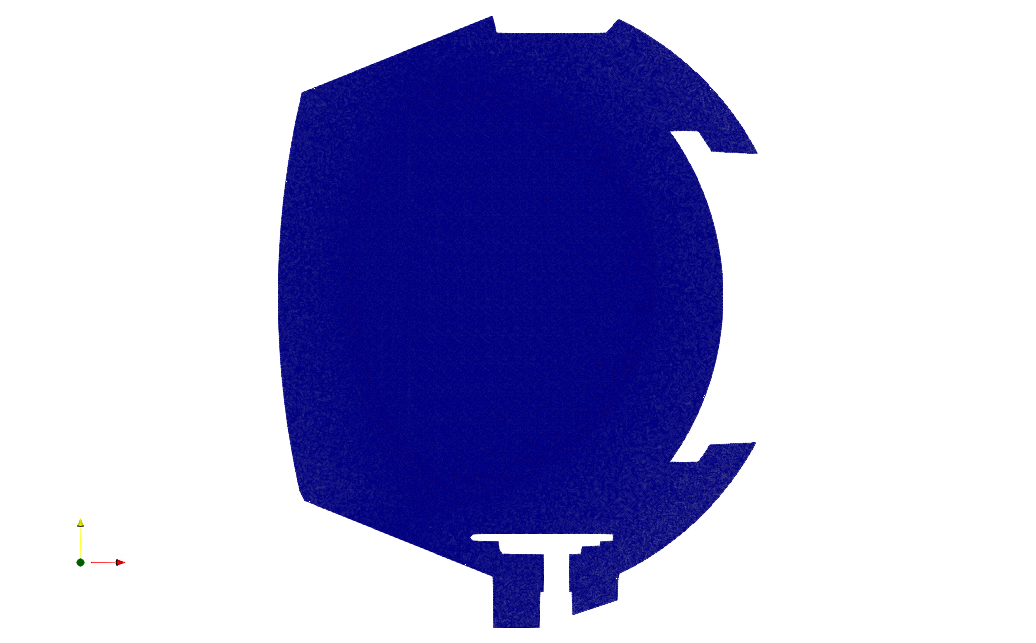}
    \caption{Target mesh}
    \label{fig:target-mesh-llcp-case}
  \end{subfigure}

  \caption{Source and target meshes on WEST reactor geometry used for field transfer. In the core region, the source mesh~(a) is aligned to the magnetic field and has only one element between each flux curve, as required for XGCm simulations. The target mesh~(b) is a general unstructured mesh over the full domain.}
  \label{fig:source-target-mesh-application}
\end{figure}

\begin{figure}[htbp]
    \centering

    \begin{subfigure}[t]{0.44\linewidth}
        \centering
        \includegraphics[
            width=\linewidth,
            trim={250pt 50pt 200pt 10pt},
            clip
        ]{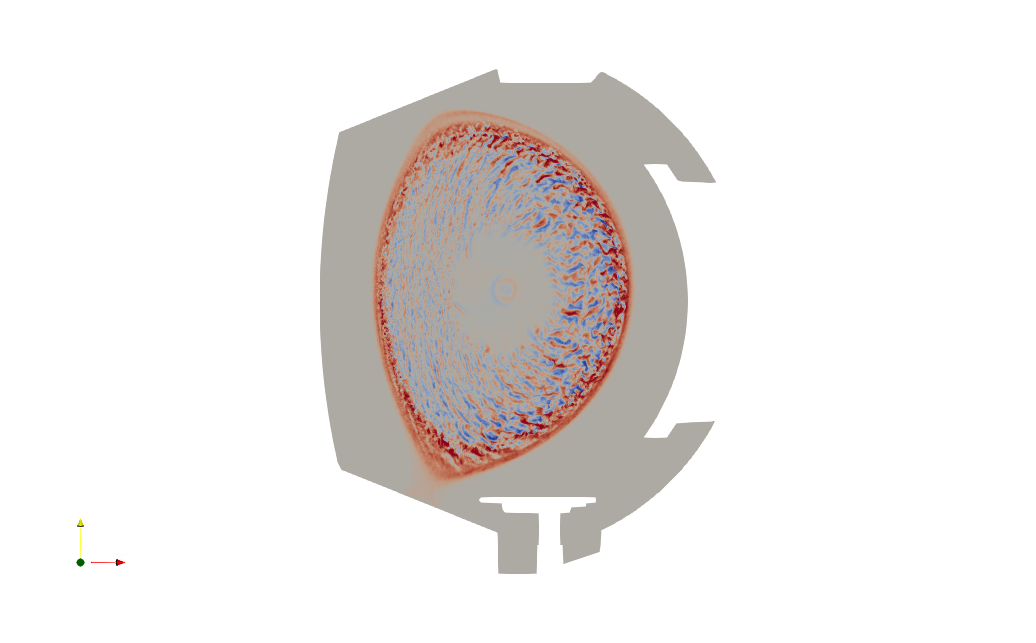}
        \caption{Source field}
        \label{fig:source_field_application}
    \end{subfigure}
    \hfill
    \begin{subfigure}[t]{0.44\linewidth}
        \centering
        \includegraphics[
            width=\linewidth,
            trim={200pt 50pt 200pt 10pt},
            clip
        ]{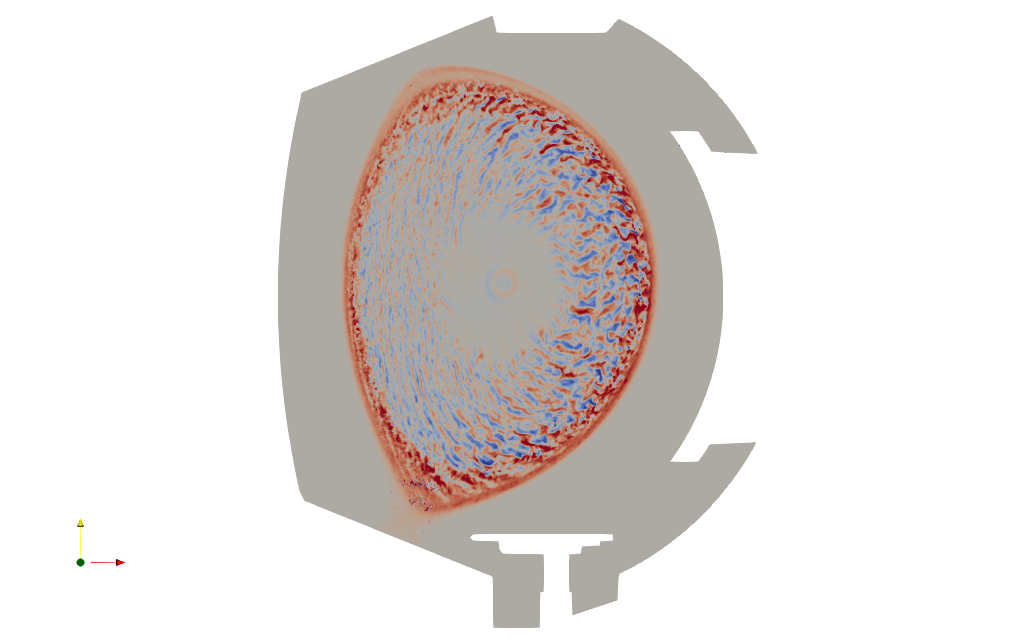}
        \caption{Radial basis}
        \label{fig:target_field_rbf_application}
    \end{subfigure}

    \par\vspace{0.5em}

    \begin{subfigure}[t]{0.44\linewidth}
        \centering
        \includegraphics[
            width=\linewidth,
            trim={200pt 50pt 200pt 10pt},
            clip
        ]{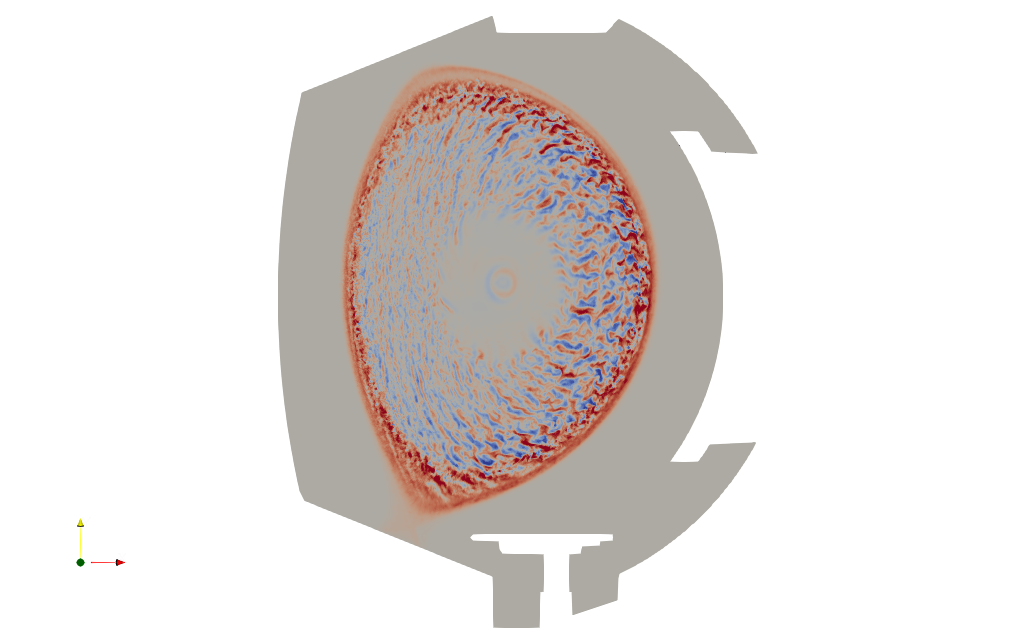}
        \caption{Mesh intersection}
        \label{fig:target_field_mi_application}
    \end{subfigure}
    \hfill
    \begin{subfigure}[t]{0.44\linewidth}
        \centering
        \includegraphics[
            width=\linewidth,
            trim={200pt 50pt 200pt 10pt},
            clip
        ]{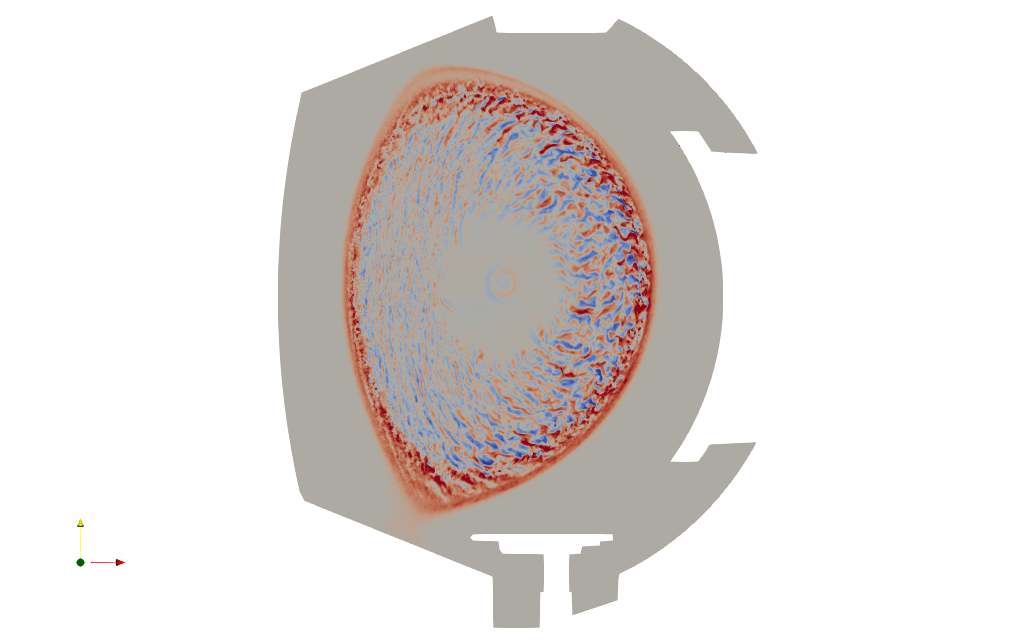}
        \caption{Monte Carlo}
        \label{fig:target_field_mc_application}
    \end{subfigure}

    \par\vspace{0.35em}

    \includegraphics[
        width=0.65\linewidth,
        trim={4pt 1pt 0pt 7pt},
        clip
    ]{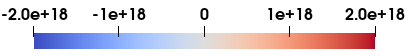}

    \caption{Comparison of field transfer of a representative ion density field from an adiabatic XGCm simulation of the WEST tokamak.}
    \label{fig:west_field_transfer_comparison}
\end{figure}

To demonstrate the application of our method to fusion simulation data, we perform a field transfer on a representative ion density field derived from an adiabatic XGCm \cite{zhang2023unstructured} simulation
of the WEST reactor (Fig.~\ref{fig:west_field_transfer_comparison}a). The source and target meshes for this field transfer example are shown in Fig.~\ref{fig:source-target-mesh-application}. The source mesh is an XGCm mesh with approximately 611,000 elements that is field
aligned in the core region, and only has a depth of a single element between each flux curve.
The target mesh is a general unstructured mesh with approximately 743,000 elements. 

Qualitatively, the fields after transfer shown in Fig.~\ref{fig:west_field_transfer_comparison} look similar. For the particular choice of parameters, the field
transfer using radial basis functions has some points near the X-point that are a visibly incorrect speckle pattern. This indicates one of the pitfalls of the radial basis function
method is that it often requires regularization, and parameter tuning to obtain high-quality field
transfers.

To quantitatively evaluate the quality of the field transfers, we show the accuracy and conservation errors for the WEST data shown in Tab.~\ref{tab:west-errors}. The accuracy errors of MI and MC are nearly identical, while the RBF error is roughly twice as large. The conservation error was close to the solver tolerance for mesh intersection. The conservation error in our Monte Carlo strategy was significantly worse than mesh intersection, for this case, however it was still more than an order of magnitude better than the RBF method.

\begin{table}[h]
\caption{Error measured on field transfer of WEST ion density dataset.}
\label{tab:west-errors}
\centering
\begin{tabular}{c c c}
\hline
\textbf{Method} & \textbf{Accuracy Error} & \textbf{Conservation Error} \\
\hline
RBF
& $1.97 \times 10^{-1}$
& $1.66 \times 10^{-3}$ \\

MI
& $9.15 \times 10^{-2}$
& $3.07 \times 10^{-12}$ \\

MC
& $(9.23 \pm 0.0003)\times 10^{-2}$
& $(4.43 \pm 1.35)\times 10^{-5}$ \\
\hline
\end{tabular}
\end{table}

\section{Conclusion}
\label{sec:conclusion}
In this paper we introduce a novel method for supporting black-box coupling that only relies on pointwise queries, but still provides asymptotic convergence of integral quantities. Such fields include machine-learned surrogates and other non-standard field representations. Our method has been compared against radial basis functions which are often used in black-box coupling as well as the mesh intersection method which cannot be used in black-box coupling due to the need for full information about the source discretization. In our computational experiments, our new method provides better accuracy and conservation error than the radial basis function approach and converges to the level of accuracy and conservation error of the mesh intersection method as the number of sample points is increased.

We also compare the performance of our new method on the GPU compared with the mesh intersection method and radial basis function method. We find that the radial basis function method is fastest for performing evaluations, and our new method is competitive with the mesh intersection method and likely to improve with further optimization and implementation of variance reduction.

One limitation of our method is that it is non-deterministic since it is a Monte Carlo method. This limitation has been somewhat mitigated by the use of the control variate. However, additional improvements are likely to stem from better sampling schemes. The extensibility of Monte Carlo integration to high dimensions also makes this method an excellent candidate to support five and six dimensional transfers needed for distribution function coupling.

\section*{Acknowledgement}
This research was supported by the U.S. Department of Energy, Office of Science Office of Advanced Scientific Computing Research, Scientific Discovery through Advanced Computing (SciDAC) Program through multiple grants including the FASTMath SciDAC Institute (DE-SC0021285 \& DE-AC52-0TNA27344), and multiple Fusion Energy SciDAC subcontracts including StellFoundry: High-fidelity Digital Models for Fusion Pilot Plant Design (DE-AC02-09CH11466), Computational Evaluation and Design of Actuators for Core-Edge Integration (CEDA) (DE-AC02-09CH11466), HifiStell: High-Fidelity Simulations for Stellarators (DE-SC0024548) and Center for Advanced Simulation of RF - Plasma - Material Interactions (DE-SC0024369). This research is also supported through a DOE Fusion Innovation Research Engine (FIRE) Collaboratives Program through a grant titled Mitigating Risks from Abrupt Confinement Loss (MIRACL) (DE-AC02-09CH11466). This research is also supported through a DOE ASCR SBIR entitled Geometry and Meshing Technologies to Support Fusion Energy System Simulations (DE-SC0024838). 
Any opinions, findings, and conclusions or recommendations expressed in this material are those of the author(s) 
and do not necessarily reflect the views of the U.S. Department of Energy.

\section*{Declaration of generative AI and AI-assisted technologies in the manuscript preparation process}
During the preparation of this work, the author(s) used ChatGPT (OpenAI) and Perplexity AI to identify and locate relevant academic papers, and Overleaf’s research‑tailored AI language feedback features for language refinement. After using these tools, the author(s) thoroughly reviewed, verified, and edited all content to ensure accuracy and maintain full responsibility for the integrity and originality of the manuscript.

\appendix

\section{Accuracy error bound}\label{app:accuracy_error_bound}
Let \(\mathcal M_t\) denote the finite element mesh partitioning the domain
\(\Omega \subset \mathbb R^d\) into elements \(\Omega_t\), such that
\[
    \Omega
    =
    \bigcup_{t\in\mathcal M_t}\Omega_t.
\]
Let \(h_t\) denote the maximal edge length of the element \(t\in \mathcal M_t\), and let
\[
    h=\max_{t\in \mathcal M_t} h_t .
\]
For a shape-regular and quasi-uniform mesh in \(d\) dimensions, the element
measure satisfies
\begin{equation}
    |\Omega_t|_{\max} \leq C h^d ,
    \label{eq:elemental_volume_equation}
\end{equation}
where \(C\) is independent of \(h\). Throughout the appendices, \(C\) denotes a generic positive constant, independent of \(h\), \(N\), and \(\alpha\), whose value may differ between occurrences.

We use the control-variate Monte Carlo load-vector approximation derived in
Section~\ref{subsec:cv_mc_galerkin_projection}. In particular, with
\(g=I_hf^s\), the sampled residual is
\begin{equation}
    \widetilde z_{\alpha}
    =
    f^s-\alpha g .
    \label{eq:app_definition_z_tilde_alpha}
\end{equation}
Using Eq.~\eqref{eq:cv_load_expanded}, the control-variate approximation of
the load-vector entry is
\begin{equation}
    \widehat b_i 
    =
    \alpha
    \int_\Omega g(x)\psi_i(x)\,dx
    +
    \sum_{t\in\mathcal M_t}
    \frac{|\Omega_t|}{N}
    \sum_{k=1}^{N}
    \widetilde z_{\alpha}(X_{t,k})\psi_i(X_{t,k}) .
    \label{eq:cv_load_vector_entry}
\end{equation}
Since the deterministic control contribution is assembled exactly, the
load-vector error is only due to the sampled residual term:
\begin{align}
    b_i-\widehat b_i
    &=
    \sum_{t\in\mathcal M_t}
    \left[
    \int_{\Omega_t}
    \widetilde z_{\alpha}(x)\psi_i(x)\,dx
    -
    \frac{|\Omega_t|}{N}
    \sum_{k=1}^{N}
    \widetilde z_{\alpha} (X_{t,k})\psi_i (X_{t,k})
    \right].
    \label{eq:load_vector_error_entry}
\end{align}

Let \(\mathbf M\) be the finite element mass matrix with entries
\begin{equation}
    M_{ij}
    =
    \int_\Omega \psi_i(x)\psi_j(x)\,dx .
    \label{eq:mass_matrix_definition}
\end{equation}
The exact and approximate Galerkin systems are
\begin{equation}
    \mathbf M\mathbf c = \mathbf b,
    \label{eq:exact_galerkin_system}
\end{equation}
and
\begin{equation}
    \mathbf M\widehat{\mathbf c}
    =
    \widehat{\mathbf b}.
    \label{eq:approximate_galerkin_system}
\end{equation}
Subtracting Eq.~\eqref{eq:approximate_galerkin_system} from
Eq.~\eqref{eq:exact_galerkin_system} gives
\begin{equation}
    \mathbf M(\mathbf c-\widehat{\mathbf c})
    =
    \mathbf b-\widehat{\mathbf b} .
    \label{eq:coefficient_error_system}
\end{equation}
Multiplying both sides by \(\mathbf M^{-1}\), we obtain
\begin{equation}
    \mathbf c-\widehat{\mathbf c}
    =
    \mathbf M^{-1}
    \left(
    \mathbf b-\widehat{\mathbf b} 
    \right).
    \label{eq:solution_coefficient_vector_error}
\end{equation}

The exact and approximate solution functions are defined as
\begin{equation}
    f^t(x)
    =
    \sum_{j=1}^{\mathcal N_t} c_j\psi_j(x),
    \label{eq:exact_solution}
\end{equation}
and
\begin{equation}
    \widehat f^t(x)
    =
    \sum_{j=1}^{\mathcal N_t} \widehat c_j\psi_j(x),
    \label{eq:approximate_solution}
\end{equation}
respectively. Subtracting Eq.~\eqref{eq:approximate_solution} from
Eq.~\eqref{eq:exact_solution}, we get
\begin{equation}
    f^t-\widehat f^t
    =
    \sum_{j=1}^{\mathcal N_t}(c_j-\widehat c_j)\psi_j .
    \label{eq:error_definition}
\end{equation}
Taking the \(L^2\) norm of Eq.~\eqref{eq:error_definition} and squaring gives
\begin{align}
    \|f^t-\widehat f^t\|_{L^2(\Omega)}^2
    &=
    \int_\Omega
    \left(
    \sum_{j=1}^{\mathcal N_t}
    (c_j-\widehat c_j)\psi_j
    \right)^2 dx
    \notag \\
    &=
    \sum_{i=1}^{\mathcal N_t}
    \sum_{j=1}^{\mathcal N_t}
    (c_i-\widehat c_i)(c_j-\widehat c_j)
    \int_\Omega \psi_i\psi_j\,dx
    \notag \\
    &=
    (\mathbf c-\widehat{\mathbf c})^T
    \mathbf M
    (\mathbf c-\widehat{\mathbf c}).
    \label{eq:error_norm}
\end{align}
Substituting Eq.~\eqref{eq:solution_coefficient_vector_error} into
Eq.~\eqref{eq:error_norm}, we obtain
\begin{equation}
    \|f^t-\widehat f^t\|_{L^2(\Omega)}^2
    =
    \left(
    \mathbf b-\widehat{\mathbf b} 
    \right)^T
    \mathbf M^{-1}
    \left(
    \mathbf b-\widehat{\mathbf b} 
    \right).
    \label{eq:error_norm_simplified}
\end{equation}

We now rewrite the load-vector error in vector form. Define the basis vector
\[
    \boldsymbol\psi(x)
    =
    \left(
    \psi_1(x),\ldots,\psi_{\mathcal N_t}(x)
    \right)^T .
\]
For each element \(\Omega_t\), define the vector-valued Monte Carlo error 
\begin{equation}
    \mathbf e_t
    =
    \int_{\Omega_t}
    \widetilde z_{\alpha}(x)\boldsymbol\psi(x)\,dx
    -
    \frac{|\Omega_t|}{N}
    \sum_{k=1}^{N}
    \widetilde z_{\alpha}(X_{t,k})
    \boldsymbol\psi(X_{t,k}) .
    \label{eq:element_vector_error}
\end{equation}
For each element, define the vector-valued integrand
\begin{equation}
     \mathbf q_t(x)
    =
    \widetilde z_{\alpha}(x)\boldsymbol\psi(x), 
    \label{eq:vector_valued_integrand}
\end{equation}
and the mean value of this integrand over the element \(\Omega_t\) 
\begin{equation}
\boldsymbol \mu_t = \frac{1}{|\Omega_t|}\int _{\Omega_t} \mathbf q_t(x)\,dx.
\label{eq:mean_value_integrand}
\end{equation}
Using this definition, the vector-valued Monte Carlo error becomes
\begin{equation}
    \mathbf e_t
    = \frac{|\Omega_t|}{N}
    \sum_{k=1}^{N}
     (\boldsymbol \mu_t - \mathbf q_t (X_{t,k})).
    \label{eq:element_vector_error_q_mu_term}
\end{equation}
Then Eq.~\eqref{eq:load_vector_error_entry} can be written compactly as
\begin{equation}
    \mathbf b-\widehat{\mathbf b} 
    =
    \sum_{t\in\mathcal M_t}\mathbf e_t .
    \label{eq:load_vector_error_element_sum}
\end{equation}
Substituting Eq.~\eqref{eq:load_vector_error_element_sum} into
Eq.~\eqref{eq:error_norm_simplified}, we get
\begin{equation}
    \|f^t-\widehat f^t\|_{L^2(\Omega)}^2
    =
    \left(
    \sum_{t\in\mathcal M_t}\mathbf e_t
    \right)^T
    \mathbf M^{-1}
    \left(
    \sum_{t'\in\mathcal M_t}\mathbf e_{t'}
    \right).
    \label{eq:error_norm_element_sum}
\end{equation}

Taking expectation on both sides gives
\begin{align}
    \mathbb E\left[
    \|f^t-\widehat f^t\|_{L^2(\Omega)}^2
    \right]
    &=
    \sum_{t\in\mathcal M_t}
    \mathbb E\left[
    \mathbf e_t^T\mathbf M^{-1}\mathbf e_t
    \right]
    \notag \\
    &\quad+
    2\sum_{t<t'}
    \mathbb E\left[
    \mathbf e_t^T\mathbf M^{-1}\mathbf e_{t'}
    \right].
    \label{eq:error_norm_expectation_expanded}
\end{align}
Since the samples are generated independently on each element and each
element-wise estimator is unbiased, we have
\[
    \mathbb E[\mathbf e_t]=\mathbf 0 .
\]
Therefore, for \(t\neq t'\),
\[
    \mathbb E\left[
    \mathbf e_t^T\mathbf M^{-1}\mathbf e_{t'}
    \right]
    =
    \mathbb E[\mathbf e_t]^T
    \mathbf M^{-1}
    \mathbb E[\mathbf e_{t'}]
    =
    0 .
\]
Thus,
\begin{equation}
    \mathbb E\left[
    \|f^t-\widehat f^t\|_{L^2(\Omega)}^2
    \right]
    =
    \sum_{t\in\mathcal M_t}
    \mathbb E\left[
    \mathbf e_t^T\mathbf M^{-1}\mathbf e_t
    \right].
    \label{eq:error_norm_expectation_element_sum}
\end{equation}

Let's define the kernel
\begin{equation}
    K_h(x)
    =
    \boldsymbol\psi(x)^T
    \mathbf M^{-1}
    \boldsymbol\psi(x)
    =
    \sum_{i=1}^{\mathcal N_t}
    \sum_{j=1}^{\mathcal N_t}
    \psi_i(x)(M^{-1})_{ij}\psi_j(x).
    \label{eq:mass_inverse_kernel}
\end{equation}
Since \(\mathbf M^{-1}\) is symmetric positive definite, we have
\[
    K_h(x)\geq 0.
\]
Using the definition of \(\mathbf e_t\) as given in Eq.~\eqref{eq:element_vector_error_q_mu_term}, 

\begin{align}
\mathbb E\left[
\mathbf e_t^T\mathbf M^{-1}\mathbf e_t
\right]
&=
\mathbb E\left[\left(\frac{|\Omega_t|}{N}\sum_{k=1}^{N}(\mathbf q_t (X_{t,k})-\boldsymbol\mu_t)\right)^T
\mathbf M^{-1}\left(\frac{|\Omega_t|}{N}\sum_{\ell=1}^{N}(\mathbf q_t(X_{t,\ell})-\boldsymbol\mu_t)\right)\right]\notag \\
&=
\frac{|\Omega_t|^2}{N^2}\sum_{k=1}^{N}\sum_{\ell=1}^{N}\mathbb E\left[(\mathbf q_t (X_{t,k})-\boldsymbol\mu_t)^T\mathbf M^{-1}(\mathbf q_t(X_{t,\ell})-\boldsymbol\mu_t)\right].
\end{align}
For \(k\neq \ell\), the random vectors
\(\mathbf q_t (X_{t,k})-\boldsymbol\mu_t\) and \(\mathbf q_t(X_{t,\ell})-\boldsymbol\mu_t\) are independent. Moreover,by the definition of \(\boldsymbol\mu_t\),
\[
    \mathbb E[\mathbf q_t (X_{t,k})-\boldsymbol\mu_t]=\mathbf 0
\]
Therefore, the corresponding cross terms vanish. Hence,
\[
\begin{aligned}
\mathbb E\left[
\mathbf e_t^T\mathbf M^{-1}\mathbf e_t
\right]
&=
\frac{|\Omega_t|^2}{N^2}
\sum_{k=1}^{N}
\mathbb E\left[
(\mathbf q_t (X_{t,k})-\boldsymbol\mu_t)^T
\mathbf M^{-1}
(\mathbf q_t (X_{t,k})-\boldsymbol\mu_t)
\right] \\
&=
\frac{|\Omega_t|^2}{N}
\mathbb E_{\Omega_t}
\left[
(\mathbf q_t(X)-\boldsymbol\mu_t)^T
\mathbf M^{-1}
(\mathbf q_t(X)-\boldsymbol\mu_t)
\right].
\end{aligned}
\]
Expanding the quadratic form and applying the expectation gives
\[
\mathbb E_{\Omega_t}
\left[
(\mathbf q_t(X)-\boldsymbol\mu_t)^T
\mathbf M^{-1}
(\mathbf q_t(X)-\boldsymbol\mu_t)
\right]
=
\mathbb E_{\Omega_t}
\left[
\mathbf q_t(X)^T
\mathbf M^{-1}
\mathbf q_t(X)
\right]
-
\boldsymbol\mu_t^T
\mathbf M^{-1}
\boldsymbol\mu_t .
\]
Since \(\mathbf M^{-1}\) is positive definite,
\[
    \boldsymbol\mu_t^T\mathbf M^{-1}\boldsymbol\mu_t \geq 0.
\]
Therefore,
\[
\mathbb E\left[
\mathbf e_t^T\mathbf M^{-1}\mathbf e_t
\right]
\leq
\frac{|\Omega_t|^2}{N}
\mathbb E_{\Omega_t}
\left[
\mathbf q_t(X)^T
\mathbf M^{-1}
\mathbf q_t(X)
\right].
\]
Using Eq.~\eqref{eq:vector_valued_integrand}, we obtain
\begin{align}
    \mathbf q_t(X)^T
    \mathbf M^{-1}
    \mathbf q_t(X)
    &=
    \left(
    \widetilde z_{\alpha}(X)\boldsymbol\psi(X)
    \right)^T
    \mathbf M^{-1}
    \left(
    \widetilde z_{\alpha}(X)\boldsymbol\psi(X)
    \right)
    \notag \\
    &=
    \widetilde z_{\alpha}(X)^2
    \boldsymbol\psi(X)^T
    \mathbf M^{-1}
    \boldsymbol\psi(X)
    \notag \\
    &=
    \widetilde z_{\alpha}(X)^2K_h(X).
    \label{eq:kernel_appearance}
\end{align}

Therefore,
\begin{align}
    \mathbb E\left[
    \mathbf e_t^T\mathbf M^{-1}\mathbf e_t
    \right]
    &\leq
    \frac{|\Omega_t|^2}{N}
    \mathbb E_{\Omega_t}
    \left[
    \widetilde z_{\alpha}(X)^2K_h(X)
    \right]
    \notag \\
    &=
    \frac{|\Omega_t|}{N}
    \int_{\Omega_t}
    \widetilde z_{\alpha}(x)^2K_h(x)\,dx .
    \label{eq:element_kernel_bound}
\end{align}
Substituting Eq.~\eqref{eq:element_kernel_bound} into
Eq.~\eqref{eq:error_norm_expectation_element_sum}, we obtain
\begin{align}
    \mathbb E\left[
    \|f^t-\widehat f^t\|_{L^2(\Omega)}^2
    \right]
    &\leq
    \sum_{t\in\mathcal M_t}
    \frac{|\Omega_t|}{N}
    \int_{\Omega_t}
    \widetilde z_{\alpha}(x)^2K_h(x)\,dx
    \notag \\
    &\leq
    \frac{|\Omega_t|_{\max}}{N}
    \|K_h\|_{L^\infty(\Omega)}
    \sum_{t\in\mathcal M_t}
    \int_{\Omega_t}
    \widetilde z_{\alpha}(x)^2\,dx
    \notag \\
    &=
    \frac{|\Omega_t|_{\max}}{N}
    \|K_h\|_{L^\infty(\Omega)}
    \|\widetilde z_{\alpha}\|_{L^2(\Omega)}^2 .
    \label{eq:solution_error_kernel_bound}
\end{align}
Using \(g = I_hf^s\) and from Eq.~\eqref{eq:app_definition_z_tilde_alpha}, we can write 
\begin{align}
    \widetilde z_{\alpha}
    &=
    f^s-\alpha g
    \notag \\
    &=
    f^s-\alpha I_hf^s
    \notag \\
    &=
    (1-\alpha)f^s+\alpha(f^s-I_hf^s).
    \label{eq:z_tilde_decomposition}
\end{align}
Therefore,
\begin{align}
    \|\widetilde z_{\alpha}\|_{L^2(\Omega)}^2
    &\leq
    C
    \left[
    (1-\alpha)^2
    \|f^s\|^2_{L^2(\Omega)}
    +
    \alpha^2
    \|f^s - I_hf^s\|^2_{L^2(\Omega)}
    \right].
    \label{eq:z_tilde_l2_first_bound}
\end{align}
Using the interpolation estimate
\[
    \|f^s-I_hf^s\|_{L^2(\Omega)}
    \leq
    C h^{p+1}\|f^s\|_{H^{p+1}(\Omega)},
\]
we obtain
\begin{equation}
    \|\widetilde z_{\alpha}\|_{L^2(\Omega)}^2
    \leq
    C
    \left[
    (1-\alpha)^2
    \|f^s\|^2_{L^2(\Omega)}
    +
    \alpha^2 h^{2p+2}
    \|f^s\|^2_{H^{p+1}(\Omega)}
    \right].
    \label{eq:z_tilde_l2_bound}
\end{equation}

Using Eq.~\eqref{eq:z_tilde_l2_bound} in
Eq.~\eqref{eq:solution_error_kernel_bound}, we get
\begin{equation}
    \mathbb E\left[
    \|f^t-\widehat f^t\|_{L^2(\Omega)}^2
    \right]
    \leq
    \frac{C|\Omega_t|_{\max}}{N}
    \|K_h\|_{L^\infty(\Omega)}
    \left[
    (1-\alpha)^2
    \|f^s\|^2_{L^2(\Omega)}
    +
    \alpha^2 h^{2p+2}
    \|f^s\|^2_{H^{p+1}(\Omega)}
    \right].
    \label{eq:solution_error_bound}
\end{equation}

Using the mesh-size bound from Eq.~\eqref{eq:elemental_volume_equation},
\[
    |\Omega_t|_{\max}
    \leq
    Ch^d,
\]
we obtain
\begin{equation}
    \mathbb E\left[
    \|f^t-\widehat f^t\|_{L^2(\Omega)}^2
    \right]
    \leq
    \frac{C h^d}{N}
    \|K_h\|_{L^\infty(\Omega)}
    \left[
    (1-\alpha)^2
    \|f^s\|^2_{L^2(\Omega)}
    +
    \alpha^2 h^{2p+2}
    \|f^s\|^2_{H^{p+1}(\Omega)}
    \right].
    \label{eq:solution_error_bound_d}
\end{equation}

The total error can be decomposed as
\begin{align}
    f^s-\widehat f^t
    =
    (f^s-f^t)+(f^t-\widehat f^t).
    \label{eq:error_decomposition}
\end{align}
By the Galerkin orthogonality condition,
\begin{equation}
    \langle f^s-f^t,w_h\rangle_{L^2(\Omega)}
    =
    0
    \qquad
    \forall w_h\in \mathcal V_t .
    \label{eq:galerkin_orthogonality}
\end{equation}
Since \(f^t\in \mathcal V_t\) and \(\widehat f^t\in \mathcal V_t\), we have
\(f^t-\widehat f^t\in \mathcal V_t\). Replacing \(w_h\) by
\(f^t-\widehat f^t\) in Eq.~\eqref{eq:galerkin_orthogonality}, we obtain
\[
    \langle f^s-f^t,f^t-\widehat f^t\rangle_{L^2(\Omega)}
    =
    0.
\]
Therefore,
\begin{equation}
    \|f^s-\widehat f^t\|_{L^2(\Omega)}^2
    =
    \|f^s-f^t\|_{L^2(\Omega)}^2
    +
    \|f^t-\widehat f^t\|_{L^2(\Omega)}^2 .
    \label{eq:error_decomposition_orthogonal}
\end{equation}

For finite elements of degree \(p\), the projection error satisfies
\begin{equation}
    \|f^s-f^t\|_{L^2(\Omega)}
    \leq
    C h^{p+1}
    \|f^s\|_{H^{p+1}(\Omega)} .
    \label{eq:projection_error_estimate}
\end{equation}
Taking expectation in Eq.~\eqref{eq:error_decomposition_orthogonal} and using
Eq.~\eqref{eq:solution_error_bound_d} and Eq.~\eqref{eq:projection_error_estimate}, we obtain
\begin{align}
    \mathbb E\left[
    \|f^s-\widehat f^t\|_{L^2(\Omega)}^2
    \right]
    &\leq
    C h^{2p+2}
    \|f^s\|^2_{H^{p+1}(\Omega)}
    \notag \\
    &\quad
    +
    \frac{C h^d}{N}
    \|K_h\|_{L^\infty(\Omega)}
    \left[
    (1-\alpha)^2
    \|f^s\|^2_{L^2(\Omega)}
    +
    \alpha^2 h^{2p+2}
    \|f^s\|^2_{H^{p+1}(\Omega)}
    \right].
    \label{eq:error_decomposition_norm}
\end{align}

From Eq.~\eqref{eq:error_decomposition_norm}, we see that the total error consists of two contributions: the discretization error associated with the finite-element approximation of \(f^s\) on the target mesh \(\mathcal M_t\)  and the sampling error introduced by the control-variate Monte Carlo approximation.
\section{Conservation error bound}\label{app:conservation_error_bound}

Let the conservation error be defined by
\begin{equation}
    E_{\mathrm{cons}}
    =
    \left|
    \int_\Omega
    \left(
    f^s-\widehat f^t
    \right)
    \,dx
    \right|.
    \label{eq:conservation_error_definition}
\end{equation}
Applying Eq.~\eqref{eq:error_decomposition} and from the conservation definition \(\int _\Omega (f^s - f^t)\, dx = 0\), Eq.~\eqref{eq:conservation_error_definition} becomes
\begin{equation}
    E_{\mathrm{cons}}
    =
    \left|
    \int_\Omega
    \left(
    f^t-\widehat f^t
    \right)
    \,dx
    \right|.
    \label{eq:conservation_error_definition_decomposed}
\end{equation}
Assume that the constant function \(1 \in \mathcal V_t\). Testing the exact and approximate Galerkin systems with \(w_h = 1\) in Eq.~\eqref{eq:galerkin_orthogonality} gives
\[
    \int_\Omega f^t(x)\,dx
    =
    I(f^s),
    \qquad
    \int_\Omega \widehat f^t(x)\,dx
    =
    \widehat I (f^s).
\]
Therefore,
\begin{equation}
    E_{\mathrm{cons}}
    =
    \left|
    I(f^s)
    -
    \widehat I (f^s)
    \right|.
    \label{eq:conservation_equals_cv_integral_error}
\end{equation}

Following the control-variate construction used in
Eq.~\eqref{eq:cv_load_expanded}, the corresponding global integral estimator is
\begin{equation}
    \widehat I (f^s)
    =
    \alpha
    \int_\Omega g(x)\,dx
    +
    \sum_{t\in\mathcal M_t}
    \frac{|\Omega_t|}{N}
    \sum_{k=1}^{N}
    \widetilde z_{\alpha} (X_{t,k}).
    \label{eq:cv_global_integral_estimator}
\end{equation}
Since the deterministic term is computed exactly, the conservation error is
only due to the Monte Carlo approximation of the residual integral. Applying
the same element-wise Monte Carlo variance argument used in the accuracy
estimate, but now for the scalar residual \(\widetilde z_\alpha\), gives
\begin{equation}
    \mathbb E\left[
    E_{\mathrm{cons}}^2
    \right]
    \leq
    \frac{|\Omega_t|_{\max}}{N}
    \|\widetilde z_{\alpha}\|_{L^2(\Omega)}^2 .
    \label{eq:conservation_residual_bound}
\end{equation}
Using Eq.~\eqref{eq:z_tilde_decomposition}, Eq.~\eqref{eq:z_tilde_l2_bound}, and Eq.~\eqref{eq:elemental_volume_equation}, we obtain
\begin{equation}
    \mathbb E\left[
    E_{\mathrm{cons}}^2
    \right]
    \leq
    \frac{C h^d}{N}
    \left[
    (1-\alpha)^2
    \|f^s\|_{L^2(\Omega)}^2
    +
    \alpha^2 h^{2p+2}
    \|f^s\|_{H^{p+1}(\Omega)}^2
    \right].
    \label{eq:conservation_bound}
\end{equation}





 \bibliography{references}

@ARTICLE{JiaoHeath2004,
  author  = {Jiao, X. and Heath, M. T.},
  title   = {Common-Refinement-Based Data Transfer between Non-Matching Meshes in Multiphysics Simulations},
  journal = {Int. J. Numer. Methods Eng.},
  volume  = {61},
  number  = {14},
  year    = {2004},
  pages   = {2402--2427},
  doi     = {10.1002/nme.1147}
}

@ARTICLE{FarrellMaddison2011,
  author  = {Farrell, P. E. and Maddison, J. R.},
  title   = {Conservative interpolation between volume meshes by local Galerkin projection},
  journal = {Computer Methods in Applied Mechanics and Engineering},
  volume  = {211--212},
  year    = {2011},
  pages   = {1171--1183}
}

@ARTICLE{Slattery2016,
  author  = {Slattery, Stuart R.},
  title   = {Mesh-Free Data Transfer Algorithms for Partitioned Multiphysics Problems: Conservation, Accuracy, and Parallelism},
  journal = {J. Comput. Phys.},
  volume  = {307},
  year    = {2016},
  pages   = {164--188},
  doi     = {10.1016/j.jcp.2015.11.055}
}

@ARTICLE{KEAST1986339,
title = {Moderate-degree tetrahedral quadrature formulas},
journal = {Computer Methods in Applied Mechanics and Engineering},
volume = {55},
number = {3},
pages = {339-348},
year = {1986},
issn = {0045-7825},
doi = {https://doi.org/10.1016/0045-7825(86)90059-9},
url = {https://www.sciencedirect.com/science/article/pii/0045782586900599},
author = {Patrick Keast}
}

@ARTICLE{Dunavant1985,
  author  = {Dunavant, David A.},
  title   = {High Degree Efficient Symmetrical Gaussian Quadrature Rules for the Triangle},
  journal = {Int. J. Numer. Methods Eng.},
  volume  = {21},
  year    = {1985},
  pages   = {1129--1148},
  doi     = {10.1002/nme.1620070316}
}

@ARTICLE{POWELL2015340,
title = {An exact general remeshing scheme applied to physically conservative voxelization},
journal = {Journal of Computational Physics},
volume = {297},
pages = {340-356},
year = {2015},
issn = {0021-9991},
doi = {https://doi.org/10.1016/j.jcp.2015.05.022},
url = {https://www.sciencedirect.com/science/article/pii/S0021999115003563},
author = {Devon Powell and Tom Abel}
}

@techreport{Powell2015R3D,
  author      = {Powell, Devon},
  title       = {R3D: Software for Fast, Robust Geometric Operations in 3D and 2D},
  number      = {LA-UR-15-26964},
  year        = {2015},
  month       = aug,
}

@book{Eberly2006,
  author    = {Eberly, David H.},
  title     = {3D Game Engine Design: A Practical Approach to Real-Time Computer Graphics},
  edition   = {2nd},
  year      = {2007},
  publisher = {CRC Press},
  address   = {Boca Raton},
  pages     = {1040},
  doi       = {10.1201/b18212},
  isbn      = {9780429176548},
}

@article{JiaoHeathCommonRefinement,
  author  = {Jiao, Xiangmin and Heath, Michael T.},
  title   = {Overlaying Surface Meshes, Part I: Algorithms},
  journal = {International Journal of Computational Geometry \& Applications},
  volume  = {14},
  number  = {6},
  pages   = {379--402},
  year    = {2004},
  doi     = {10.1142/S0218195904001512}
}

@article{geuzaine2009gmsh,
  author  = {Geuzaine, Christophe and Remacle, Jean-François},
  title   = {Gmsh: A Three-Dimensional Finite Element Mesh Generator with Built-in Pre- and Post-Processing Facilities},
  journal = {International Journal for Numerical Methods in Engineering},
  year    = {2009},
  volume  = {79},
  number  = {11},
  pages   = {1309--1331},
  doi     = {10.1002/nme.2579}
}

@article{merson2025pcms,
  author        = {Merson, Jacob S. and Smith, Cameron W. and Shephard, Mark S. and Hasan, Fuad and Paudel, Abhiyan and Castillo-Crooke, Angel and Mathew, Joyal and Elahi, Mohammad},
  title         = {{PCMS}: {P}arallel {C}oupler for {M}ultimodel {S}imulations},
  journal       = {arXiv preprint arXiv:2510.18838},
  year          = {2025},
  month         = oct,
  archivePrefix = {arXiv},
  eprint        = {2510.18838},
  primaryClass  = {cs.DC},
  doi           = {10.48550/arXiv.2510.18838},
  url           = {https://arxiv.org/abs/2510.18838},
  note          = {Submitted on 21 Oct 2025}
}

@article{bungartzPreCICEFullyParallel2016,
    title = {{preCICE} – {A} fully parallel library for multi-physics surface coupling},
    volume = {141},
    issn = {00457930},
    url = {https://linkinghub.elsevier.com/retrieve/pii/S0045793016300974},
    doi = {10.1016/j.compfluid.2016.04.003},
    language = {en},
    urldate = {2022-07-01},
    journal = {Computers \& Fluids},
    author = {Bungartz, Hans-Joachim and Lindner, Florian and Gatzhammer, Bernhard and Mehl, Miriam and Scheufele, Klaudius and Shukaev, Alexander and Uekermann, Benjamin},
    month = dec,
    year = {2016},
    pages = {250--258},
}

@article{chourdakisPreCICEV2Sustainable2022,
    title = {{preCICE} v2: {A} sustainable and user-friendly coupling library},
    volume = {2},
    issn = {2732-5121},
    shorttitle = {{preCICE} v2},
    url = {https://open-research-europe.ec.europa.eu/articles/2-51/v1},
    doi = {10.12688/openreseurope.14445.1},
    language = {en},
    urldate = {2022-07-06},
    journal = {Open Research Europe},
    author = {Chourdakis, Gerasimos and Davis, Kyle and Rodenberg, Benjamin and Schulte, Miriam and Simonis, Frédéric and Uekermann, Benjamin and Abrams, Georg and Bungartz, Hans-Joachim and Cheung Yau, Lucia and Desai, Ishaan and Eder, Konrad and Hertrich, Richard and Lindner, Florian and Rusch, Alexander and Sashko, Dmytro and Schneider, David and Totounferoush, Amin and Volland, Dominik and Vollmer, Peter and Koseomur, Oguz Ziya},
    month = apr,
    year = {2022},
    pages = {51},
}

@article{keyesMultiphysicsSimulationsChallenges2013,
    title = {Multiphysics simulations: {Challenges} and opportunities},
    volume = {27},
    issn = {1094-3420, 1741-2846},
    shorttitle = {Multiphysics simulations},
    url = {http://journals.sagepub.com/doi/10.1177/1094342012468181},
    doi = {10.1177/1094342012468181},
    language = {en},
    number = {1},
    urldate = {2023-03-06},
    journal = {The International Journal of High Performance Computing Applications},
    author = {Keyes, David E and McInnes, Lois C and Woodward, Carol and Gropp, William and Myra, Eric and Pernice, Michael and Bell, John and Brown, Jed and Clo, Alain and Connors, Jeffrey and Constantinescu, Emil and Estep, Don and Evans, Kate and Farhat, Charbel and Hakim, Ammar and Hammond, Glenn and Hansen, Glen and Hill, Judith and Isaac, Tobin and Jiao, Xiangmin and Jordan, Kirk and Kaushik, Dinesh and Kaxiras, Efthimios and Koniges, Alice and Lee, Kihwan and Lott, Aaron and Lu, Qiming and Magerlein, John and Maxwell, Reed and McCourt, Michael and Mehl, Miriam and Pawlowski, Roger and Randles, Amanda P and Reynolds, Daniel and Rivière, Beatrice and Rüde, Ulrich and Scheibe, Tim and Shadid, John and Sheehan, Brendan and Shephard, Mark and Siegel, Andrew and Smith, Barry and Tang, Xianzhu and Wilson, Cian and Wohlmuth, Barbara},
    month = feb,
    year = {2013},
    pages = {4--83},
}

@article{farrellConservativeInterpolationUnstructured2009,
    title = {Conservative interpolation between unstructured meshes via supermesh construction},
    volume = {198},
    issn = {00457825},
    url = {https://linkinghub.elsevier.com/retrieve/pii/S0045782509001315},
    doi = {10.1016/j.cma.2009.03.004},
    language = {en},
    number = {33-36},
    urldate = {2022-03-19},
    journal = {Computer Methods in Applied Mechanics and Engineering},
    author = {Farrell, P.E. and Piggott, M.D. and Pain, C.C. and Gorman, G.J. and Wilson, C.R.},
    month = jul,
    year = {2009},
    pages = {2632--2642},
}

@article{deboerComparisonConservativeConsistent2008,
    title = {Comparison of conservative and consistent approaches for the coupling of non-matching meshes},
    volume = {197},
    issn = {0045-7825},
    url = {https://www.sciencedirect.com/science/article/pii/S0045782508001916},
    doi = {10.1016/j.cma.2008.05.001},
    number = {49},
    urldate = {2025-08-25},
    journal = {Computer Methods in Applied Mechanics and Engineering},
    author = {de Boer, A. and van Zuijlen, A. H. and Bijl, H.},
    month = sep,
    year = {2008},
    pages = {4284--4297},
}

@article{jaimanAssessmentConservativeLoad2005,
    title = {Assessment of conservative load transfer for fluid-solid interface with non-matching meshes},
    volume = {64},
    issn = {0029-5981, 1097-0207},
    url = {https://onlinelibrary.wiley.com/doi/10.1002/nme.1434},
    doi = {10.1002/nme.1434},
    language = {en},
    number = {15},
    urldate = {2022-03-19},
    journal = {International Journal for Numerical Methods in Engineering},
    author = {Jaiman, R. K. and Jiao, X. and Geubelle, P. H. and Loth, E.},
    month = dec,
    year = {2005},
    pages = {2014--2038},
}

@article{shashkovRemappingMeshesIsoparametric,
    title = {Remapping between meshes with isoparametric cells: a case study},
    language = {en},
    author = {Shashkov, Mikhail and Lipnikov, Konstantin},
}

@article{hermesHighorderSolutionTransfer2025,
    title = {High-order {Solution} {Transfer} between {Curved} {Triangular} {Meshes}},
    volume = {20},
    issn = {2157-5452, 1559-3940},
    url = {http://arxiv.org/abs/1810.06806},
    doi = {10.2140/camcos.2025.20.1},
    number = {1},
    urldate = {2025-12-14},
    journal = {Communications in Applied Mathematics and Computational Science},
    author = {Hermes, Danny and Persson, Per-Olof},
    month = jan,
    year = {2025},
    note = {arXiv:1810.06806 [math]},
    pages = {1--27},
}

@phdthesis{ibanezCONFORMALMESHADAPTATION,
    address = {Troy, NY},
    type = {{PhD}},
    title = {conformal mesh adaptation on heterogeneous supercomputers},
    language = {en},
    school = {Rensselaer Polytechnic Institute},
    author = {Ibanez, Daniel Alejandro},
    month = nov,
    year = {2016},
}

@article{osada2002shape,
  author    = {Osada, Robert and Funkhouser, Thomas and Chazelle, Bernard and Dobkin, David},
  title     = {Shape Distributions},
  journal   = {ACM Transactions on Graphics (TOG)},
  volume    = {21},
  number    = {4},
  pages     = {807--832},
  year      = {2002},
  month     = {October},
  publisher = {ACM New York, NY, USA},
  doi       = {10.1145/571647.571648}
}

@article{adamsProjectionMethodParticle2026,
    title = {A {Projection} {Method} for {Particle} {Resampling}},
    volume = {321},
    issn = {0010-4655},
    url = {https://www.sciencedirect.com/science/article/pii/S0010465526000068},
    doi = {10.1016/j.cpc.2026.110024},
    language = {en},
    journal = {Computer Physics Communications},
    publisher = {Elsevier},
    author = {Adams, Mark  F. and Knepley, Matthew and Finn, Daniel  S. and Pusztay, Joseph  V.},
    year = {2026},
}

@misc{hasanGPUAccelerationMonte2025,
    title = {{GPU} {Acceleration} of {Monte} {Carlo} {Tallies} on {Unstructured} {Meshes} in {OpenMC} with {PUMI}-{Tally}},
    url = {http://arxiv.org/abs/2504.19048},
    doi = {10.48550/arXiv.2504.19048},
    urldate = {2025-06-20},
    publisher = {arXiv},
    author = {Hasan, Fuad and Smith, Cameron W. and Shephard, Mark S. and Churchill, R. Michael and Wilkie, George J. and Romano, Paul K. and Shriwise, Patrick C. and Merson, Jacob S.},
    month = apr,
    year = {2025},
    note = {Submitted.},
}

@inproceedings{mersonSpatiallyContinuousFunctional,
    address = {Torino, Italy},
    title = {Spatially {Continuous} {Functional} {Expansion} {Tallies} on {Unstructured} {Meshes}},
    author = {Merson, Jacob S. and Belanger, Hunter and Singh, Parth},
    note = {Under Review.},
}

@inproceedings{slatteryDataTransferKit2013,
    address = {Sun Valley, ID},
    title = {The {Data} {Transfer} {Kit}: {A} {Geometric} {Rendezvous}-{Based} {Tool} for {Multiphysics} {Data} {Transfer}},
    language = {en},
    publisher = {American Nuclear Society},
    author = {Slattery, S R and Wilson, P P H and Pawlowski, R P},
    month = may,
    year = {2013},
    pages = {11},
}

@article{novakCoupledMonteCarlo2022,
    title = {Coupled {Monte} {Carlo} and thermal-fluid modeling of high temperature gas reactors using {Cardinal}},
    volume = {177},
    issn = {03064549},
    url = {https://linkinghub.elsevier.com/retrieve/pii/S0306454922003450},
    doi = {10.1016/j.anucene.2022.109310},
    language = {en},
    urldate = {2026-02-02},
    journal = {Annals of Nuclear Energy},
    author = {Novak, A.J. and Andrs, D. and Shriwise, P. and Fang, J. and Yuan, H. and Shaver, D. and Merzari, E. and Romano, P.K. and Martineau, R.C.},
    month = nov,
    year = {2022},
    pages = {109310},
}

@article{zienkiewiczSuperconvergentPatchRecovery1992a,
    title = {The superconvergent patch recovery anda posteriori error estimates. {Part} 2: {Error} estimates and adaptivity},
    volume = {33},
    issn = {0029-5981, 1097-0207},
    shorttitle = {The superconvergent patch recovery anda posteriori error estimates. {Part} 2},
    url = {http://doi.wiley.com/10.1002/nme.1620330703},
    doi = {10.1002/nme.1620330703},
    language = {en},
    number = {7},
    urldate = {2019-05-22},
    journal = {International Journal for Numerical Methods in Engineering},
    author = {Zienkiewicz, O. C. and Zhu, J. Z.},
    month = may,
    year = {1992},
    pages = {1365--1382},
}

@article{zienkiewiczSuperconvergentPatchRecovery1992b,
    title = {The superconvergent patch recovery ({SPR}) and adaptive finite element refinement},
    volume = {101},
    issn = {00457825},
    url = {http://linkinghub.elsevier.com/retrieve/pii/004578259290023D},
    doi = {10.1016/0045-7825(92)90023-D},
    language = {en},
    number = {1-3},
    urldate = {2018-12-10},
    journal = {Computer Methods in Applied Mechanics and Engineering},
    author = {Zienkiewicz, O.C. and Zhu, J.Z.},
    month = dec,
    year = {1992},
    pages = {207--224},
}

@article{dingPerformanceEvaluationGPUAccelerated2018,
    title = {Performance {Evaluation} of {GPU}-{Accelerated} {Spatial} {Interpolation} {Using} {Radial} {Basis} {Functions} for {Building} {Explicit} {Surfaces}},
    volume = {46},
    issn = {1573-7640},
    url = {https://doi.org/10.1007/s10766-017-0538-6},
    doi = {10.1007/s10766-017-0538-6},
    language = {en},
    number = {5},
    urldate = {2023-09-27},
    journal = {International Journal of Parallel Programming},
    author = {Ding, Zengyu and Mei, Gang and Cuomo, Salvatore and Xu, Nengxiong and Tian, Hong},
    month = oct,
    year = {2018},
    pages = {963--991},
}

@inproceedings{schneiderDataParallelRadialBasisFunction2023,
    title = {Data-{Parallel} {Radial}-{Basis} {Function} {Interpolation} in {preCICE}},
    url = {https://www.scipedia.com/public/2023r},
    doi = {10.23967/c.coupled.2023.016},
    language = {en},
    urldate = {2025-03-08},
    booktitle = {10th edition of the {International} {Conference} on {Computational} {Methods} for {Coupled} {Problems} in {Science} and {Engineering}},
    publisher = {CIMNE},
    author = {Schneider, D. and Shrader, T. and Uekermann, B.},
    year = {2023},
}

@article{morricalAttributeAwareRBFsInteractive2023,
    title = {Attribute-{Aware} {RBFs}: {Interactive} {Visualization} of {Time} {Series} {Particle} {Volumes} {Using} {RT} {Core} {Range} {Queries}},
    issn = {1077-2626, 1941-0506, 2160-9306},
    shorttitle = {Attribute-{Aware} {RBFs}},
    url = {https://ieeexplore.ieee.org/document/10296023/},
    doi = {10.1109/TVCG.2023.3327366},
    language = {en},
    urldate = {2023-11-27},
    journal = {IEEE Transactions on Visualization and Computer Graphics},
    author = {Morrical, Nate and Zellmann, Stefan and Sahistan, Alper and Shriwise, Patrick and Pascucci, Valerio},
    year = {2023},
    pages = {1--11},
}

@article{zhang2023unstructured,
  title   = {Development of an unstructured mesh gyrokinetic particle-in-cell code for exascale fusion plasma simulations on GPUs},
  author  = {Zhang, C. and Diamond, G. and Smith, C. W. and Shephard, M. S.},
  journal = {Computer Physics Communications},
  volume  = {291},
  pages   = {108824},
  year    = {2023},
  doi     = {10.1016/j.cpc.2023.108824},
  url     = {https://doi.org/10.1016/j.cpc.2023.108824}
}

@article{gilesMultilevelMonteCarlo,
    title = {Multilevel {Monte} {Carlo} methods},
    language = {en},
    author = {Giles, Michael B},
}






\end{document}